%% file: main_arxiv.tex
\documentclass[aps,prl,twocolumn,superscriptaddress]{revtex4-2}

\usepackage{graphicx,amssymb,amstext,amsmath,amsbsy,greekctr,color}
\usepackage{bbm}
\usepackage{listings}

\usepackage[utf8]{inputenc} 
\usepackage{hyperref}       
\usepackage{url}            
\usepackage{booktabs}       
\usepackage{amsfonts}       
\usepackage{nicefrac}       
\usepackage{microtype}      
\usepackage{lipsum}		
\usepackage{natbib}

\begin{document}

\title{Understanding Reaction Mechanisms from Start to Finish}
\author{Rik S. Breebaart}
\affiliation{Van 't Hoff Institute for Molecular Sciences, Universiteit van Amsterdam, Science Park 904, 1098 XH Amsterdam, The Netherlands}
\author{Gianmarco Lazzeri}
\affiliation{Institute of Computer Science, Goethe University Frankfurt}
\affiliation{Frankfurt Institute for Advanced Studies}
\author{Roberto Covino}
\affiliation{Institute of Computer Science, Goethe University Frankfurt}
\affiliation{Frankfurt Institute for Advanced Studies}
\email{covino@fias.uni-frankfurt.de}

\author{Peter G. Bolhuis*}
\affiliation{Van 't Hoff Institute for Molecular Sciences, Universiteit van Amsterdam, Science Park 904, 1098 XH Amsterdam, The Netherlands}
\email{p.gbolhuis@uva.nl}

\begin{abstract}
Understanding mechanisms of rare but important events in complex molecular systems, such as protein folding or ligand (un)binding, requires accurately mapping  transition paths from an initial to a final state. The committor is the ideal reaction coordinate for this purpose, but calculating it for high-dimensional, nonlinear systems has long been considered intractable. Here, we introduce an iterative path sampling strategy for computing the committor function for systems with high free energy barriers. We start with an initial guess
to define isocommittor interfaces for transition interface sampling. The resulting path ensemble is then reweighted and used to train a neural network, yielding a more accurate committor model. This process is repeated until convergence, effectively solving the long-standing circular problem in enhanced sampling where a good reaction coordinate is needed to generate efficient sampling, and vice-versa. The final, converged committor model can be interrogated to extract  mechanistic insights. We demonstrate the power of our method on a benchmark 2D potential and a  more complex   host-guest (un)binding process in explicit solvent.

\end{abstract}


\maketitle

Molecular dynamics (MD) simulations enable prediction and  understanding of  transitions between metastable states in complex biomolecular and condensed matter systems, but are hampered by the long timescales involved. 
Enhanced sampling techniques can overcome these limitations, but usually require a good reaction coordinate, capable of faithfully capturing the transition dynamics. 
The latter is not always available, leading to inaccurate estimates of free energy barriers, kinetics, and mechanisms \cite{Chandler1998}.
In the last decades it has become clear that the  optimal reaction coordinate is the committor function, 
$p_B({x})$, which predicts for a configuration ${x}$ somewhere in the barrier region the probability of reaching the final product state B before A \cite{Bolhuis2000,E2010}. 
Knowing the committor not only leads to more effective enhanced  sampling, but also improved rate estimates  and above all creates mechanistic understanding along the  reactive process \cite{Lazzeri2025}.

Obtaining a full description of the committor function has long been deemed intractable, as direct evaluation of the committor by shooting off swarms of MD trajectories, and establish whether they reach state B before state A, is extremely costly. However, in the past few decades several approaches have been developed that aim to approximate the committor based on limited trajectory sampling data, e.g., using transition path sampling (TPS)~\cite{Dellago1998,Dellago2002,Bolhuis2002,Peters2006,lechnerNonlinearReactionCoordinate2010,Mullen2014,Peters2016}. Moreover, the rise of machine learning techniques has made accurate modeling of the high dimensional committor function and extracting mechanistic interpretation  a more tractable target~\cite{maAutomaticMethodIdentifying2005,vanErp2016,Khoo2018,Frassek2021,Mori2020,Wang2020,
Rotskoff2022,Jung2023,lazzeriMolecularFreeEnergies2023a,kangComputingCommittorCommittor2024,Chen2023,Mitchell2024,Megas2025}.
Still, learning the full committor function poses significant challenges. 

\begin{figure}[t]
    \centering
    \includegraphics[width=1\linewidth]{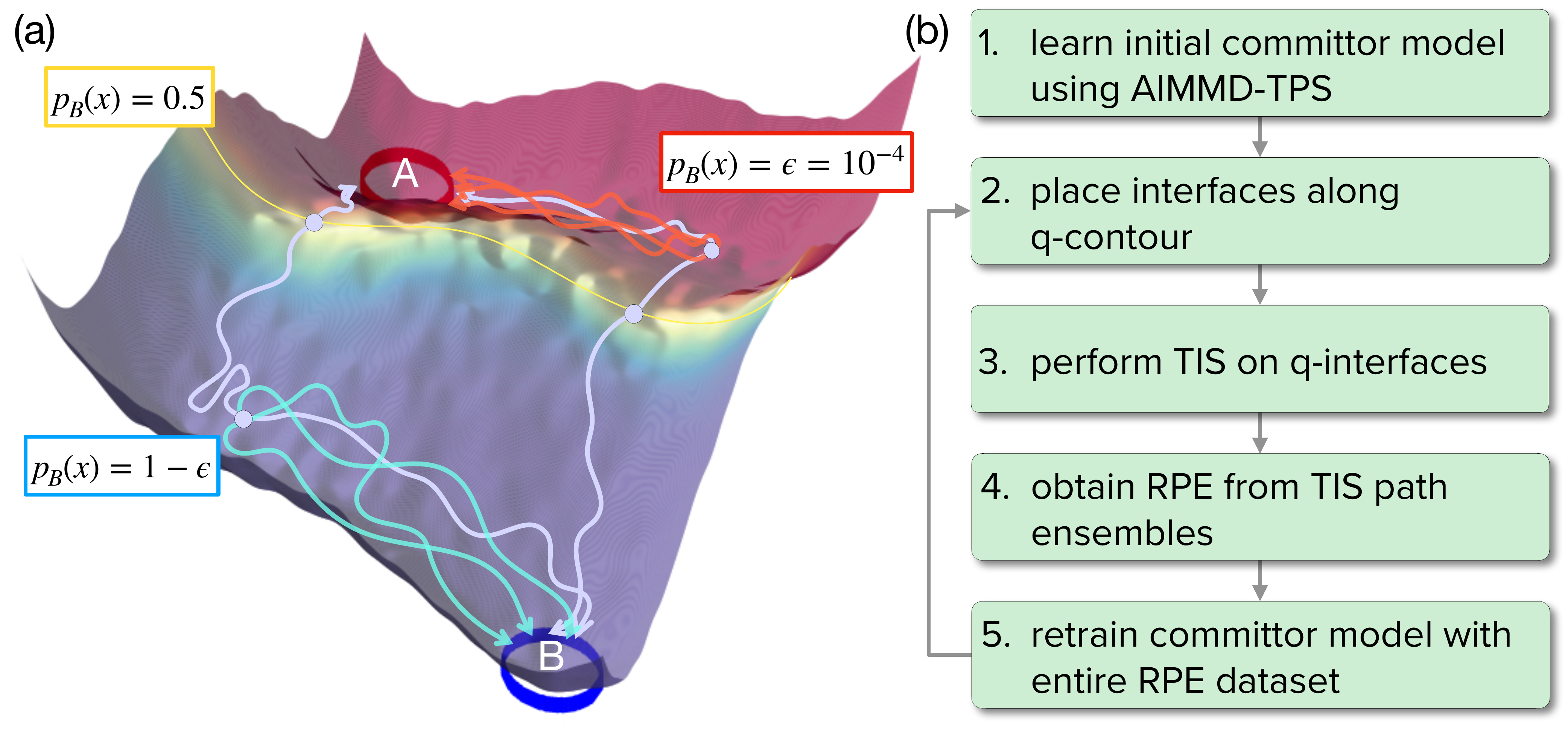}
    \caption{(a) schematic  of the problem in terms of a rough  energy landscape:  away from the TS (yellow region) reactive paths can encounter interesting mechanistic intermediates and changes in relevant variables, characterized by very low  committor values (red, blue regions). (b)  AIMMD-TIS algorithm. }
    \label{fig:scheme}
\end{figure}

First, the committor $p_B(x)$ in general behaves like a sigmoidal function:
it starts at very low values near the initial state, gradually rises, and then rapidly increases as it approaches the transition state (TS), where it crosses \( p_B = 0.5 \). After this point, it continues to slowly rise towards unity while approaching the final state \cite{Bolhuis2002,E2010}.
In particular, for  high barriers ($>10 k_BT$), such functions become essentially step-like, with sharp changes near the transition state and little variation near the stable basins where $p_B\approx0 $ or $1$. This makes it difficult to model across the full configuration space. 
Second, the dimensionality  can be high, as many features might play a role in the committor---in principle even up to the $3N$ dimensional cartesian coordinates. Hence, machine learning and dimensionality reduction are crucial for committor modeling.
A third challenge is the acquisition of data to be used to train the committor function. Direct evaluation of the committor by shooting off many MD trajectories for the entire configuration space would require near infinite computational resources. 
As a result, many approaches  prioritize the transition region, where information gain per sample is high, to obtain mechanistic information, also because the TS traditionally is a major objective for mechanistic studies. 
However, this may lead to missing important dynamics away from the TS. 
While the TS is undoubtedly important, (much) more information can be gleaned along the   entire reactive process. Reactive trajectories may traverse regions of phase space on their way to or from the dynamical bottleneck  along different channels, linger briefly in a transient state, or move in directions orthogonal to the prospect reaction coordinate at the TS (see Fig.~\ref{fig:scheme}a for an illustration).  
The committor function, particularly the isocommittor surfaces, encapsulates  this information about the dynamics beyond the TS.
However,  estimating accurately committor values far away from the  TS, 
where $p_B$ approaches 0 or 1, 
is often prohibitively expensive.  For example, 
estimating $p_B\approx10^{-3}$ 
reliably would  require   thousands of independent  trajectories for a single configuration alone. 

Here, we aim  to learn the committor and extract its important ingredients  along the {\em entire} reactive process over a high barrier, based on the sampling of unbiased MD trajectories.
To do so, we address the above  challenges by building upon recent innovations in 
TPS-based techniques, such as  
Artificial Intelligence for Molecular Mechanistic Discovery (AIMMD), which 
leverages machine learning to enhance TPS efficiency and simultaneously gain mechanistic understanding \cite{Jung2023}. AIMMD learns 
 a neural network-based logit-committor model $q(x|\theta)$ to describe the committor 
 $\tilde{p}_B(x) = \left[ 1 + e^{-q(x|\theta)} \right]^{-1}$, where $\theta$ is a set of parameters.
In particular, we propose here to significantly augment AIMMD by  employing transition interface sampling (TIS) \cite{vanErp2003,Cabriolu2017,swensonOpenPathSamplingPythonFramework2019} with isocommittor surfaces as interfaces. 
These isocommittor interfaces are defined as $q(x|\theta)=q$ in the  logit-committor. 
 Next, we  compute 
  the so-called reweighted path ensemble (RPE) \cite{Rogal2010}, 
which combines trajectory sets obtained in TIS and reweights them to estimate the full equilibrium path ensemble, the equivalent to  a (very) long simulation.
As the fate of each trajectory (reactive or not) is known, all configurations $x$ in the trajectory can serve as shooting points for  committor learning, increasing the training data by a factor proportional to the average path length.
We  use
the RPE 
to train the committor model with all these
configurations, yielding a  
much more accurate representation of the system's behavior, including in regions far away from the TS.
This  model is then used to redefine   interfaces in q-space. Subsequent TIS simulations and retraining leads to a converged  model $q(x|\theta)$.   
To bootstrap the iterative process we 
start with a short AIMMD-TPS procedure to train an initial committor model, 
avoiding the need to select a prior interface CV manually.

{\bf The AIMMD-TIS algorithm}. 
Below, we explain the algorithm, summarized in Fig.~\ref{fig:scheme}b, in more detail, followed by  exemplary applications on  a 2D potential and a complex molecular  host-guest (un)binding process. 

{\it TIS (step 2-3):}  TIS places $n$ non-intersecting interfaces  $\lambda_0, \lambda_1,\ldots,\lambda_{n}$  between states A and B \cite{vanErp2003,Cabriolu2017,swensonOpenPathSamplingPythonFramework2019}, defined  by a monotonically increasing progress variable $\lambda(x)$ 
 where $\lambda_A = \lambda_0$ and $\lambda_B=\lambda_n$ describe the boundary interfaces for state A and B respectively.
With an initial committor model, obtained with AIMMD-TPS \cite{Jung2023},
interfaces can be placed along isocommittor surfaces $\lambda(x)=q(x|\theta)$. 
Stable state boundaries  $\lambda_{A,B}$,  defined  in
the descriptor (CV) space, must not intersect with the interfaces.
An exact committor model ensures this  
automatically, but with approximate models, overlap must be carefully avoided.
We define bounds in committor space based on stable state data:
$q_{\text{max}}^{A} = \max_{x \in A} q(x|\theta), \quad q_{\text{min}}^{B} = \min_{x \in B} q(x|\theta).$

Interfaces $\lambda_i = q_i$ are placed uniformly between these bounds to ensure good overlap of consecutive ensembles.
Performing TIS   
employing the 2-way shooting move for successive interfaces $\lambda$ leads to the all important  crossing probability, $P_{A}(\lambda_{i+1}|\lambda_{i})$, 
the probability that  a trajectory starting in A and crossing $\lambda_{i}$ also crosses $\lambda_{i+1}$.  The reverse B to A transition is treated in a similar way.
 
{\it RPE (step 4)}:  Reweighting the forward and backward interface path ensembles  based on matching the crossing probabilities $P_{A,B}(\lambda|\lambda_i)$ with WHAM \cite{ferrenbergOptimizedMonteCarlo1989,shirtsStatisticallyOptimalAnalysis2008} yields the overall crossing probability $P_{A,B}(\lambda|\lambda_1)$, as well as weights $w(\bf x )$ for each sampled path  ${\bf{x}}=\{x_0,x_1,\dots x_L \}$, with $x_i$ being the $i$-th configuration in the trajectory of length $L$. 
Scaling the forward and backward ensemble weights by matching reactive paths weights $(w_{AB}=w_{BA})$, finally leads to the RPE, representing the 
complete equilibrium path ensemble.
Assigning all configurations $x_i$ of each pathway $\bf x$ of the RPE a  weight $w_i = w[\bf x]$,  allows to evaluate
ensemble averages over the RPE of any observable  as $\langle O \rangle\approx \sum_{x_i\in \text{RPE}}w_i O(x_i)$. This includes time dependent averages over the trajectories $\langle O({\bf x}(t))\rangle$. 

The stable state simulations contribute to the RPE by  matching   the crossings through the first $\lambda_{1}$ or last $\lambda_{n-1}$  interface. They also
provide respectively  
 the flux $\phi_{A1},~\phi_{B(n-1)}$ of trajectories going out of A, B.  
We then obtain the rate constants  as $k_{AB}=\phi_{A1}P_A(\lambda_B|\lambda_{1})$ and $k_{BA}=\phi_{B(n-1)}P_B(\lambda_A|\lambda_{n-1})$.

{\it Training the committor model (step 5):}
All reweighted configurations serve as training data points, 
each with their respective weight contributing to the loss function. Each data-point $v_i = \{x_i, w_i, (n_A^i, n_B^i)\}$ includes the configuration $x_i$, the RPE weight $w_i$ and the 2-way shot result of its path
$(n_A^i,n_B^i)=\{0,1,2\}$.
Configurations from reactive AB paths have much lower weights compared to those from non-reactive AA (or BB) paths, 
which dominate near stable states, but still contribute, leading to low but  accurate committor values close to the stable states.
The loss function for  (re)training the committor model on the RPE data is a weighted likelihood:
\begin{align}\hspace{-0.3cm}
        \mathcal{L}_{\rm wl} = -\hspace{-0.2cm}\sum_{v_i\in \text{RPE}}  \hspace{-0.2cm} w_i \left[n_A^i\ln(1\hspace{-0.05cm}-\hspace{-0.05cm}\tilde{p}_B(x_i|\theta))
        +n_B^i\ln \tilde{p}_B(x_i|\theta)\right] \hspace{-0.3cm}
\end{align}
The committor should monotonically increase from A to B \cite{E2010}. 
Therefore, an additional loss term is added
\begin{equation}
\mathcal{L}_{\rm smooth} = \sum_{x_i\in RPE}\frac{1}{N}\left|\nabla_\xi q(x_i|\theta)\right|^2
\end{equation}
that enforces smoothness in the model\footnote{This term  is reminiscent of the variational principle
for solving the backward Kolmogorov equation in the overdamped regime \cite{E2010,kangComputingCommittorCommittor2024}.}, where $\xi$ denotes a reduced  descriptor-space, or the full coordinate space $x$ itself. 
L1 regularization is  applied to 
reduce the influence of irrelevant degrees of freedom.
The  total loss function is: 
\begin{equation}
    \mathcal{L} = \mathcal{L}_{\rm wl}+\alpha \mathcal{L}_{\rm smooth} + \eta\mathcal{L}_{l1}
\end{equation}
where $\alpha$ and $\eta$ are hyper-parameters balancing the contribution of the different loss terms. 
We find each point in the RPE roughly contributes equally to $\mathcal{L}_{\rm wl}$  (See SI). 

{\it Iterating the process:}
The process is iterated to refine the committor model. 
Each iteration uses an improved RPE-trained committor model 
to place interfaces, enhancing the overlap between consecutive interface ensembles. 

\begin{figure}[t]
    \centering
        \includegraphics[width=\linewidth]{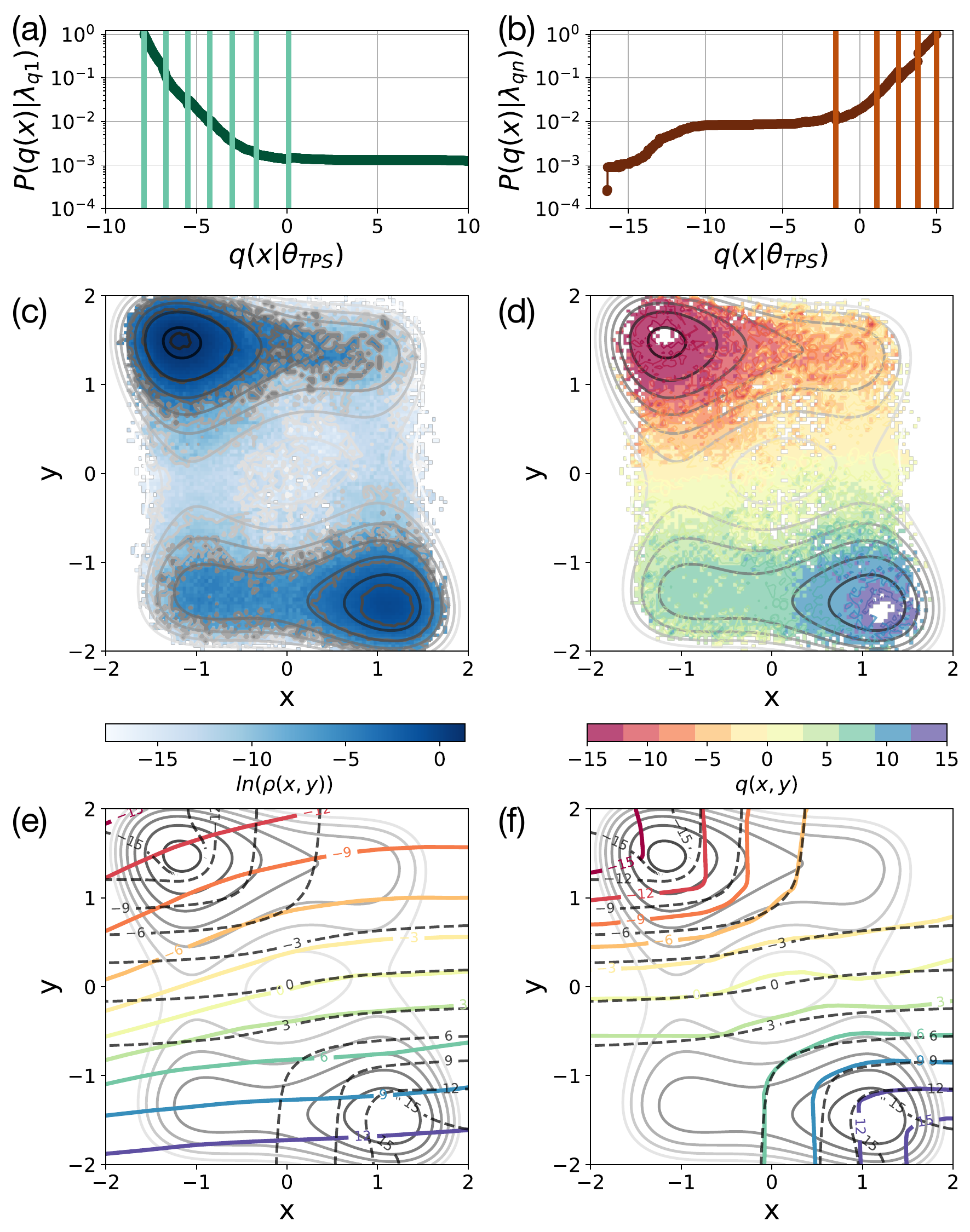}
     \caption{
     WQ potential results. 
    (a,b) TIS  crossing probabilities. (c) RPE distribution  $\ln\rho(x,y)$, with underlying WQ potential contour in gray. (d) RPE projection of $q(x,y)$ computed as  the logit of Eq.~\ref{eq:p_B_RPE_data_projection}.
    (e) Committor model  $q(x,y)$ slice (colored contours) 
    after initial AIMMD-TPS 
    compared to the theoretical committor (dashed lines) \cite{covinoMolecularFreeEnergy2019}. 
    (f) After the second  iteration the model quantitatively agrees with theory, up to $q=12$. 
     }. 
\label{fig:WQ_sampling_committor}
\end{figure}

{\bf Wolfe-Quapp potential}.
We illustrate our method on a 22-dimensional model system, where the first two dimensions are described by the Wolfe-Quapp (WQ) potential \cite{Wolfe1975,quappGrowingStringMethod2005}, chosen for its change in direction 
orthogonal to the TS reaction coordinate close to the stable states and its two reactive channels. The energy surface 
with a barrier height of $\Delta G=10 k_B T$ is shown in Fig.~\ref{fig:WQ_sampling_committor})c. The remaining 20  dimensions are governed by harmonic potentials.
Langevin dynamics were used to propagate the system.
Stable states A and B are defined as circles in $x$ and $y$ with radius $r=0.25$ at $(-1.15,1.5)$ and $(1.15,-1.5)$, respectively.
An initial linear trajectory from A to B was used for the AIMMD-TPS procedure, performing 
1000 shots with a uniform selector in $q$-space to explore both near and away from the TS.
The resulting committor, projected 
onto the $x,y$ plane in  Fig.~\ref{fig:WQ_sampling_committor}e, captures the key transition dynamics (in $y$) but struggles near the stable states (in $x$). 
TIS interfaces were based on the learned initial $q(x|\theta)$. 
Direct simulations starting from the stable states yielded the boundary values for the interfaces, so that 
interfaces ranged from $q_{\text{max}}^{A}=-7.875$ to $q_{\text{min}}^{B}=5.0$.
For each interface 1000 shots are performed using OpenPathSampling~\cite{swensonOpenPathSamplingPythonFramework2019} with a Gaussian selector in $q(x|\theta)$ close to the interface $q$.
Fig.~\ref{fig:WQ_sampling_committor}a,b shows the full conditional crossing probability of this first iteration.

Combining the  crossing probabilities and their WHAM weights with stable state data yields the RPE. 
From the  RPE data we can extract  the conditional distribution of configurations coming from or going to A  
\begin{equation}
    \rho_A(x,y) = \frac{1}{Z}\sum_{v_i\in \text{RPE}}w_i n_A^i \delta(x_i-x)\delta(y_i-y),
\end{equation}
with $Z=\sum_{v_i\in\text{RPE}} w_i$, and a similar expression for $\rho_B(x,y)$.
From these we can compute the (projected) committor 
\begin{equation}
    p_B(x,y) = \frac{\rho_B(x,y)}{\rho_A(x,y)+\rho_B(x,y)}.
    \label{eq:p_B_RPE_data_projection}
\end{equation}
as well as the logit-committor  $q(x,y) = \ln(p_B(x,y))-\ln(1-p_B(x,y))$.
In addition, we obtain the free energy as  
$F(x,y) = -k_b T \ln(\rho(x,y))$, 
with  $\rho(x,y)= \rho_A(x,y)+\rho_B(x,y)$.
Fig \ref{fig:WQ_sampling_committor}c,d  plots the RPE-based free energy and $q(x,y)$. 

The neural network logit-committor model used a feedforward architecture with 22 input features, two hidden layers (128 and 64 nodes), ReLU activations, and a linear output layer.
The model was trained using batches of $n_{\text{batch}} = 32,768$ data points, uniformly sampled from the full RPE dataset of  $\sim 10^6$ points. The weights for the regularization components for smoothness and L1 regularization are set to $\alpha=10^{-1}$ and $\eta=10^{-3}$. This ensures that smoothness is an important factor in the training of the model, but allows for enough variability in the model to reflect the data. These hyper-parameters may be tuned depending on the system and height of the barrier. 
The dataset is split into a train and test set with a ratio of $8:2$. 
A stopping criteria is used to prevent overfitting, tracking the improvement of the model on the test set and stopping if no new improvements in the loss is obtained after \texttt{n\_conv} steps.
We stress that the procedure does not amount to a simple fit of the $xy$ projected RPE data, but instead faithfully represents the full 22 dimensional space. 

\begin{figure}[t]
    \centering
    \includegraphics[width=\linewidth]{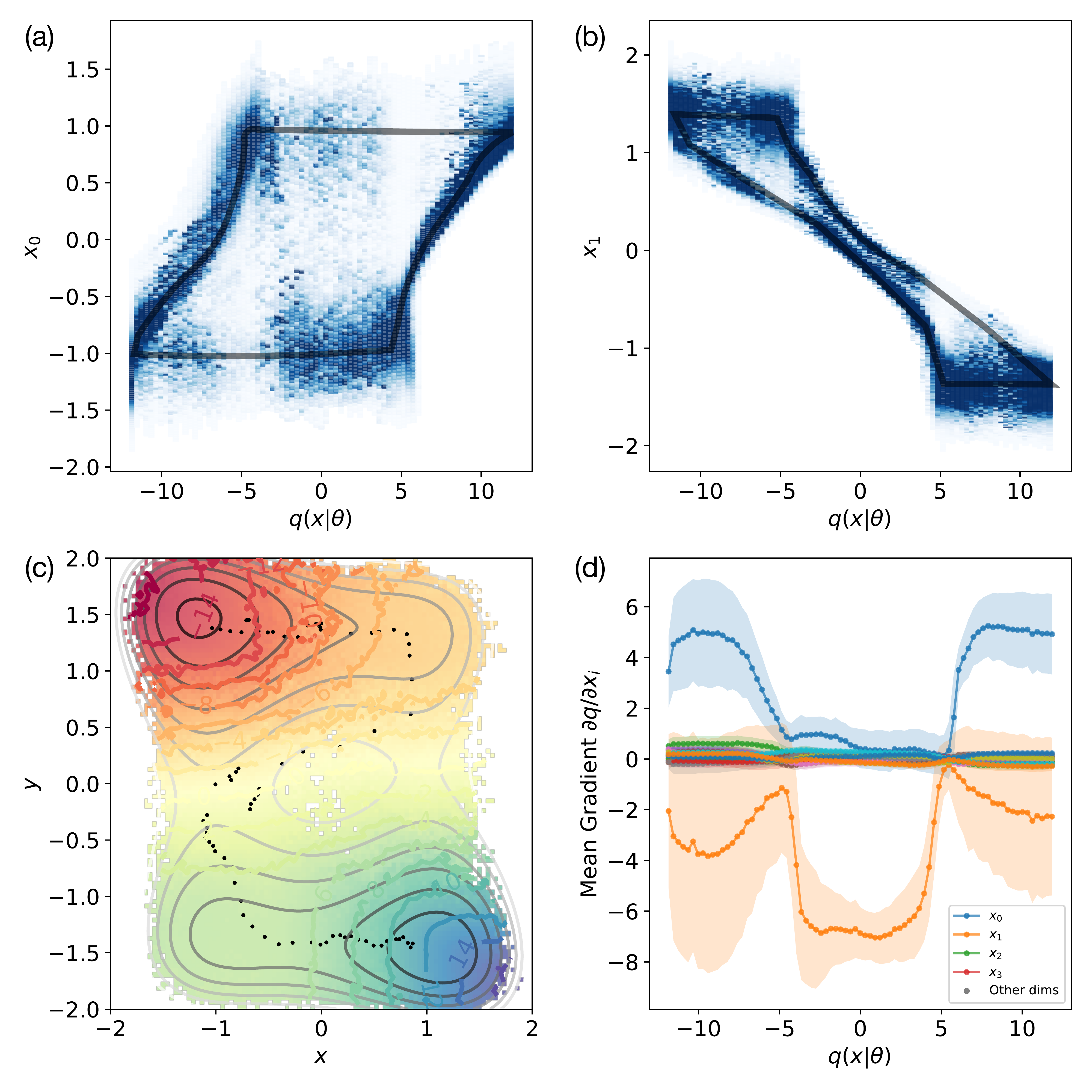}
    \caption{
    Mechanistic insights from the RPE and committor model.  
        (a, b) Distribution $\rho(x|\tilde{q})$ and $\rho(y|\tilde{q})$ shows two reactive channels.
        (c) Most probable path $\arg\max_x \rho_{\text{RPE}}(x|\tilde{q})$ across $\tilde{q}$ based on the RPE (colors as in Fig.~\ref{fig:WQ_sampling_committor}d).  
        (d) Average gradient $\langle \partial_x q(x|\theta) \rangle_{RPE} |_{\tilde{q}}$, highlighting the relevance of $x$ and $y$.
}
    \label{fig:WQ_q_model_analysis}
\end{figure}

This first iteration 
 improves  the committor model  closer to the stable states up to  $q\approx 3-6$ (see SI).
Repeating the same procedure, including the stable state analysis, determining  q-space interfaces, and performing TIS, a new RPE is constructed by which the model is retrained.
Fig.~\ref{fig:WQ_sampling_committor}f  shows  
this second iteration improves the model 
up to $q = 12$, i.e., a committor accurate up to $p_B\sim 10^{-6}$.
With a neural net model describing the committor and the RPE representing the equilibrium distribution  
we can follow how the system progresses along the reaction coordinate $q(x|\theta)$ from start to finish.
Fig.~\ref{fig:WQ_q_model_analysis}a,b 
shows the distribution $\rho(x|q(x|\theta))$ of configurations in the first two dimensions (those related to the WQ potential) as a function of their $q(x|\theta)$-values, normalized for each value of $q$.
 Fig.~\ref{fig:WQ_q_model_analysis}a indicates how the $x$ component indeed progresses up along the two channels, with one channel more  populated at the transition state, as expected.  The y-component (Fig.~\ref{fig:WQ_q_model_analysis}b) also shows two channels,  much  closer to each other at the transition state.

\begin{figure*}
    \centering
    \includegraphics[width=1\linewidth]{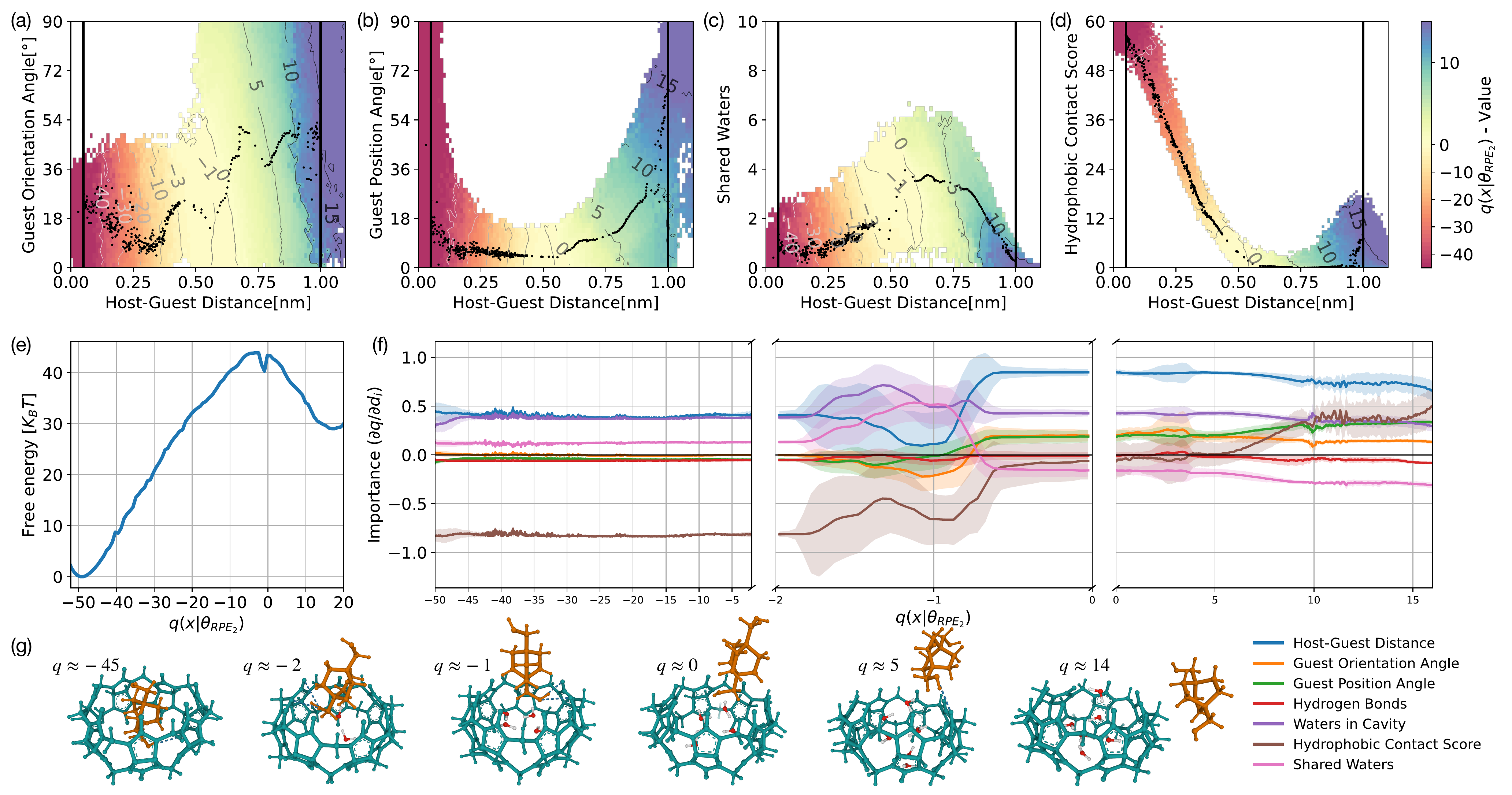}
    \caption{{ (a)-(d)}: committor projected onto the plane of four descriptors versus distance. Colored regions show the expectation value of the learned $q$-model, projected from RPE data. Stable states are defined as bound ($|r|<0.05$ nm) and unbound ($|r|>1.0$ nm), {(e)} Free energy computed along the logit committor. {(f)} gradient-based feature importance along the $q$-coordinate, highlighting dominant descriptors across the reaction. The middle panel focuses on the metastable state at $q\approx -1$. {(g)} representative snapshots along isocommittor surfaces, illustrating five key stages of the unbinding transition. 
    }
    \label{fig:HG_committor-result_iteration_2}  
\end{figure*}

In Fig.~\ref{fig:WQ_q_model_analysis}c we plot 
$\arg \max_x \rho(x|\tilde{q})$  for  $\tilde{q} \in [-14,14]$  
illustrating a mean path moving from one side of the barrier to the other side. Note that since there are  two channels  this allows identifying a change of preferred channel along the committor, 
also visible in Fig.~\ref{fig:WQ_q_model_analysis}a.
Fig.~\ref{fig:WQ_q_model_analysis}d shows the mean gradient 
$\langle \partial_{x} q(x|\theta) \rangle_{\rho(x|\tilde{q})}$ for all system dimensions, highlighting their relevance along $\tilde{q}$. Positive values suggest  the descriptor should increase to reach state B, while negative values suggest it should  decrease. Only the first two dimensions contribute significantly, while the others show negligible gradients. The large variance in the $x$ and $y$ gradients arises from the two reactive channels, as  the process can occur in either $x$-then-$y$ or $y$-then-$x$ order.

{\bf Host-Guest System}. To illustrate how our method performs for  a  complex system that exhibits non-linear behavior away from the transition region, we examine  a solvated host-guest (un)binding system, in which a B2 molecule (guest) binds to a CB7 ring (host)~\cite{Moghaddam2011,Fenley2014,Henriksen2015}. 
The primary descriptor in this system is the center-of-mass 
distance, with shorter distances representing the bound state ($r<0.05\text{nm}$) and larger distances indicating the unbound state ($r>1.0\text{nm}$). While this distance is obviously a key component, other structural and dynamic features may also influence the transition.  
We defined a set of 14 structural descriptors that capture key geometric, hydrogen bonding, hydrophobic, and solvation properties. These descriptors serve as inputs to the committor model and will provide mechanistic insight into (un)binding  pathways. A full description of the simulation setup, descriptor definitions, and computational methodology is provided in the SI.

After an initial AIMMD-TPS run using these components  we could reduce the number of dimensions to 7. 
Next, the AIMMD-TIS pipeline is applied iteratively to refine our mechanistic understanding of the entire binding transition.  
Fig.~\ref{fig:HG_committor-result_iteration_2} shows 
the model predictions  after two TIS iterations. 
The model applied on the ${\sim} 10^6$ RPE data points shows 
 how the committor changes along the process for each of  four (out of 7) key descriptors relative to the center distance. The dotted path indicates the average descriptor position for a given q value based on the RPE distribution, showing  how the guest moves out of  (or into) the host cavity. 
 Fig.~\ref{fig:HG_committor-result_iteration_2}e gives the free energy barrier along q, revealing a metastable state at $q\approx-1$.
 
The relative gradient of the model along $q$ in  Fig.~\ref{fig:HG_committor-result_iteration_2}f shows the  importance of   key ingredients   during the entire transition, from $q\approx-50$ to $q\approx 15$, indicating that committor can be  accurately computed  up to $p_B\sim 10^{-15}$.
The process can be  split into  distinct events. First from $q=-50$ to $-2$, the distance, number of waters in the cavity and hydrophobic contact score are dominant, driving the guest outward. Between $-2$ and $-1$, distance becomes less relevant, but shared waters,  orientational angle and  hydrogen bonding  become more important, 
In the metastable region ($q\approx -1$), all gradients approach zero (see SI). 
Leaving the metastable state, distance regains importance, the guest position and orientation angle become prevalent, while the hydrophobic contact and shared waters becomes less important. Between $q=0$ and $15$, the hydrophobic contact and the shared waters, and hydrogen bonding regain importance.
This analysis obtained from the learned committor model aligns with the physical mechanism observed in the trajectories: initially, the guest must move upward through the host, 
reducing its distance between the hydrophobic cores and allowing space for water to enter.
Entry into the metastable region requires a slight reorientation of the guest and water molecules entering the cavity. In exiting the metastable state, the host cavity is fully filled with water. 
Afterward, it continues to reorient, loses its hydrogen bond and breaks free from the host at its hydrophobic outer region.

Besides mechanistic information we obtain the rate constants $k_{\rm BU} = 4 \times 10^{-9} s^{-1}$, $k_{\rm UB} = 4\times  10^3 s^{-1}$, 
leading to  $\Delta G = - k_B T\ln k_{BU}/ k_{UB} \approx 27.6 k_BT$, which slightly differs from   Fig.~\ref{fig:HG_committor-result_iteration_2}e, due to undersampling of the unbound state.
Our result  agrees reasonably with a previously calculated free energy of dissociation, but is too high compared  to the experimental values, likely due to  force field issues \cite{Moghaddam2011,Fenley2014,Henriksen2015}. 

{\bf Conclusion}.
Building upon  innovations in transition path sampling techniques, notably TIS and AIMMD, 
we  have introduced an iterative scheme  to 
 construct  
the RPE  efficiently from TIS along isocommittor surfaces.  This allowed to  model the committor along the entire transition from initial to final state  in a neural network  up to very low values. 
Imposing smoothness in the committor training process is key. 
Our approach yields valuable information: the relevant variables that play a role in the transition process, the mechanism, possible intermediates, but also the free energy along the relevant CVs and the rate constants of the process.

The current implementation of AIMMD-TIS requires sufficient data:
the  computed RPE determines the quality of the model. For instance, the model can describe an RPE covering multiple channels, but 
if only one channel is sampled, we can only expect an improved single-channel model.
To enhance  switching between channels, AIMMD-TIS could  employ (infinite swap) replica exchange TIS \cite{vanerpReactionRateCalculation2007,bolhuisRareEventsMultiple2008,Roet2022}. Combination with  rejection free path sampling~\cite{Lazzeri2025} 
would pair  efficient sampling with full committor learning.

Our committor models are based on unbiased explicit MD trajectories that have fully relaxed to the stable states, and thus include global information on the process.
In contrast, other approaches, e.g. Refs.~\cite{Chen2023,Megas2025},
 employ  swarms of trajectories to compute the local drift in a committor consistent  string 
described by neural networks.  Also Kang et al.~\cite{kangComputingCommittorCommittor2024} employ  a variational principle for the committor which only uses local information. 
One advantage of sampling unbiased dynamical trajectories is that they allow direct inspection and enable discovery.
Moreover, they permit a posteriori  retraining or modifying
a new committor model, 
without additional sampling~\cite{Jung2023}. 

The final model gives much insight, but in  the  feature selection step correlation between descriptors  could be important. In the future  different regularization approaches should be explored. 

{\bf Acknowledgments}
G.L. and R.C. acknowledge the support of Goethe University Frankfurt, the Frankfurt Institute for
Advanced Studies, the LOEWE Center for Multiscale Modelling in Life Sciences of the state of Hesse, the CRC 1507: Membrane-associated Protein Assemblies, Machineries, and
Supercomplexes (P09), and the International Max Planck Research School on Cellular Biophysics.

{\bf Supplementary information} In the SI we give details on the methodology, simulations settings, and present additional results.

\bibliographystyle{apsrev4-2}
\bibliography{references} 

\end{document}


\title{Supplementary Information:\\ Understanding Reaction Mechanisms from Start to Finish}
\author{Rik S. Breebaart}
\affiliation{Van 't Hoff Institute for Molecular Sciences, Universiteit van Amsterdam, Science Park 904, 1098 XH Amsterdam, The Netherlands}

\author{Gianmarco Lazzeri}
\affiliation{Institute of Biochemistry, Goethe University Frankfurt}
\affiliation{Frankfurt Institute for Advanced Studies}
\author{Roberto Covino}
\affiliation{Institute of Computer Science, Goethe University Frankfurt}
\affiliation{Frankfurt Institute for Advanced Studies}
\author{Peter G. Bolhuis}
\affiliation{Van 't Hoff Institute for Molecular Sciences, Universiteit van Amsterdam, Science Park 904, 1098 XH Amsterdam, The Netherlands}
\email{p.gbolhuis@uva.nl}

\maketitle

\section{Methods}
\subsection{Background}

\input{general}

\subsection{The AIMMD-TIS loss function}
\label{A:loss_Function}
\input{Sections/SI/Loss_function}

\subsection{Interfaces in committor space}
\label{A:interface_placement_in_committor_space}
\input{Sections/SI/Interface_placement_in_committor_space}

\subsection{Uniform sampling of shooting points in q-space}
\label{A:uniform_sampling_in_q}
\input{Sections/SI/Uniform_sampling_in_q_v2}

\subsection{Specific AIMMD-TIS setup}
\label{A:General_Settings_AIMMD-TIS}

\input{Sections/SI/general_AIMMD_TIS_setup}

\section{Simulation details}

\input{Sections/SI/Simulation_settings}

\section{Results}
\subsection{The Wolfe-Quapp system}
\label{A:Additional_info_WQ}

\input{Sections/SI/Additional_info_WQ}

\subsection{Host Guest system}
\input{Sections/SI/Host-Guest}

\section{Balance loss function}
\label{A:balance_log-likelihood_loss}
\input{Sections/SI/Loss_function_theory_observation}

\bibliographystyle{apsrev4-2}
\bibliography{references}

%% file: general.tex
In this section we briefly summarize the general principles of 
the employed path sampling methodology.

\subsubsection{TPS and TIS}

Transition path sampling (TPS)  harvests an ensemble of unbiased dynamical trajectories connecting the initial A  to the final state B,  employing a Markov Chain Monte Carlo algorithm\cite{Bolhuis2002}. 
A trajectory ${\bf x}= \{x_1, x_2 \dots x_L\}$ consists of $L$ frames or  slices  separated by a time interval  $\Delta \tau$. 
Each configuration time slice or frame $x=\{r,p\}$ contains all positions $r$ and momenta $p$ of the system.
A trajectory is generated through the molecular dynamics of interest. 

Starting from an initial trajectory,  TPS employs the shooting move algorithm to generate  new trial pathways using an MD  engine, and accepts reactive paths according to an appropriate detailed balance rule involving indicator functions $h_{A,B}(x)$ that identify when a state has been reached\cite{Dellago2002}.
While TPS  also can give reaction rate constants, a significant improvement came with its extension transition interface sampling \cite{vanerpNovelPathSampling2003}. 

 Transition interface sampling  (TIS) \cite{vanerpNovelPathSampling2003,Cabriolu2017} 
 places a sequence of $n+1$ hypersurfaces, so-called interfaces,  $\lambda_0, \lambda_1,\ldots,\lambda_n$  
 between the metastable states A and B. $\lambda$ is a suitable order parameters which is capable of distinguishing state A and B. $\lambda_0=\lambda_{A}$  is the initial state boundary and $\lambda_n=\lambda_{B}$ the final state boundary. 

TIS path ensembles consist of trajectories 
that leave state A, cross a  predefined interface  and then return to A (non reactive ) or go on to end in B (reactive). 
Sampling these path ensembles via the shooting move allows for the evaluation of the unbiased kinetic rate constants. 

Assuming the interfaces increase monotonically ($\lambda_{i-1}<\lambda_{i}<\lambda_{i+1}$ ), then the rate constant is \cite{vanerpNovelPathSampling2003} 
$$k_{AB}=\phi_{01}P_{A}(\lambda_{B}|\lambda_1)=\phi_{01}\prod_{i=1}^{n-1}P_{A}(\lambda_{i+1}|\lambda_{i}).$$
where $\phi_{01}$ is the flux through the first interface
A similar expression is defined for the B to A transition. 
The crossing probabilities play a central role in the reweighted path ensemble (RPE).

\subsubsection{AIMMD}

Artificial Intelligence for Molecular Mechanistic Discovery (AIMMD) 
leverages machine learning to enhance path sampling efficiency and gain mechanistic understanding\cite{Jung2023}, by optimizing a model of the committor function based on the transition path sampling outcome\cite{Peters2006,Hummer2004}. 

The committor function, $p_B(x)$, describes the probability that a configuration $x$ will first reach (meta)stable state $B$ before state $A$. If we launch $N = n_A + n_B$ unbiased shooting trajectories from a configuration $x$ in a system with states $A$ and $B$, the number of trajectories reaching $A$ and $B$, respectively $n_A$ and $n_B$, are in the case of diffusive dynamics distributed binomially \cite{Peters2006,Hummer2004}:
\begin{equation}
P(n_A,n_B|x) = \binom{N}{n_A} (1-p_B(x))^{n_A} p_B(x)^{n_B},
\end{equation}
with $p_A(x) = 1 - p_B(x)$. Note that we ignore the influence of the  momenta\cite{Peters2006}. 

Using TPS, AIMMD treats two-way shooting moves as Bernoulli trials, where trajectories succeed or fail to reach state $B$.
To estimate the most likely committor function model $\tilde{p}_B(x)$ that best fits observed trajectory outcomes from $N$ configurations $x_i$ with shooting results $(n_i^A, n_i^B)$, we maximize the associated binomial log-likelihood by minimizing the negative log-likelihood loss function \cite{maAutomaticMethodIdentifying2005,Jung2023}:

\begin{equation}
\mathcal{L}_{\text{lh}} = - \sum_i \left[ n_i^A \ln(1 - \tilde{p}_B(x_i)) + n_i^B \ln \tilde{p}_B(x_i) \right].
\end{equation}

In the AIMMD framework \cite{jungArtificialIntelligenceAssists2019,Jung2023} (termed AIMMD-TPS here and in the main text), the committor function $\tilde{p}_B(x)$ is learned by minimizing  this loss function $\mathcal{L}_{\text{lh}}$, 
using the TPS shooting moves outcomes, where each two-way shooting move  provides a configuration $x_{sp}$ along with its trajectory outcomes $(n_A, n_B)$.
More precisely,
AIMMD learns a neural network-based committor model, $q(x|\theta)$, where
$\tilde{p}_B(x) = \left[ 1 + e^{-q(x|\theta)} \right]^{-1}.$
The logit-committor model $q(x|\theta)$ can  be defined in full space or in terms of a low- dimensional feature representation $\xi(x)$ that captures the essential characteristics, yielding a parametric form $q(\xi(x)|\theta)$, where $\tilde{p}_B(x) = \left[ 1 + e^{-q(\xi(x)|\theta)} \right]^{-1}.$
The final trained committor model contains crucial information about which combination of input coordinate are important for the RC, by  e.g. symbolic regression\cite{Jung2023}, or by autoencoders\cite{Frassek2021}.

\subsubsection{The reweighted path ensemble}
If one had access to an infinitely long trajectory which transits between A and B an infinite amount of times, one could learn the committor using all the configurations in this infinite path 
as the fate of each sub-trajectory -- either going to A or to B --- is known.  
Therefore, one could include all configurations in committor analysis. 
However, in practice this is unachievable as 
it would require an unfeasible long simulation time to capture enough 
rare event transitions. Nevertheless, it is possible to approximate the statistics of an infinite MD trajectory  by applying TIS and reweighting the interface path ensemble trajectories according to their respective WHAM weights  \cite{Rogal2010}
The Weighted Histogram Analysis (WHAM) Method allows to reweight histograms such that the error in the combined histogram is minimized \cite{ferrenbergOptimizedMonteCarlo1989}.
To obtain the RPE, WHAM is applied to all  TIS crossing probability histograms  $P_{A}(\lambda|\lambda_i)$,
leading to an estimate for the weight for each path based on how far that path traversed in the $\lambda$ \cite{Rogal2010}.
Assigning each TIS path its designated   weight $w_i$ leads to a path statistics that should converge to that  of an infinite  MD trajectory.

%% file: Sections/SI/Loss_function.tex
In the AIMMD-TPS scheme, only the shooting points contribute to the committor training, as each shot offers an unbiased binary outcome useful for likelihood-based learning. However, the method introduced in the main text broadens this scope by leveraging all configurations along the trajectories, through the use of the Reweighted Path Ensemble (RPE) \cite{Rogal2010}, constructed via transition interface sampling (TIS) \cite{vanErp2003}.

In the RPE, each trajectory is given a weight according to its likelihood of occurring under equilibrium statistics and since we know where each path ends, this allows all time slices —not just shooting points— to contribute meaningfully to committor training.

The weighted log-likelihood loss is formulated as:

\begin{equation}
   \mathcal{L}_{wl} = -\frac{1}{Z}\sum_i w_i \left[ n_i^A \ln(1- \tilde{p}_B(x_i)) + n_i^B\ln \tilde{p}_B(x_i)\right],
\end{equation}
where each weight $w_i$ accounts for the contribution of this configuration, ensuring that the configurations collectively represent the equilibrium distribution and $Z = \sum_i^{\text{RPE}} w_i$ the normalization constant. 

The committor should monotonically increase from A to B\cite{eTransitionPathTheoryPathFinding2010}. 
Therefore, an additional loss term is added
\begin{equation}
\mathcal{L}_{\rm smooth} = \sum_{x_i\in RPE}\frac{1}{N}\left|\nabla_\xi q(x_i|\theta)\right|^2
\end{equation}
that enforces smoothness in the model, where $\xi$ denotes the reduced  descriptor-space, or the full coordinate space $x$ itself. 
L1 regularization is also applied to 
reduce the influence of irrelevant degrees of freedom.
The resulting total loss function is: 
\begin{equation}
    \mathcal{L} = \mathcal{L}_{\rm wl}+\alpha \mathcal{L}_{\rm smooth} + \eta\mathcal{L}_{l1}
\end{equation}
where $\alpha$ and $\eta$ are hyper-parameters balancing the contribution of the different loss terms. 
We find each point in the RPE roughly contributes equally to $\mathcal{L}_{\rm wl}$ (see section \ref{A:balance_log-likelihood_loss}).

%% file: Sections/SI/Interface_placement_in_committor_space.tex
TIS can be performed using a model committor $\tilde{p}_B(x)$. In the AIMMD-TIS algorithm, we begin with an approximate committor model (obtained through AIMMD shooting or a previous cycle of RPE training) to construct interfaces on the isocommittor surfaces. These interfaces are defined in $q$-space: the logit of the committor. 
To allow for interfaces in q-space two requirements are needed: (i) interfaces cannot intersect the stable states (ii) interfaces must monotonically increase from A to B. In case we would have access to the exact committor, the first criteria would be automatically satisfied as the committor by definition is 0 in A and 1 in B. However, since we are dealing with an approximate committor model for interfaces and iteratively improve upon this model, careful consideration has to be made to ensure that this overlap does not happen. 
While TIS would in principle also work for  interfaces intersecting with the states, in practice this would favor trajectories  towards the regions just outside the state definition, or even to regions where trajectories jump ahead further in q-space, leading to longer decorrelation times.  

To ensure the interfaces do not intersect the stable states, separate stable state simulations have to be performed, from which stable state data is obtained that (within the stable state volume by which the state is defined in the TPS/TIS scheme) can be analyzed for their corresponding committor value according to the trained NN model. We then choose  interfaces outside of this region by taking the maximum in q for the A state $q_{\text{max}}^{A} = \max_{x \in A} q(x|\theta)$ and minimum in q at the B state $\quad q_{\text{min}}^{B} = \min_{x \in B} q(x|\theta)$.

Note however that, as the  dimensionality of feature space increases, the accuracy of the predicted starting bound in q-space for which interfaces are allowed decreases, which may result in the first interfaces overlapping with the stable state. 
While this does not prohibit performing TIS, it is best if such overlap is avoided, for instance by making the state definitions smaller. 
To this end careful analysis of the TIS ensembles near the states has to be performed.

After interfaces in q-space are chosen such that they do not cross through the stable state boundary, we can make smart choices of the interface placements  by utilizing that we have knowledge of the committor.
As shown in Refs.{\cite{lazzeriMolecularFreeEnergies2023a}} and \cite{vanden-eijndenAssumptionsUnderlyingMilestoning2008}, the committor can be directly related to the crossing probability, with the conditional crossing probability given by
\begin{equation}
    P_A(\lambda_j|\lambda_i)= \frac{\lambda_i}{\lambda_j} \text{ for } \lambda_j\geq\lambda_i,
    \label{eq.A Perfect commitor cross_prob}
\end{equation}
 when using $\lambda\equiv p_B$ as isocommittor surfaces\cite{lazzeriMolecularFreeEnergies2023a}. 
 
For the construction of the full crossing probability and RPE we want to have an adequate overlap between the consecutive interface ensembles such that WHAM can accurately be performed. Using the knowledge of the crossing probability for an exact committor, we can  choose interfaces such that there is adequate overlap between them (assuming that the estimate of the committor is reasonable), at the same time reducing the number of interfaces needed.
More precise, we can place interfaces on isocommittor surfaces to ensure a desired conditional  probability $P_A(\lambda_j|\lambda_i)$ of crossing from interface $\lambda_i$ to interface $\lambda_j$ (where $\lambda_i<\lambda_j$; most relevant is $\lambda_j=\lambda_i+1$) 
To relate the crossing probability to interfaces in $q$-space, we write Eq.\ref{eq.A Perfect commitor cross_prob} as
\begin{equation}
    P_A(q_j|q_i)=\frac{1+e^{-q_j}}{1+e^{-q_i}},
\end{equation}
where we can now select interfaces $q_j$ given $q_i$ such that we have a desired overlap factor $\alpha$ between them as 
\begin{equation}
    \label{eq:interfaces_ideal_committor}
    q_j(q_i) = -\ln(\alpha_c(1+e^{-q_j})-1),
\end{equation}
for interfaces going from $A$ and similarly can be constructed for trajectories leaving $B$.

With Eq.(\ref{eq:interfaces_ideal_committor}) we can construct a set of TIS isocommittor interfaces, given a certain overlap  probability $\alpha_c$. As the model improves, it starts to  match the predicted overlap and fewer additional interfaces may be needed to ensure proper overlap. To initialize this iterative procedure  it is therefore good to have a slightly higher overlap probability $\alpha_c$ (e.g. $\alpha_c=0.5$ for $50\%$ overlap) between the consecutive interfaces and as the model becomes more representative of the exact committor, this may be reduced.
Note that the number of interfaces scales with the barrier height.

%% file: Sections/SI/Uniform_sampling_in_q_v2.tex
While the standard AIMMD-TPS approach employs a Lorentzian to select shooting points, sometimes it is beneficial to use
a shooting point selector that uniformly selects in q-space, ensuring balanced coverage and exploration rather than exploitation for the committor model. This algorithm is similar to choosing a Lorentzian with large scaling factor $\gamma\gg1$ in AIMMD-TPS\cite{jungArtificialIntelligenceAssists2019,Jung2023} and the uniform selector in q-space based on the transition path ensemble in Ref.\cite{lazzeriMolecularFreeEnergies2023a}.
Given a trajectory $\mathbf{x}$, each point $j$ is mapped to its corresponding q-model value $q(x_j|\theta)$.
To uniformly sample a point $ x_j $ from path $\mathbf{x}$ in $ q $-space, we divide the $ q $-range into $ N $ equally spaced bins of width $ \Delta q $. Each non-empty bin is selected with equal probability, and a point is then uniformly chosen from within that bin.

Let $n_i[\mathbf{x}] $ denote the number of points in bin $ i $, i.e.,
\begin{equation}
n_i[\mathbf{x}] = \sum_{j=1}^{L_x} \mathbf{1}_{[q_i,q_{i+1}]}(q(x_j)),
\end{equation}
where $ \mathbf{1}_{[q_i,q_{i+1}]} $ is the indicator function, equal to 1 if $ q(x_j) $ lies in the bin and 0 otherwise.
The number of non-empty bins is:
\[
N_{\text{nz}} = \left| \left\{ i \mid n_i > 0 \right\} \right|.
\]
To avoid selecting empty bins, we restrict sampling to the $ N_{\text{nz}} $ bins with $ n_i > 0 $. 
Let $\text{idx}(x_j) = i$ denote the bin index of $x_j$.
The selection probability of a point $ x_{\text{sp}} $ is then:
\begin{equation}
p_{\text{sel}}(x_{\text{sp}} | \mathbf{x}) = \frac{1}{N_{\text{nz}}} \cdot \frac{1}{n_\text{idx}(x_{sp})[\mathbf{x}]}, 
\end{equation}
for the bin idx such that $x_{sp}\in [q_i,q_{i+1}]$.
This two-step process—uniformly choosing among non-empty bins and then uniformly choosing a point within a selected bin—ensures detailed balance and avoids rejection due to empty bins. This was implemented using a single random variable $ \zeta \in [0,1] $ by computing the cumulative distribution over the non-zero bins and selecting the first index where the cumulative sum is greater than $ \zeta $, $\sum_{k=0}^{j} p_{\text{sel}}(x_k) > \zeta.$

In TPS, this selection probability enters the acceptance criterion for a proposed path:
\begin{equation}
\begin{split}
p_{\text{acc}}(\mathbf{x}^{(n)}|\mathbf{x}^{(o)}) =& h_A(x_0^{(n)}) h_B(x_L^{(n)}) \\&\min \left[1, \frac{N_{\text{nz}}^{(n)}n_\text{idx}(x_{sp})[\mathbf{x}^{(n)}]}{N_{\text{nz}}^{(o)}n_\text{idx}(x_{sp})[\mathbf{x}^{(o)}]} \right].
\end{split}
\end{equation}
where $h_A(x)$ (or $h_B(x)$) are indicator functions which are 1 if configuration $x$ is in region A (or B) and 0 otherwise.

If all selection probabilities are zero (i.e., no meaningful prediction), a point is chosen at random from the trajectory as all points fall within the same bin.

%% file: Sections/SI/General_AIMMD_TIS_setup.tex
\subsubsection{Initializing committor model with AIMMD-TPS.}
For the initial AIMMD-TPS run the emphasis is on obtaining a good initial committor model, capable of  distinguishing  the two states, and that can be used to place interfaces along. 
The goal is thus exploration rather than exploitation of shooting points and state space. In  AIMMD-TPS  therefore a selector is chosen which focuses on exploration. This can be done by either using the uniform selector as described in section \ref{A:uniform_sampling_in_q} or by using a Lorentzian selector as introduced by Jung et al. \cite{jungArtificialIntelligenceAssists2019,Jung2023} with large scaling factor $\gamma\gg1$. When using the Lorentzian selector one can also choose to focus more on the transition state region to emphasize a model that can 
describe the main barrier region, this may be more beneficial in the cases where the  barrier is larger than 10 $k_B T$.

The efficiency factor $\alpha_{\rm eff}$ introduced by Jung et al. \cite{Jung2023} scales the learning rate based on the expected and generated number of transition paths. In the original AIMMD  work Ref.~\cite{Jung2023}, this factor was used to adjust the learning rate when sufficient sampling occurred around the transition state and the expected and obtained number of reactive paths are close, prioritizing exploitation over exploration. However, this approach seems to limit the model’s ability to learn beyond identifying the transition region as it decreases its learning when the transition region is properly identified. To ensure continuous learning rather than over-exploitation of transition path sampling, $\alpha_{\rm eff}$ has been disabled in all AIMMD-TPS procedures. This is done with the aim of focusing on the objective of developing an accurate committor model rather than improving sampling acceptance for TPS.

It is beneficial to have a committor model in which dropout is used when performing AIMMD-TPS based training, to prevent overfitting of the model due to the low number of data-points in the training set of shooting points compared to the parameter space of the NN model. During RPE training dropout can be disabled, as the significantly larger dataset reduces the risk of overfitting. 
In all AIMMD-TPS runs presented the committor model was trained for one epoch every five shooting attempts, using a fixed learning rate of $lr_{0}=10^{-3}$.

\subsubsection{TIS with interface based selector.}
When performing TIS, two-way shooting is used to create new trajectories, allowing for a higher decorrelation between MC steps. To ensure that most shots satisfy the TIS ensemble criteria, we use a Gaussian shooting point selector of the form 
\begin{equation}
    p_{sel}(x_j|\mathbf{x})\propto e^{-\alpha_g (q(x_j)-l_0)^2}, 
\end{equation}
where $\alpha_g$ relates to the width of the Gaussian and $l_0$ the mean in $q$-space. 
The mean $l_0 = q_i + \Delta l_0$ is chosen for each interface to be close to the interface $q_i$ such that most shooting attempts satisfy the TIS ensemble criterion. The width of the gaussian is chosen dependent on the current knowledge  of the barrier steepness  and accuracy of the model, in general a $\alpha_g=1/2$ results in sufficient spread in selection points.
However, in the case the barrier is much steeper an increase in the variance (and decrease in $\alpha_g$) may be used to ensure enough samples exist in the gaussian selector region. As discussed in section \ref{A:interface_placement_in_committor_space}, interfaces may be placed to ensure sufficient overlap between consecutive interfaces, however as the initial model does not fully reflect the exact committor of the system, an increased number of interfaces may be needed further from the TS to ensure adequate overlap. 

\subsubsection{RPE based model training}
From the TIS ensembles the RPE can be obtained using WHAM\cite{Rogal2010}. For each configuration the descriptors, shooting results and RPE weight are collected. Each data-point $v_i = \{x_i, w_i, (n_A^i, n_B^i)\}$ includes the configuration $x_i$ (or descriptor set $\xi(x_i))$, the RPE weight $w_i$ and the 2-way shot result of its path.  The RPE dataset is split into a train and test set with a ratio of 8/2. 
For all RPE training done in this work a batch size of $n_{\rm batch}=32768$ points is used which where randomly selected from the train data with uniform sampling of data points. This was done to have a batch of adequate size such that each batch roughly contains data-points in all space regions, reflecting the equilibrium distribution due to their weights.

A stopping criteria is used to prevent overfitting, tracking the improvement of the model on the test-set and stopping if no new improvements in the test loss is obtained after \texttt{n\_conv} epochs. An Adam\cite{Kingma2015Adam:Optimization} optimizer is used with a learning rate with initial fixed learning rate $lr$ which decays with a step decay with decay rate of $\alpha_{\rm lr_{decay}}$ after a certain number of epochs $n_{\rm lr_{decay}}$.

%% file: Sections/SI/Simulation_settings.tex
All TPS and TIS simulations were performed using the OpenPathSampling package (OPS)\cite{swensonOpenPathSamplingPythonFramework2019} and the AIMMD package\cite{Jung2023} for AIMMD based TPS sampling. 
Committor model based interfaces where implemented in OPS collective variable functions and used as interfaces for TIS. The RPE was computed from the crossing probabilities which was computed using the WHAM implementation of OPS. Committor models were trained using the AIMMD workflow. Python based notebooks and analysis scripts are available online.

\subsection{Wolfe-Quapp Potential}
The Wolfe-Quapp (WQ) potential energy surface\cite{Wolfe1975,quappGrowingStringMethod2005} with 20 additional noise dimensions is defined as: 

\begin{align}
        U(\vec{x}) =& \frac{\Delta G}{5}(x_0^4 +x_1^4 - 2x_0^2 +  x_0x_1-4x_1^2 + 0.3x_0 + 0.1x_1) \notag \\
        &+ \sum_{j=0}^{D-1} \omega_j^2x_j^2/2,
    \label{eq:Wolfe-Quape_potential}
\end{align}
where $x_0,x_1,\ldots,x_{D-1}$ are cartesian coordinates, $\Delta G=10 k_B T$ sets the barrier height in the WQ potential determined by the first two coordinates $x_0,x_1$, and the sum defines harmonic oscillators in the 20 additional noise dimensions, with 
 harmonic oscillator angular velocities $\omega_j$: [0.00, 0.00, 1.00, 1.50, 7.57, 4.29, 3.81, 6.41, 7.76, 5.38, 9.85, 7.48, 5.85, 5.14, 4.75, 7.83, 5.51, 2.48, 5.18, 7.90, 3.46, 3.40] in units of $\sqrt{k_B T}$. 
In this potential a single particle with unit mass is propagated according to overdamped Langevin dynamics,
employing the BAOAB integrator\cite{leimkuhlerRationalConstructionStochastic2012} implemented in the toy system module of OpenPathSampling \cite{swensonOpenPathSamplingPythonFramework2019}, with $\Delta t = 0.05 \tau$, $\gamma=2.5 \tau^{-1}$.
Stable states A and B are defined as circular regions in $x$ and $y$ with radius $r=0.25$ at $(-1.15,1.5)$ and $(1.15,-1.5)$, respectively.

To compare the obtained committor to the true committor of the Wolfe-Quapp potential we establish the latter by solving the backward Kolmogorov equation:
\begin{equation}
    \nabla \cdot (\nabla U_{WQ}(x,y)+\nabla)p_B(x,y)=0.
\end{equation}
for the 2D Wolfe-Quapp potential using the relaxation method \cite{covinoMolecularFreeEnergy2019}, with $p_B(x,y)=0$  for $x,y \in A$ and $p_B(x,y)=1$ for $x,y \in B$ as boundary conditions.  The higher dimensional noise dimensions are ignored as they do not play a role in the behavior of the committor. 

\begin{table}[t]
\centering
\begin{tabular}{|l|l|}
\hline
\textbf{Parameter}       & \textbf{Value} \\ \hline
Potential Name           & Wolfe-Quapp    \\
Temperature              & 1              \\
Time Step                & 0.05  $\tau$       \\
Gamma                    & 2.5   $\tau^{-1}$\\ \hline

\multicolumn{2}{|c|}{\textbf{TPS Settings}} \\ \hline
n\_frames\_max            & 5000           \\
n\_steps\_per\_frame       & 1              \\ \hline

\multicolumn{2}{|c|}{\textbf{AIMMD Settings}} \\ \hline
Distribution             & Uniform        \\ \hline

\multicolumn{2}{|c|}{\textbf{Parameters AIMMD-TPS}} \\ \hline
Efficiency Factor        & False          \\
Learning Rate (lr)  & 1e-3          \\
Epochs per Train         & 1              \\
Interval                 & 5              \\
\hline
\multicolumn{2}{|c|}{\textbf{RPE training Parameters AIMMD}} \\ \hline
RPE training Learning Rate (lr) & 1e-3 \\
Smoothness Penalty Weight & 1e-1         \\
L1 Regularization        & 1e-3          \\ \hline

\multicolumn{2}{|c|}{\textbf{Network Layers}} \\ \hline
Layer 1                  & 128            \\
Layer 2                  & 64             \\ \hline

\multicolumn{2}{|c|}{\textbf{Dropout Rates}} \\ \hline
Layer 1                  & 0.05           \\
Layer 2                  & 0.01           \\ \hline

Activation Function      & ReLU           \\ \hline

\multicolumn{2}{|c|}{\textbf{TIS Settings}} \\ \hline
Shooting Move            & Two Way         \\
Selector                 & Gaussian in $q$ space      \\
Modification Method      & Random Velocities \\ \hline
\end{tabular}
\caption{Simulation Settings for Wolfe-Quapp Potential}
\label{tab:A:WQ simulation settings}

\end{table}

\textbf{AIMMD model settings:}
Table \ref{tab:A:WQ simulation settings} presents the AIMMD settings for the WQ system. We used a fixed learning rate of $lr_{0}=10^{-3}$, and trained the model for one epoch every five shooting attempts during the AIMMD-TPS procedure. The neural network had an input of the 22 potential dimensions, feeding into a first layer with 128 nodes, followed by a second layer with 64 nodes,  mapping to a single output. A ReLU activation function was applied between layers, while the final output layer was a linear feedforward layer without activation.

During AIMMD-TPS and AIMMD-TIS RPE training, dropout rates of 0.05 and 0.01 were applied to the first and second layers, respectively. 

\subsection{Host-Guest system}

\textbf{System Overview}
The host-guest binding mechanism of Cucurbit[7]uril (CB[7]) and the B2 guest molecule in explicit water solvent was investigated\cite{Moghaddam2011,Fenley2014,Henriksen2015}. The primary objective was to analyze the binding and unbinding pathways and identify mechanistic characteristics along the entire mechanism. The system consists of one CB[7] host molecule, one B2 guest molecule, and 1445 water molecules. The center-of-geometry distance $r_{CG}$ between the host and guest was used as the primary order parameter to distinguish the bound(  $r_{CG} < 0.05$ nm )  and unbound state ($r_{CG} > 1.0$ nm). 

\textbf{Simulation Setup}
All simulations were performed with the  OpenPathSampling package using the OpenMM engine with the Amber GAFF force field\cite{Wang2004} with periodic boundary conditions and explicit solvent. The initial structure was obtained from the openmmtools GitHub repository\cite{chodera2018choderalab}. Non-bonded interactions were treated with the Particle Mesh Ewald (PME) method using a cutoff of 1.0 nm and an Ewald error tolerance of 0.005. Rigid water models were employed, and hydrogen bond constraints were applied.
Molecular dynamics simulations were performed in the NPT ensemble using the OpenMM VVVRIntegrator at a temperature of 300 K, with a time step of 0.002 ps saving every 50th step and a friction coefficient of 1.0 ps$^{-1}$. Pressure was maintained at 1.0 bar using the OpenMM MonteCarloBarostat \cite{eastmanOpenMM8Molecular2024}, applied every 25 integration steps.

\textbf{System Analysis and Descriptors:}
\begin{figure}[b!]
    \centering\includegraphics[width=0.6\linewidth]{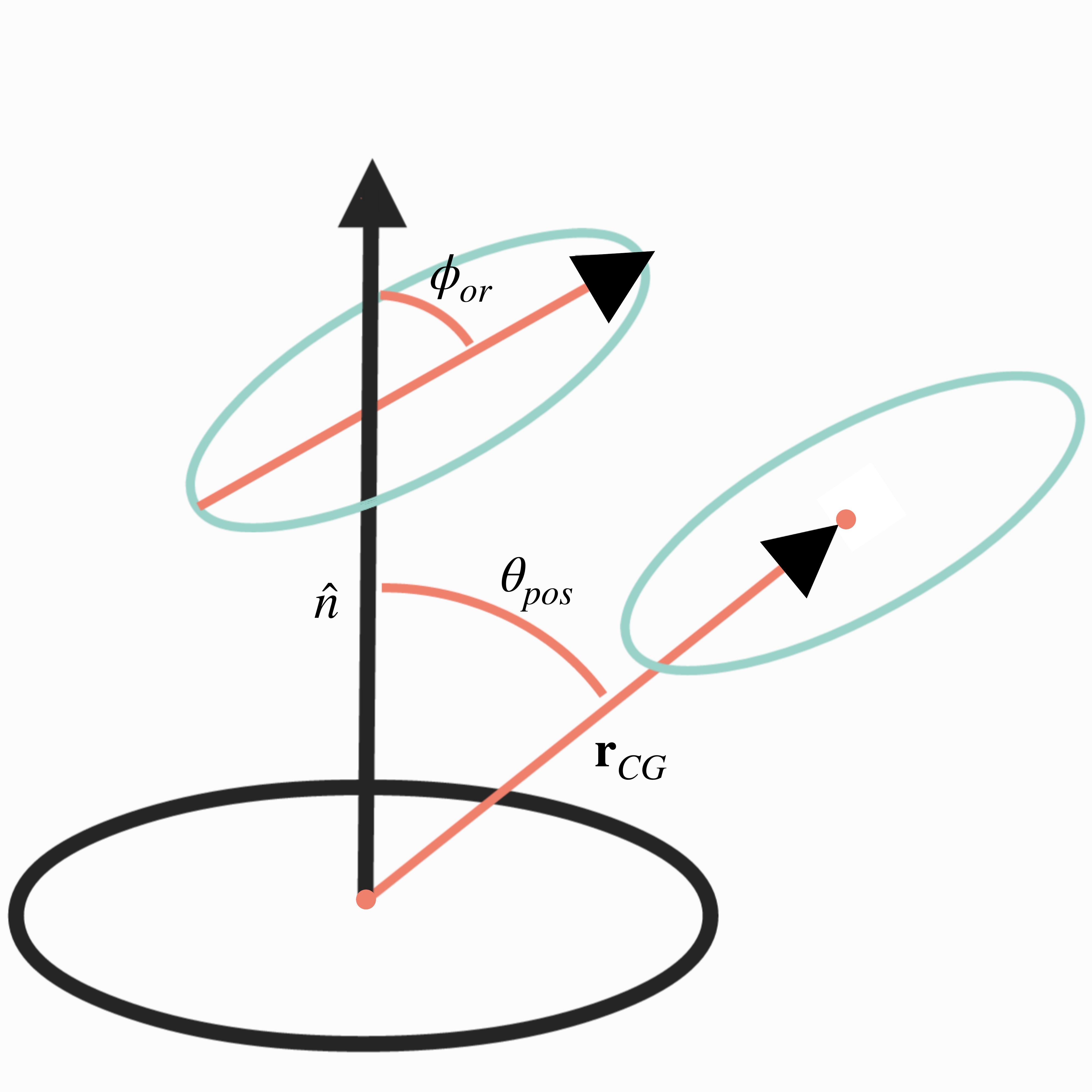}
    \caption{Schematic illustration of the geometric descriptors of the Host Guest system. Here the black ring represents the Host molecule and its normal vector $\hat{n}$, the green ring the guest molecule with its orientation angle $\phi_{\text{or}}$ and position angle $\theta_{\text{pos}}$ relative to the host's normal vector with a distance $r_{CG}$ from the host Center-Of-Geometry.}
    \label{fig:A:Schematic Host Guest descriptors}
\end{figure}

\begin{table*}[hbt!]
\centering
\caption{Descriptors used to characterize the CB[7]-B2 host–guest system.   For details on descriptor calculations, see text.} 
\label{tab:A descriptor-overview-HG}
\begin{tabular}{ll}
\toprule
\textbf{Descriptor} & \textbf{Description} \\
\midrule
$r_{CG}$ & Center-of-Geometry distance between host and guest \\
$\phi_{\text{or}}$ & Guest alignment relative to host's normal vector \\
$\theta_{\text{pos}}$ & Guest's spatial position angle relative to host \\
$n_{\text{HB}}$ & Number of hydrogen bonds between host and guest \\
$n_{\text{waters}}$ & Continuous water molecules count within $r_c=0.5~\text{nm}$ of host COM \\
$S_{\text{hydrophobic}}$ & Score of hydrophobic contacts between host and guest \\
$\psi_{\text{OH,top}}$ & Dihedral of guest's top OH group (H15–O1–C9–C4) \\
$\psi_{\text{OH,bot}}$ & Dihedral of guest's bottom OH group (H18–O2–C10–C3) \\
$n_{\text{water,head}}$ & Continuous water molecules count near guest’s head group (cutoff $r_c=0.35~\text{nm}$) \\
$n_{\text{water,center}}$ & Continuous water molecules count near guest’s central region (cutoff $r_c=0.35~\text{nm}$)\\
$n_{\text{water,tail}}$ & Continuous water molecules count near guest’s tail group (cutoff $r_c=0.35~\text{nm}$)\\
$n_{\text{water,shared}}$ & Water molecules close to both host and guest (cutoff $r_c=0.5~\text{nm}$) \\
$\Delta r_{\text{mouth}}$ & Radial deformation across host cavity mouth \\
$\Delta n_{\text{HB,asym}}$ & Asymmetry in H-bonds between guest head/tail \\
\bottomrule
\end{tabular}
\end{table*}

To systematically characterize the CB[7]-B2 host–guest system and analyze its binding/unbinding pathways, a set of 14 structural descriptors were defined. These descriptors were used as inputs to the AIMMD committor model and are designed to capture geometric, hydrophobic, hydrogen bonding, and solvation features of the complex. Table \ref{tab:A descriptor-overview-HG} provides an overview of these descriptors, and Figure \ref{fig:A:Schematic Host Guest descriptors} presents a schematic illustration of the key spatial geometric definitions.

\textbf{Descriptor Details:} 
Here, we provide a more detailed description of the 14 structural descriptors. The host–guest center-of-geometry distance ($r_{CG}$) was computed using the minimum image convention under periodic boundary conditions and served as the principal measure of spatial separation in which the states are defined. To describe relative orientations, we calculated the guest orientation angle ($\phi_{\text{or}}$) as the angle between a vector along the guest's axis and the normal vector of the host (see Fig.~\ref{fig:A:Schematic Host Guest descriptors}). 
The guest position angle ($\theta_{\text{pos}}$) captured the spatial alignment of the guest relative to the host normal and was symmetrized as $\min(\theta, \pi - \theta)$. Hydrogen bonding interactions between host and guest were identified using the Wernet–Nilsson criteria\cite{wernetStructureFirstCoordination2004}, resulting in the total hydrogen bond count ($n_{\text{HB}}$). Additionally, we quantified hydrogen bond asymmetry ($\Delta n_{\text{HB,asym}}$) as the difference in the number of hydrogen bonds formed with the guest’s head versus its tail, highlighting directional preferences in binding. 
To ensure smoothness, the solvent and contact based descriptors were implemented using continuous sigmoidal switching functions. 
The hydrophobic contact score $S_{\text{hp}}$ was defined as:
\begin{equation}
    S_{\text{hp}} = \sum_{i \in \text{guest}} \sum_{j \in \text{host}} \frac{1 - \left( \frac{r_{ij}}{r_0} \right)^n}{1 - \left( \frac{r_{ij}}{r_0} \right)^m}
\end{equation}
where $r_{ij}$ is the minimum-image distance between hydrophobic atoms $i$ and $j$, and $(r_0 = 0.5~\text{nm}, n = 6, m = 12)$ define the smooth cutoff behavior.
Water coordination numbers (e.g., $n_{\text{waters}}$, $n_{\text{water,head}}$, $n_{\text{water,center}}$, $n_{\text{water,tail}}$) were computed using a sigmoidal function:
\begin{equation}
    f(r) = \frac{1}{1 + \exp[\alpha (r - r_c)]}
\end{equation}
with steepness parameter $\alpha = 30~\text{nm}^{-1}$ and cutoff $r_c$ specific to each descriptor (typically in the range $0.35$–$0.6$~nm). The contribution of a water molecule was determined by computing $f(r)$ for the distances $r$ from relevant regions (e.g., host cavity, guest head/tail) to the molecule and summing over all water molecules.
The number of \textit{shared} waters close to both host and guest was computed by evaluating the product of two sigmoids:
\begin{equation}
    f_{\text{shared}}(r_h, r_g) = \frac{1}{1 + \exp[\alpha (r_h - r_c)]} \cdot \frac{1}{1 + \exp[\alpha (r_g - r_c)]}
\end{equation}
where $r_h$ and $r_g$ are distances from the water molecule to the host and guest centers of geometry, respectively and cutoff $r_c=0.5$~nm and $\alpha=30~\text{nm}^{-1}$.
The conformational flexibility of the guest hydrophilic region was monitored using two torsional angles: the top hydroxyl dihedral ($\psi_{\text{OH,top}}$) and the bottom hydroxyl dihedral ($\psi_{\text{OH,bot}}$), each computed from standard four-atom definitions involving the H–O–C–C backbone. Lastly, structural deformation of the host cavity mouth ($\Delta r_{\text{mouth}}$) was measured as the difference between the maximum and minimum radial distances from the host’s center of geometry to the centers of the seven symmetric microcycles forming its portal, capturing dynamic variations in cavity opening. 

\textbf{AIMMD Model Settings:}
The committor model for the Host–Guest system was trained using the AIMMD framework with the settings summarized in Table~\ref{tab:A:CB7B2_simulation_settings}. A fully connected neural network architecture was used with 14 normalized input descriptors (see Table~\ref{tab:A descriptor-overview-HG}), a first hidden layer of 128 nodes, followed by a second hidden layer of 64 nodes and a third of 16 nodes, after which a linear output layer. Each of the input descriptors was scaled between 0 and 1 before being used as input into the NN committor model. ReLU activations were used between hidden layers, and dropout was applied with rate of 0.05 for the first and second layers, respectively, during TPS training. The learning rate for the initial TPS stage was set to \(lr_0 = 10^{-3}\), and model updates were performed once every five shooting attempts.

\begin{table}[t!]
\centering
\begin{tabular}{|l|l|}
\hline
\textbf{Parameter}       & \textbf{Value} \\ \hline
System Name              & CB7:B2         \\
\hline

\multicolumn{2}{|c|}{\textbf{States}} \\ \hline
Bound State              & $r_{CG} < 0.05$ nm          \\
Unbound State            & $r_{CG} > 1.0$  nm          \\ \hline

\multicolumn{2}{|c|}{\textbf{Integrator Settings}} \\ \hline
Temperature              & 300 K          \\
Time Step (dt)           & 0.002 ps       \\
Friction                 & 1.0 $\text{ps}^{-1}$ \\ 
Pressure                 & 1.0 bar\\
Barostat frequency       & 25 \\

\hline
\multicolumn{2}{|c|}{\textbf{TPS Settings}} \\ \hline
n\_frames\_max            & 2000           \\
n\_steps\_per\_frame       & 50             \\ \hline

\multicolumn{2}{|c|}{\textbf{AIMMD Settings}} \\ \hline
Distribution             & Lorentzian     \\
Scale                    & 5.0            \\\hline

\multicolumn{2}{|c|}{\textbf{Parameters AIMMD-TPS}} \\ \hline
Efficiency Factor        & False          \\
Initial Learning Rate (lr\_0)  & 1e-3          \\
Epochs per Train         & 1              \\
Interval                 & 5              \\
\hline
\multicolumn{2}{|c|}{\textbf{RPE training Parameters AIMMD}} \\ \hline
RPE training Learning Rate (lr) & $it_1$\ 1e-3\ \ $it_2$\ 1e-4 \\
Smoothness Penalty Weight & $it_1$\ 1e-1\ \ $it_2$ 1e-1\          \\
L1 Regularization        & $it_1$\ 0\ \  $it_2$\ 0         \\ \hline

\multicolumn{2}{|c|}{\textbf{Network Layers}} \\ \hline
Layer 1                  & 128             \\
Layer 2                  & 64             \\
Layer 3                  & 32             \\ \hline

\multicolumn{2}{|c|}{\textbf{Dropout Rates }} \\ \hline
Layer 1                  & 0.05           \\
Layer 2                  & 0.05           \\
Layer 3                  & 0.0            \\ \hline

Activation Function      & ReLU           \\ \hline

\multicolumn{2}{|c|}{\textbf{TIS Settings}} \\ \hline
Shooting Move            & Two Way         \\
Selector                 & Gaussian in $q$ space      \\
Modification Method      & Random Velocities \\ \hline
\end{tabular}
\caption{Simulation Settings for CB7:B2 Host-Guest System}
\label{tab:A:CB7B2_simulation_settings}
\end{table}

%% file: Sections/SI/Additional_info_WQ.tex
\textbf{AIMMD-TPS}
Starting from a linear trajectory from state $A$ to $B$    the AIMMD-TPS procedure performed 1000 shots with a uniform selector in $q$-space (see section \ref{A:uniform_sampling_in_q}) to explore both near and away from the TS. The  AIMMD-TPS scheme saves every frame of the integrator and restricts the  maximum path-length to 5000 frames.

\begin{figure}[t!]
    \centering
    \includegraphics[width=0.6\linewidth]{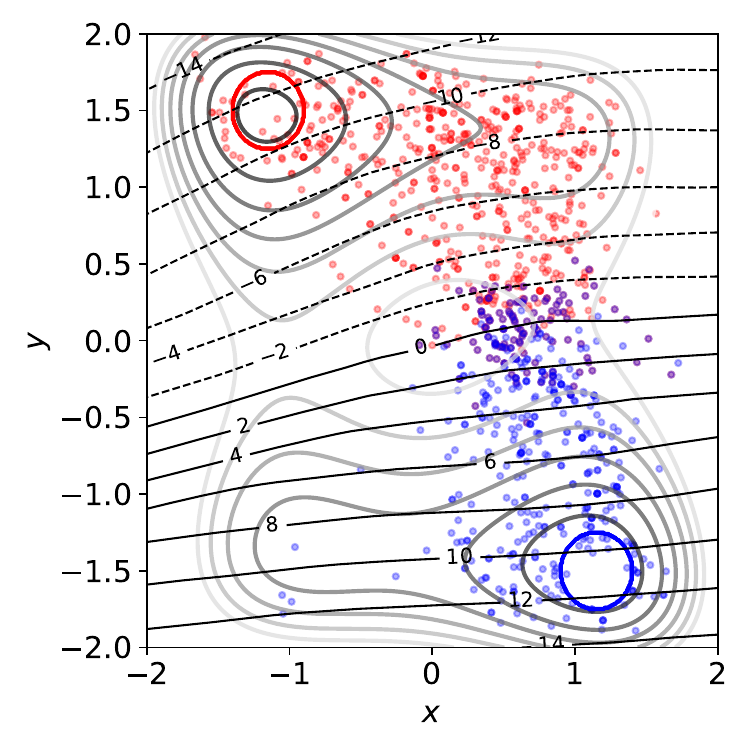}
   \caption{AIMMD-TPS Shooting points colored by their shot result (red going to A, blue going to B) and learned committor model q-contours slice when performing AIMMD-TPS with uniform selector in q-space. The red and blue rings indicate the A and B state, respectively.}
    \label{fig:A. ShootingPointsWQ}    
\end{figure}

Figure \ref{fig:A. ShootingPointsWQ} shows
the obtained shooting points in the initial AIMMD-TPS procedure.
The model mostly learned about the right-hand-side channel on the WQ potential. The TPS data does not contain information on the other x-dependencies of the committor near the stable states in the extremely low $p_B$ regions near the states.

\textbf{TIS sampling on $q(x|\theta_{TPS})$:} The learned initial $q(x|\theta_{TPS})$ model from AIMMD-TPS was used to define interfaces after a $5000 \tau$ simulation starting from both stable states, by which the boundary values $q_{\text{max}}^{A}=-7.875$ and  $q_{\text{min}}^{B}=5.0$ are determined.
The forward interfaces in $q(x|\theta_{TPS})$-space were: [-7.875, -6.67, -5.46, -4.25, -3.01, -1.68, 0.09], and the backward interfaces: [5.0, 3.78, 2.52, 1.11, -1.56]. These interfaces are used to perform TIS calculations on the initial committor model and from it generate an RPE for retraining.
 
 1000 two way shooting MC steps were performed per TIS interface, using a
 Gaussian selector as specified in  section \ref{A:General_Settings_AIMMD-TIS}, with $\alpha_g=4$ and mean $l_0=\lambda_{qi} \mp0.2$ for each interface ensemble $i$, where -0.2 is used for forward and + 0.2 for backward ensembles. 

\begin{figure}[h!]
    \centering
    \includegraphics[width=\linewidth]{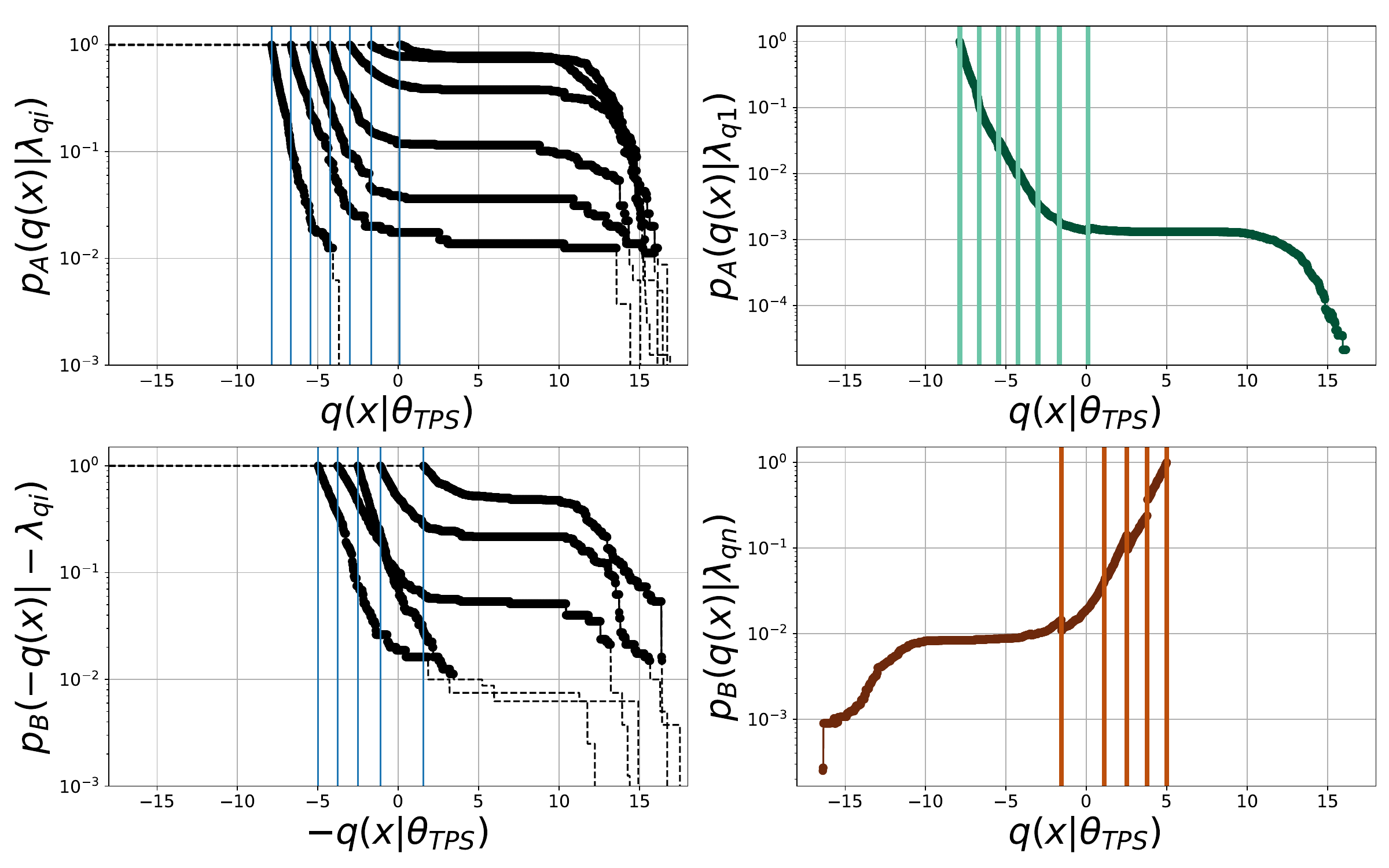}
    \caption{Crossing probabilities on committor model $q(x|\theta_{TPS})$ trained via AIMMD-TPS for the Wolfe-Quapp system. Top left: Forward crossing probabilities $p_A(q(x)|\lambda_{q_i})$ conditioned on crossing forward interface $\lambda_{q_i}$. Top right: Combined crossing probability from initial interface $\lambda_{q_1}$ through WHAM with a cut-off of 0.01. Bottom left: Backward crossing probability $p_B(-q(x)|-\lambda_{q_i})$ for each interface. Bottom right: Combined crossing probability from the backward ensemble, final interface $\lambda_{q_n}$. }
    \label{fig:A:crossing iteration1}
\end{figure}

 Figure\ref{fig:A:crossing iteration1} shows the crossing probabilities and WHAM combined crossing probability for the TIS ensembles. 
For the WHAM procedure performed using OpenPathSampling\cite{swensonOpenPathSamplingPythonFramework2019},    crossing probability histograms below  a cut-off of $10^{-2}$   were discarded.

From the stable state data the flux $\phi_{A1}$ was computed as $\phi_{A1}=\frac{C_1}{C_A}$ where $C_1$ is the number of crossing through the first interface and $C_A$ through the stable state boundary.
Combining the different crossing probabilities and their WHAM weights we obtain the RPE, which consisted of $9\times10^5$ configurations in total. 
The resulting RPE projection on the x,y plane is shown in the main text Fig.~2c. 

\begin{figure}[t!]
    \centering
    \includegraphics[width=\linewidth]{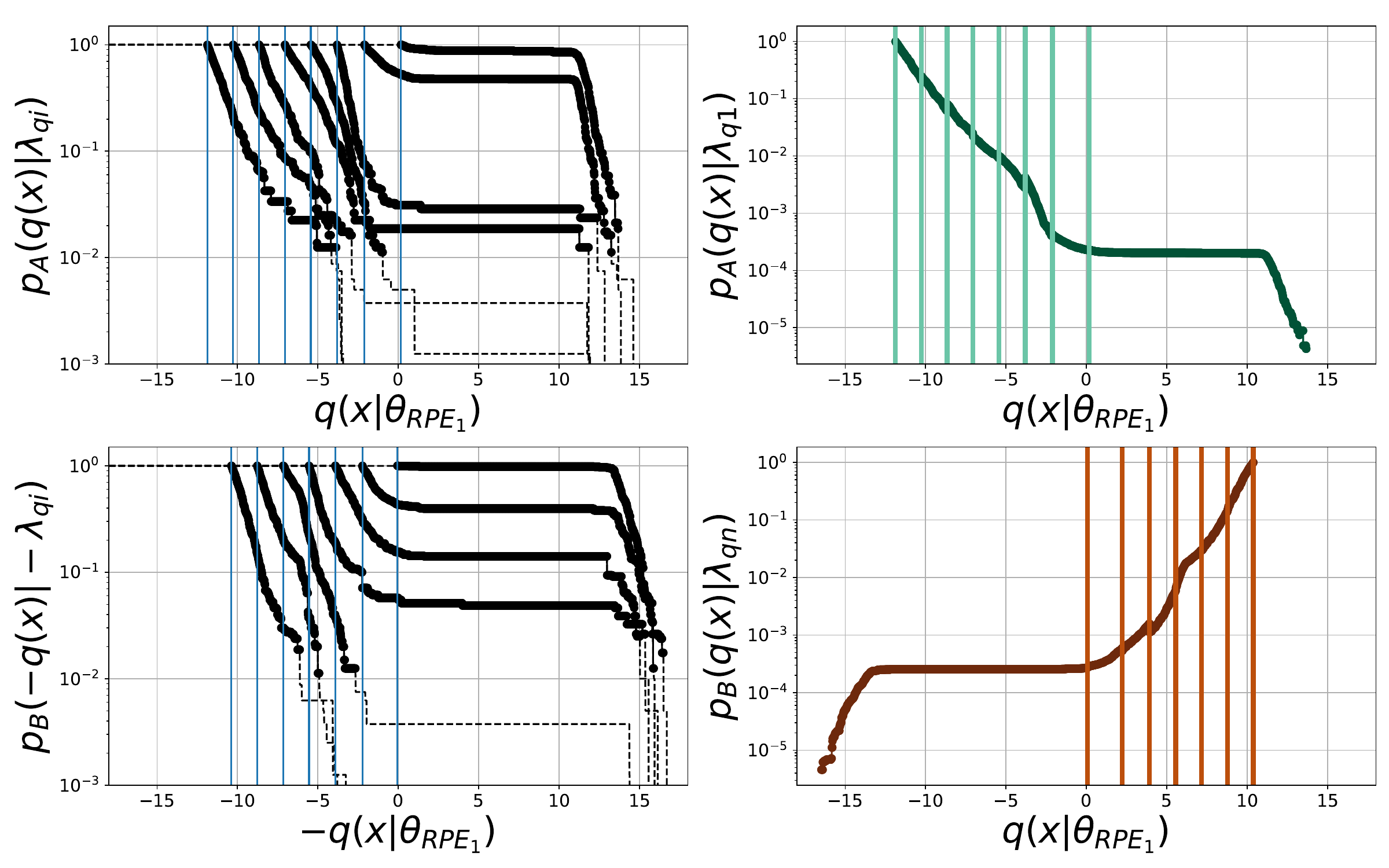}
    \caption{Crossing probabilities on committor model $q(x|\theta_{RPE_1})$ obtained from the first iteration RPE training for the Wolfe-Quapp system. Top left: Forward crossing probabilities $p_A(q(x)|\lambda_{q_i})$ conditioned on crossing forward interface $\lambda_{q_i}$. Top right: Combined crossing probability from initial interface $\lambda_{q_1}$ through WHAM with a cut-off of 0.01. Bottom left: Backward crossing probability $p_B(-q(x)|-\lambda_{q_i})$ for each interface. Bottom right: Combined crossing probability from the backward ensemble, final interface $\lambda_{q_n}$. }
    \label{fig:A:crossing iteration2}
\end{figure}

\begin{figure*}
    \centering
    \includegraphics[width=1\linewidth]{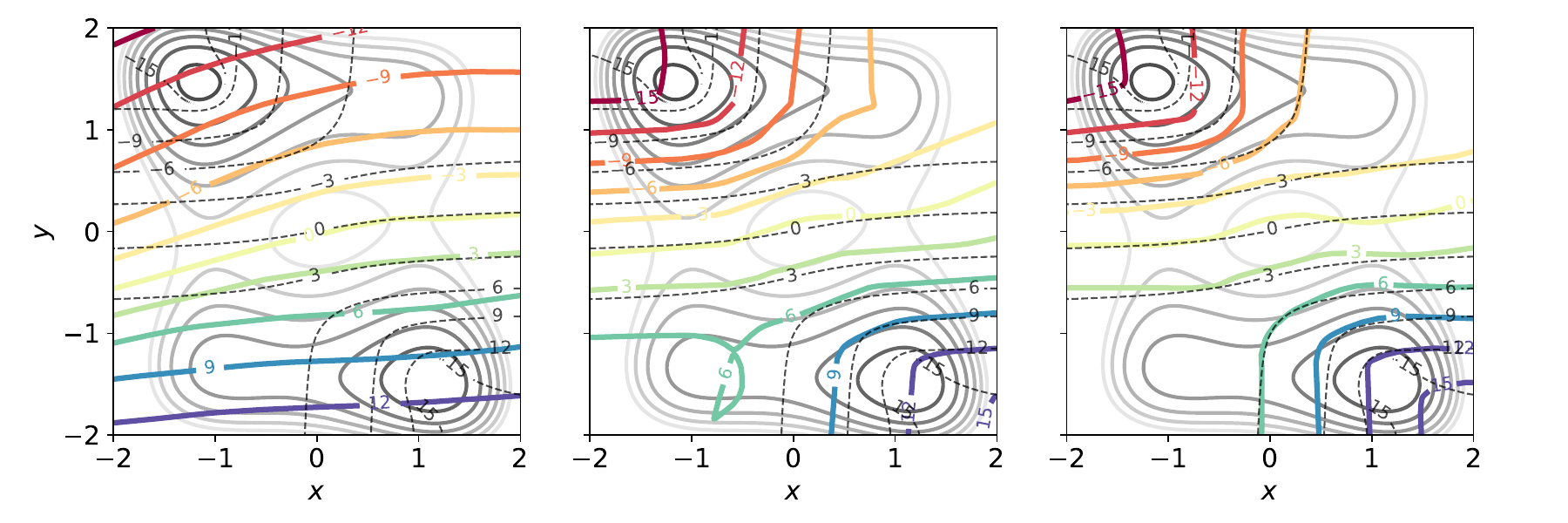}
    \caption{Left: Logit committor model  $q(x,y|\theta)$ slice (colored contours) 
    after initial AIMMD-TPS 
    compared to the theoretical committor (dashed lines). 
    Middle: After the first TIS iteration RPE committor learning and
    Right: After the second TIS iteration the committor quantitatively agrees with theory, up to $q=12$. }
    \label{fig:WQ_committor_projection_result}
\end{figure*}

\begin{figure}[t!]
    \centering
\includegraphics[width=1\linewidth]{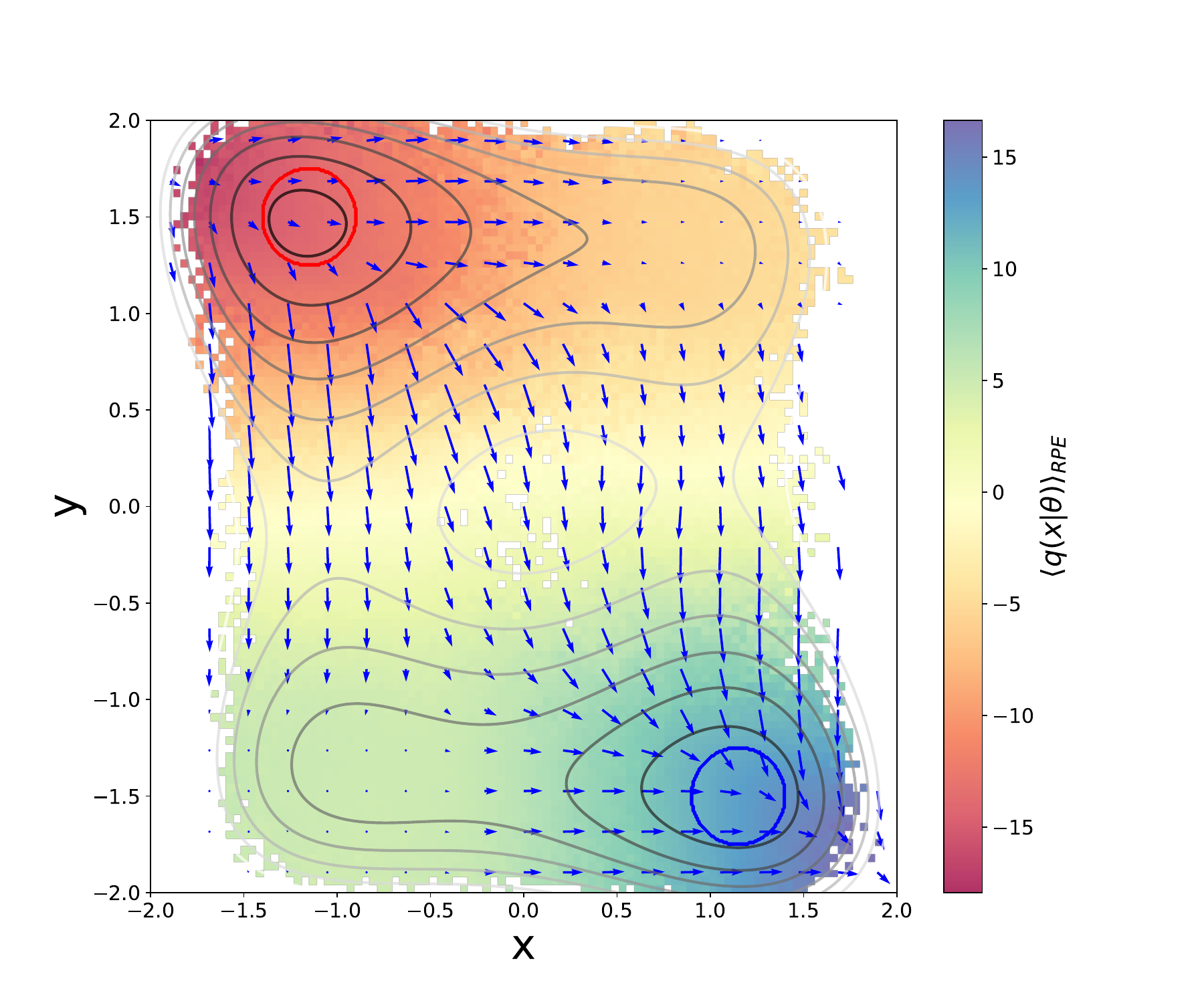}
    \caption{Gradient field of the logit committor model $q(x|\theta)$ projected on the x,y-plane. Arrows indicate gradient direction and relative magnitude; background color shows the projected committor value. Red and blue rings denote the A and B states, respectively.}
    \label{fig:A WQ_gradient_field}
\end{figure}
\begin{figure*}[t!]
    \centering \includegraphics[width=1\linewidth]{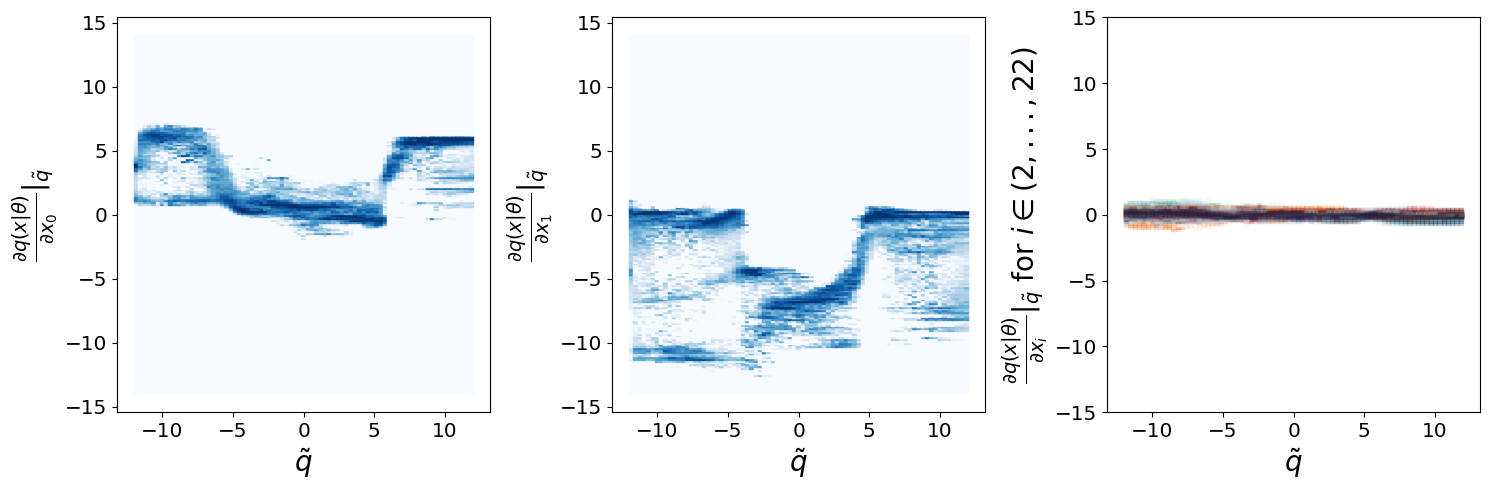}
    \caption{Distribution of gradient contribution across RPE configurations, binned by logit committor model value. Left: x-dimension. Middle: y-dimension and Right: remaining nuisance dimensions. 
    }
    \label{fig:A:gradient_distribution_WQ}
\end{figure*}
\textbf{RPE based weighted committor training:} With the   obtained RPE the committor model is trained in batches of $n_{batch}=32768$ points, randomly selected from the data with uniform sampling of configurations. The weights for the regularization components for smoothness and L1 regularization are set to $\alpha=1/10$ and $\eta=10^{-3}$.  The dataset is split into a train and test set with a ratio of 8/2. 
A stopping criteria is used to prevent overfitting, tracking the improvement of the model on the test-set and stopping if no new improvements in the test loss is obtained after \texttt{n\_conv}=100 epochs. An Adam optimizer is used with a learning rate of $lr=10^{-3}$ and learning rate scheduler with a decay of 0.95 after 2000 epochs.

As shown in Fig.~\ref{fig:WQ_committor_projection_result}b 
the new committor model greatly improves the models contours, retaining the y-component behavior near the TS and gaining more x-dependent behavior near the stable states.

\begin{figure*}[t!]
    \centering
    \includegraphics[width=1\linewidth]{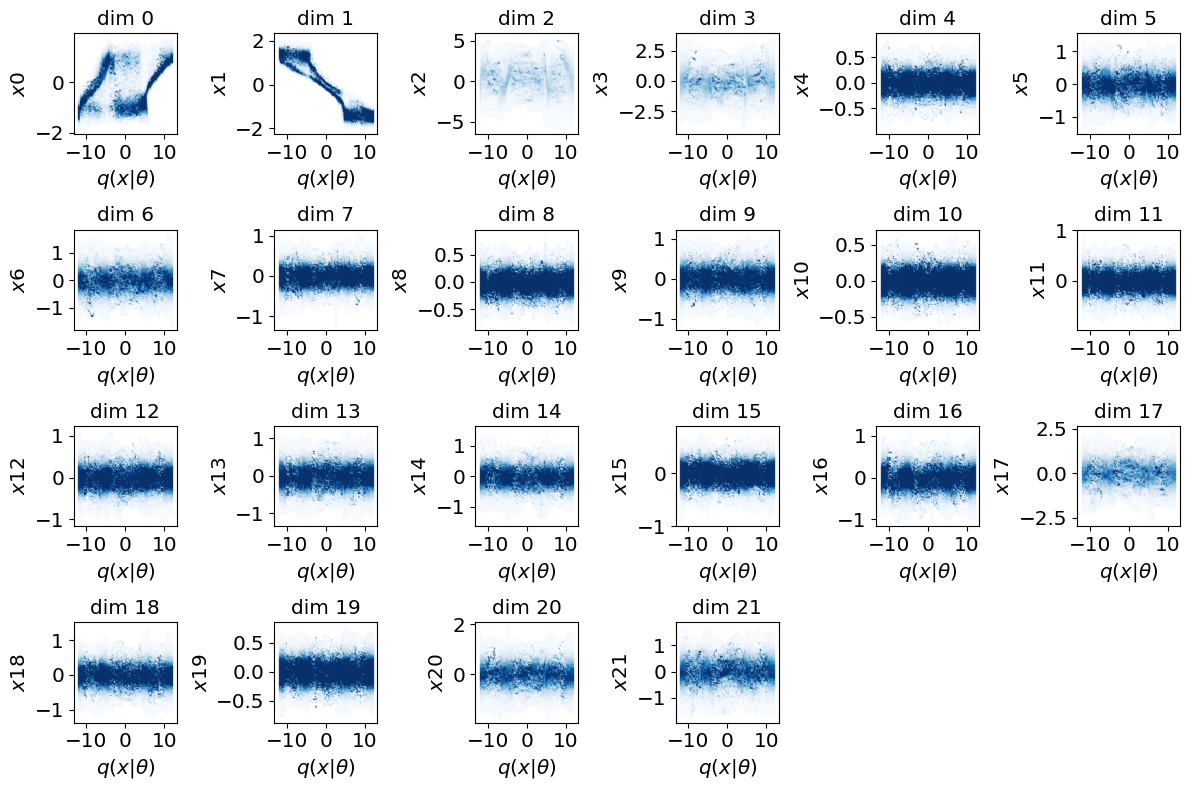}
    \caption{Distribution $\rho(x|q(x|\theta_{RPE_2})=\tilde{q})$ of all 22 input dimensions of the Wolfe-Quapp system as a function of logit committor value $\tilde{q}$ normalized independent for each $\tilde{q}$.}
    \label{fig:A: all_distributions_descriptors_along_q}
\end{figure*}
\textbf{TIS sampling on $q(x|\theta_{RPE_1})$:}
After the first iteration of RPE training a new and improved committor model $q(x|\theta_{RPE_1})$ is obtained on which new interfaces can be placed. Using the Stable state data new boundary values $q_{\text{max}}^{A}=-11.875$ and  $q_{\text{min}}^{B}=10.375$ are determined. 
The forward interfaces on this newly trained model $q(x|\theta_{RPE_1})$ were: [-11.875, -10.27, -8.66, -7.05, -5.44, -3.81, -2.11, 0.16] and the backward interfaces: [10.375, 8.77, 7.16, 5.55, 3.92, 2.23, 0.06]. For the TIS sampling the same Gaussian selector was used as for the previous iteration ($\alpha_g=4$ and mean $l_0=\lambda_{qi} \mp0.2$).
In Fig.~\ref{fig:A:crossing iteration2}, the interface conditional crossing probabilities and WHAM combined crossing probabilities are shown. For the WHAM procedure again a cut-off of $10^{-2}$ was used.

\textbf{RPE iteration 2:}
After extracting the RPE from TIS on the $q(x|\theta_{RPE_1})$ interfaces (consisting of $1.2\times10^6$ configurations), training is continued using the same batch size and loss function combination (L1 regularization and smoothness term; see table \ref{tab:A:WQ simulation settings}, but with a reduced learning rate $lr=10^{-4}$. The resulting committor slice on the x,y,-plane is shown in the main text Fig.~2f,
illustrating a further refinement of the model and its qualitative agreement with the exact committor.

\begin{figure*}[t]
    \centering
    \includegraphics[width=1\linewidth]{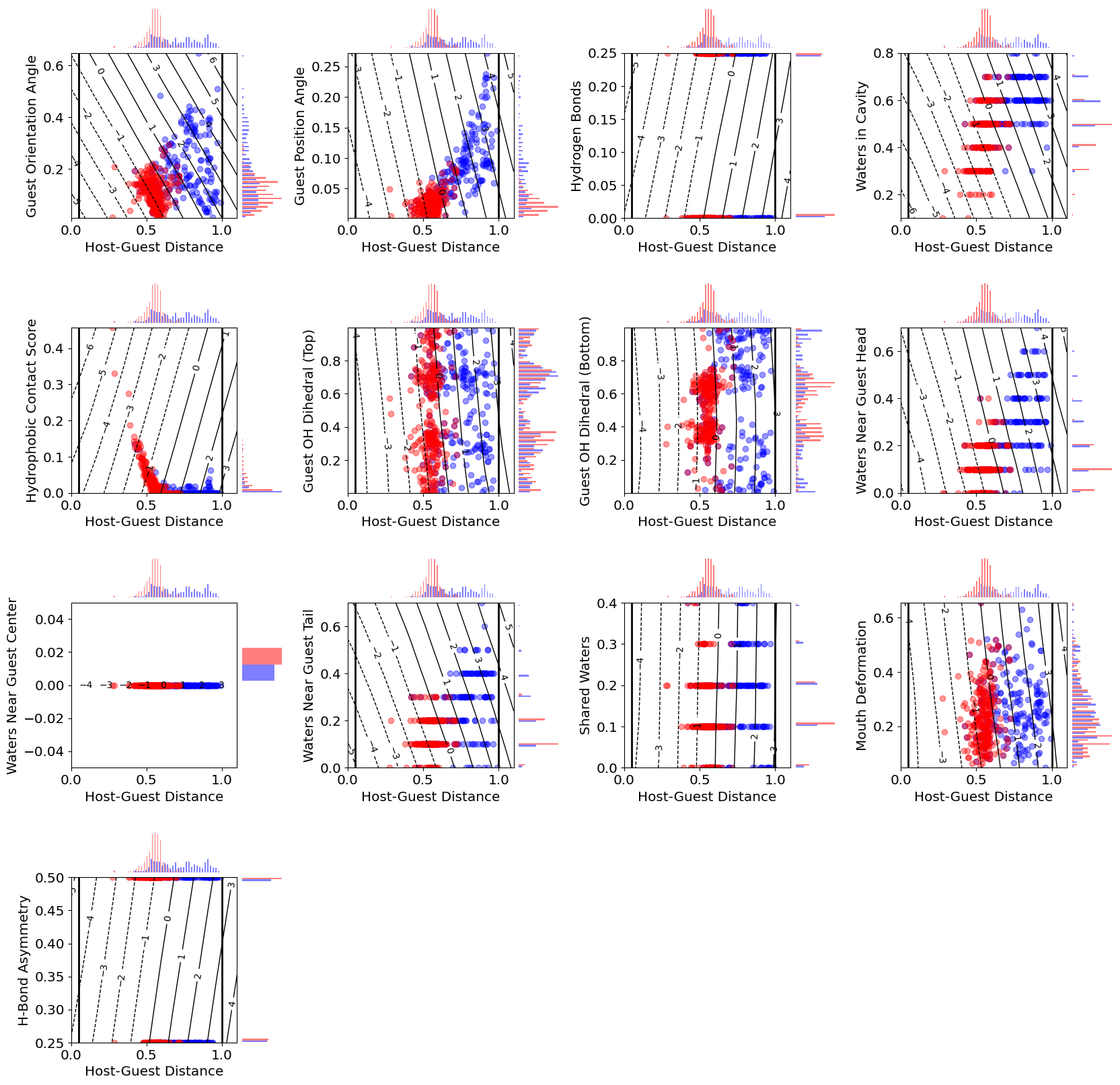}
    \caption{AIMMD-TPS shooting points for the Host Guest system colored by shot result (red: transition to bound state, blue: transition to unbound state), overlaid with committor model contours. Shown as a slice across the primary distance descriptor ($x$-axis) and the remaining scaled collective variables ($y$-axis). AIMMD-TPS was performed using a Lorentzian selector in committor space with $\gamma = 5$.}
    \label{fig:A:HG AIMMD-TPS model result}
\end{figure*}

\textbf{Model analysis:}
With the updated committor model $q(x|\theta)$ and the RPE, we can now analyze the transition mechanisms in great detail. 
To identify the descriptor dimensions that govern the process, we compute the gradient  $\partial_i q(x|\theta)$ with respect to each input coordinate $x_i$. Dimensions with negligible gradient are not involved in the reaction mechanism, whereas large gradients magnitudes indicate features which strongly influence the transition. 

In contrast to earlier work using HIPR\cite{Jung2023}, which perturbs inputs randomly and globally, the gradient-based analysis employed here captures the local sensitivity of the committor. HIPR identifies dominant components across the full transition but may overlook subtle, localized contributions due to its global perturbation strategy. In our approach, the gradient directly reveals how small changes in individual dimensions affect the committor locally along the path ensemble.

The spatial distribution of gradients also informs us about the smoothness and monotonicity of the model. In particular, vanishing gradient $|\nabla q(x|\theta)|<\epsilon$ with $\epsilon=10^{-5}$ may indicate saddle points or "bubbles" - non-monotonic regions where the committor does not increase or decrease consistently in all directions. No such unphysical behaviour was observed in the current model.

Figure \ref{fig:A WQ_gradient_field} shows the expection value of the gradient projected onto the x,y-plane:
\begin{equation}
    \langle \partial_x q(x|\theta)\delta(x-\tilde{x})\delta(y-\tilde{y})\rangle_{RPE},
\end{equation}
visualized as a vector field overlaid on the logit committor model surface.

The analysis reveals two distinct transition channels and demonstrates that the model captures the directional sensitivity of the committor, highlighting the mechanistic pathways between the  A and B states.

Figure \ref{fig:A:gradient_distribution_WQ} shows the distribution of gradient contribution per descriptor dimensions across the RPE as a function of the logit  committor model value $\tilde{q}$. The first two panels correspond to the $x$ and $y$ coordinates of the WQ potential, while the right panel shows the harmonic nuisance dimensions.
While the $x$ and $y$ gradients show large, structural variations with $\tilde{q}$, the nuisance dimensions exhibit near-zero gradients across the board, indicating their irrelevance to the reaction mechanism. A notable feature is the bimodality of the gradient distribution at fixed $\tilde{q}$, reflecting the presence of the two mechanistic pathways.

\textbf{Descriptor behavior along  $q(x)$:}
To further investigate the mechanism, we can examine the distribution $\rho(x|\tilde{q})$ of each input coordinate as a function of the committor values. This allows us to identify features that evolve along the reaction coordinate and distinguish them from static or nuisance dimensions.

Figure~\ref{fig:A: all_distributions_descriptors_along_q} shows that only the first two dimensions exhibit strong shifts with $\tilde{q}$, consistent with their mechanistic relevance. The nuisance dimenions maintain a harmonic distribution across the full range of $\tilde{q}$, validating the model's ability to supress irrelevant features. Some deviation in dimensions 2 and 3, corresponding to broader harmonic wells, may reflect some small model contribution due to insufficient sampling in these dimensions.

From these distributions, mechanistic features can be extracted either via the mean $\langle x \rangle_{\rho(x|\tilde{q})}$ or the mode $\arg \max_x \rho(x|\tilde{q})$ as a function of $\tilde{q}$, providing a detailed picture of how relevant coordinates evolve during the transition.

%% file: Sections/SI/Host-Guest.tex
\begin{figure}[t]
    \centering
    \includegraphics[width=1\linewidth]{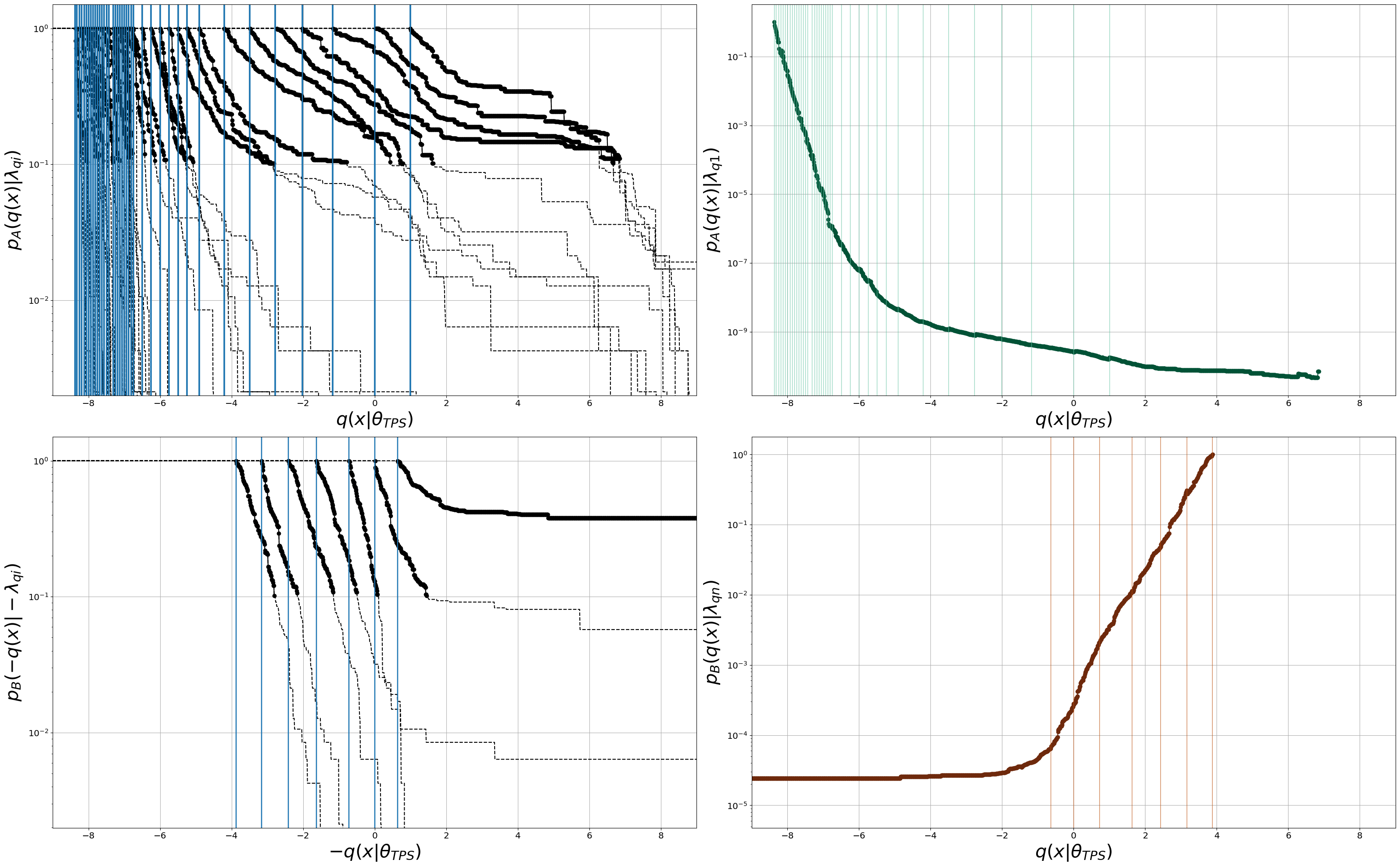}
    \caption{Crossing probabilities based on the AIMMD-TPS-trained committor model $q(x|\theta_{\text{TPS}})$ for the Host–Guest system. Top left: Forward crossing probabilities $p_A(q(x)|\lambda_{q_i})$ (black dotted lines indicates region above cutoff, cutoff = 0.1). Top right: Combined WHAM-based crossing probability . Bottom left: Backward crossing probabilities $p_B(-q(x)|-\lambda_{q_i})$. Bottom right: Combined backward crossing probability.}
    \label{fig:A: Crossing Probabilities HG it 1}
\end{figure}

\textbf{Stable State Simulations}
5 ns Equilibrium simulations of the bound and unbound states were performed to generate a reference dataset, which was used for initial interface placement and to include stable state data in the reweighted path ensemble. These simulations also served to pre-train the initial committor model.

\textbf{Initial AIMMD-TPS Procedure}
AIMMD-based transition path sampling (TPS) was performed to generate reactive trajectories connecting the bound and unbound states. The goal of this initial sampling stage was to produce a preliminary committor model $q(x|\theta_{\text{TPS}})$ that distinguishes between the two metastable basins.

TPS simulations were performed with a maximum path length of $0.2$~ns. A total of $500$ shooting moves were attempted, using a Lorentzian selector~\cite{Jung2023} in committor space with scale parameter $\gamma = 5$. This selector focuses sampling of configurations near the transition region while still allowing exploration further from the TS. Of the 500 attempted moves, 54 resulted in accepted transition paths, corresponding to an acceptance rate of 10.4\%.
The initial trajectory used to seed the TPS simulation was generated by performing a committor analysis from a selected configuration and connecting one trajectory leading to the bound state and one leading to the unbound state, thus constructing a complete reactive path.

The resulting committor model $q(x|\theta_{\text{TPS}})$, trained on the outcome of attempted TPS shooting moves, is shown in Fig~\ref{fig:A:HG AIMMD-TPS model result}. The committor is visualized as a slice across the center-of-geometry distance $r_{CG}$ and other scaled descriptors.

\begin{figure}
    \centering
    \includegraphics[width=1\linewidth]{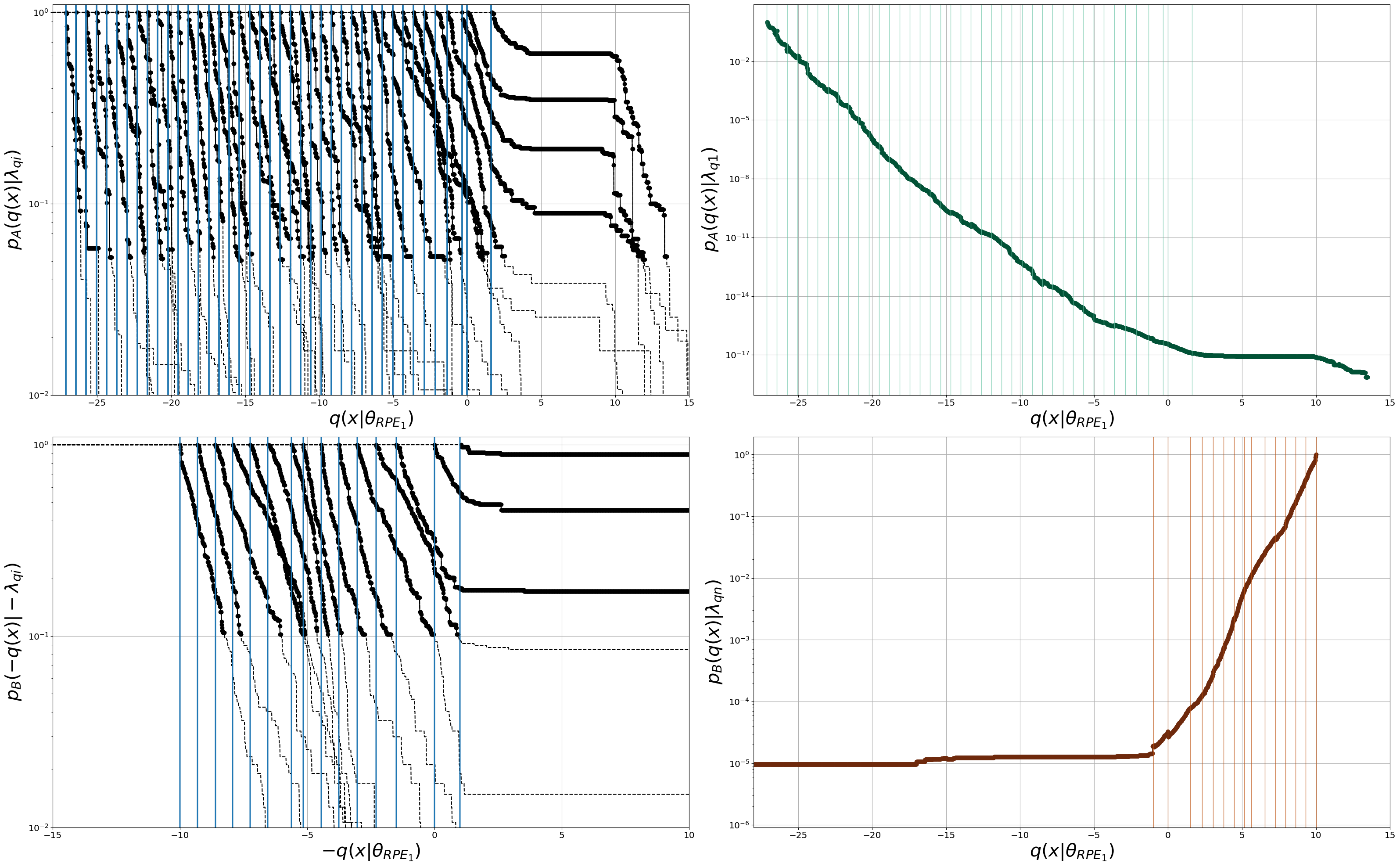}
    \caption{Crossing probabilities based on the first iteration AIMMD RPE trained committor model $q(x|\theta_{\text{RPE}_1})$ for the Host–Guest system. Top left: Forward crossing probabilities $p_A(q(x)|\lambda_{q_i})$ (black dotted lines indicates region above cutoff, cutoff = 0.05). Top right: Combined WHAM-based crossing probability . Bottom left: Backward crossing probabilities $p_B(-q(x)|-\lambda_{q_i})$ (with cutoff=0.1). Bottom right: Combined backward crossing probability.}
    \label{fig:A: Crossing Probabilities HG it 2}
\end{figure}

\begin{figure*}[t!]
    \centering
    \includegraphics[width=1\linewidth]{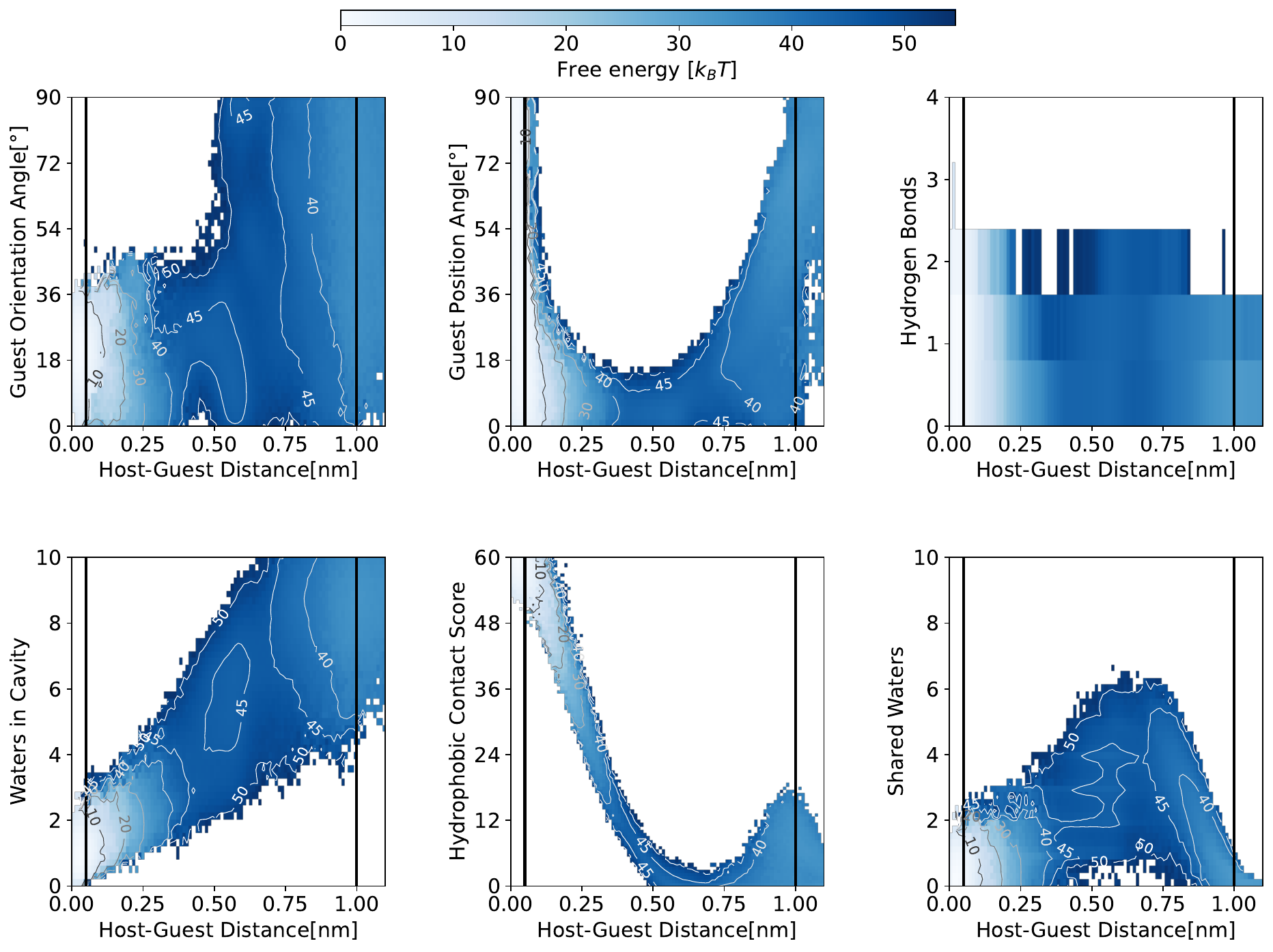}
    \caption{Free energy surface projected onto six descriptors versus the host–guest distance. The free energy was computed using the reweighted path ensemble constructed via TIS performed on $q(x|\theta_{RPE_1})$ interfaces.}
    \label{fig:RPE_HG_iteration_2}
\end{figure*}

\begin{table}[t!]
\caption{TIS info forward ensembles $q(x|\theta_{RPE_1})$ interfaces}
\label{tab:TIS_info_forward}
\scriptsize
\resizebox{0.5\textwidth}{!}{\begin{tabular}{lrrrr}
\toprule
$\lambda_q$ & avg path length [ps] & acc ratio & MC steps & Selector $\alpha_g$ \\
\midrule
-27.125 & 0.850 & 0.158 & 1000 & [0.0078125, 0.02] \\
-26.430 & 1.020 & 0.195 & 1000 & [0.0078125, 0.02] \\
-25.740 & 0.900 & 0.183 & 1000 & [0.0078125, 0.02] \\
-25.050 & 0.840 & 0.195 & 1000 & [0.0078125, 0.02] \\
-24.360 & 1.060 & 0.219 & 1000 & [0.0078125, 0.02] \\
-23.670 & 1.250 & 0.213 & 1000 & [0.0078125, 0.02] \\
-22.980 & 1.040 & 0.165 & 1000 & [0.0078125, 0.02] \\
-22.290 & 1.310 & 0.198 & 1000 & [0.0078125, 0.02] \\
-21.600 & 1.260 & 0.181 & 1000 & [0.0078125, 0.02] \\
-20.910 & 1.250 & 0.179 & 1000 & [0.0078125, 0.02] \\
-20.220 & 1.260 & 0.204 & 1000 & [0.0078125, 0.02] \\
-19.530 & 1.370 & 0.229 & 1000 & [0.0078125, 0.02] \\
-18.840 & 1.570 & 0.238 & 1000 & [0.0078125, 0.02] \\
-18.150 & 1.600 & 0.260 & 1000 & [0.0078125, 0.02] \\
-17.460 & 1.690 & 0.274 & 1000 & [0.0078125, 0.02] \\
-16.770 & 1.770 & 0.241 & 1000 & [0.0078125, 0.02] \\
-16.080 & 1.460 & 0.210 & 1000 & [0.0078125, 0.02] \\
-15.390 & 2.140 & 0.258 & 500 & [0.0078125] \\
-14.700 & 1.860 & 0.232 & 500 & [0.0078125] \\
-14.010 & 2.130 & 0.258 & 500 & [0.0078125] \\
-13.320 & 2.780 & 0.204 & 500 & [0.0078125] \\
-12.630 & 2.280 & 0.226 & 500 & [0.0078125] \\
-11.940 & 2.160 & 0.204 & 500 & [0.0078125] \\
-11.250 & 2.620 & 0.222 & 500 & [0.0078125] \\
-10.560 & 2.440 & 0.172 & 500 & [0.0078125] \\
-9.870 & 2.930 & 0.218 & 500 & [0.0078125] \\
-9.180 & 3.800 & 0.240 & 500 & [0.0078125] \\
-8.490 & 2.860 & 0.200 & 500 & [0.0078125] \\
-7.800 & 2.690 & 0.228 & 500 & [0.0078125] \\
-7.110 & 3.230 & 0.194 & 500 & [0.0078125] \\
-6.420 & 3.200 & 0.264 & 500 & [0.0078125] \\
-5.730 & 8.270 & 0.250 & 500 & [0.0078125] \\
-5.030 & 18.720 & 0.312 & 500 & [0.0078125] \\
-4.330 & 11.360 & 0.326 & 500 & [0.0078125] \\
-3.620 & 5.920 & 0.278 & 500 & [0.0078125] \\
-2.900 & 24.950 & 0.442 & 500 & [0.5] \\
-2.150 & 43.480 & 0.362 & 500 & [0.5] \\
-1.330 & 68.160 & 0.382 & 500 & [0.5] \\
-0.330 & 96.460 & 0.391 & 445 & [0.5] \\
0.000 & 95.860 & 0.340 & 415 & [0.5] \\
1.630 & 146.910 & 0.178 & 500 & [0.5] \\
\bottomrule
\end{tabular}
}
\end{table}

\begin{table}[t]
\caption{TIS info backward ensembles $q(x|\theta_{RPE_1})$ interfaces}
\label{tab:TIS_info_backward}
\scriptsize
\resizebox{0.5\textwidth}{!}{\begin{tabular}{lrrrr}
\toprule
$\lambda_q$ & avg path length [ps] & acc ratio & MC steps & Selector $\alpha_g$\\
\midrule
-1.000 & 125.920 & 0.102 & 433 & [0.5] \\
0.000 & 126.610 & 0.230 & 500 & [0.5] \\
0.550 & 82.510 & 0.346 & 500 & [0.5] \\
1.500 & 74.780 & 0.405 & 359 & [0.5] \\
2.290 & 52.580 & 0.440 & 500 & [0.5] \\
3.040 & 39.260 & 0.440 & 500 & [0.5] \\
3.760 & 40.010 & 0.404 & 500 & [0.5] \\
4.460 & 35.580 & 0.434 & 500 & [0.5] \\
5.160 & 30.560 & 0.476 & 500 & [0.5] \\
5.620 & 27.400 & 0.510 & 500 & [0.5] \\
6.550 & 24.130 & 0.522 & 500 & [0.5] \\
7.240 & 19.530 & 0.464 & 500 & [0.5] \\
7.930 & 20.380 & 0.554 & 500 & [0.5] \\
8.610 & 12.040 & 0.468 & 500 & [0.5] \\
9.310 & 6.220 & 0.464 & 500 & [0.5] \\
10.000 & 9.120 & 0.438 & 500 & [0.5] \\
\bottomrule
\end{tabular}
}
\end{table}

\textbf{RPE from $q(x|\theta_{TPS})$ based interfaces:}
For the first iteration TIS calculations, a Gaussian selector was applied in $q(x|\theta_{TPS})$ with $\alpha_g=1/2$ and $l_0 = \lambda_{q_i}$ for backward and forward interface ensembles, respectively. A two-way shooting move scheme was used, guiding the selection of shooting points close to the interface. For each interface 500 MC steps are performed of which the first 30 are discarded to remove corelation from the initial path.
Interfaces were placed to ensure proper overlap, with a denser distribution near the A state for the forward interfaces (from $ q = -8.525 $ to $ q = -5.0$) to capture the steep crossing probabilities in the highly stable basin. For these interfaces a more narrow gaussian selector was needed as the trajectories did not traverse far in q-space. For the low q-values ($q=-8.525$ till $q=-6.875$) the gaussian selector was set to $\alpha_g=4$ with $l_0=\lambda_{q_i}\pm0.1$ where -0.1 is used for forward and + 0.1 for backward ensembles.
Fewer interfaces were needed further from the bound state as crossing probabilities increased ensuring adequate overlap between interfaces. In total 43 forward and 7 backward interfaces are sampled.
Figure \ref{fig:A: Crossing Probabilities HG it 1} shows the full set of interfaces, along with their crossing probabilities.
Note the steep crossing  probability of the bound to unbound transition.
For the combined crossing probability a WHAM cut-off of $0.1$ is used, as highlighted by the black plotted parts of the crossing probabilities.

\textbf{RPE training iteration 1:}
Using the reweighted path ensemble from the interface ensembles, the committor model $q(x|\theta_{\text{TPS}})$ was refined to capture behavior across the full transition range, including regions with extremely small transition probabilities ($p_B(x) \sim 10^{-15}$) near the stable states.
The initial AIMMD-TPS model has been trained on 14 descriptors, for the RPE training a reduced descriptor set of 7 descriptors could be identified. The descriptors by which the model is trained are $r_{CG},\phi_{\text{or}},\theta_{\text{pos}},n_{\text{HB}},n_{\text{waters}},S_{\text{hydrophobic}}$ and $n_{\text{water,shared}}$. 
The model was trained using an Adam optimizer~\cite{Kingma2015Adam:Optimization} with a batch size of $n_{\text{batch}} = 32{,}768$, randomly sampled from an RPE dataset of approximately $n_{\text{points}} \sim 3.7 \times 10^6$. To prevent the logit committor model from diverging toward infinity in distant unbound regions, stable-state data beyond $r_{CG} = 1.1\,\mathrm{nm}$ was excluded from training. By definition, all unbound stable-state data goes to its state and thus contributes uniformly with $p_B = 1$, driving the logit committor $q \rightarrow \infty$. At the edge of the state reactive sampled trajectories in the RPE also influence the model however including data much further beyond the cutoff would have biased the model due to the extensive diffusive spread.
The data was split 80/20 into training and validation subsets. Loss weights and optimizer parameters are listed in Table~\ref{tab:A:CB7B2_simulation_settings} under \textit{RPE training parameters AIMMD}.

For RPE training, the smoothness and L1 loss terms contribution weights have to be carefully selected to balance the small likelihood contribution in the total loss. This was necessary due to the large weights near the stable basins because of the extremely peaked behavior in the stable-state regions. The overall contribution of the loss scales with the height of the barrier. A barrier of $e^{-\Delta F}$ leads to an weighted-log-likelihood loss of the same order. For the training a ratio between the log likelihood loss and smoothness  loss was set to be 1/10th the log-likelihood loss and no l1 loss was applied. 
The resulting committor model $q(x|\theta_{\text{RPE}_1})$ now captures the behavior up to $p_B\sim 10^{-13}$.

\begin{figure*}
    \centering
    \includegraphics[width=1\linewidth]{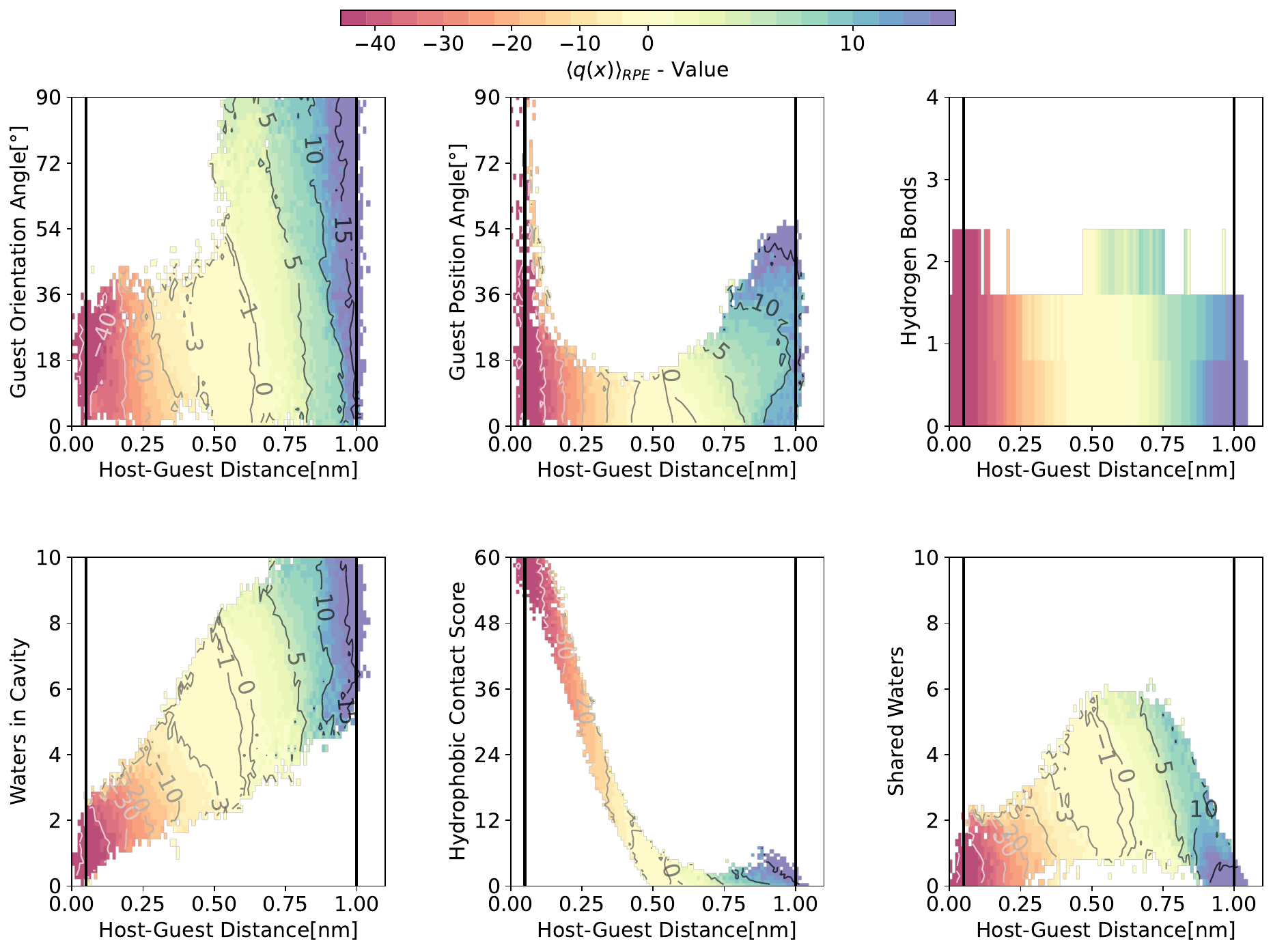}
    \caption{Marginalized $q(x,y)$ logit committor as a function of six descriptors relative to the host–guest distance. The RPE data used was based on TIS in $q(x|\theta_{RPE_1})$ interfaces.}
    \label{fig:SI:HG_RPE_q_marginalization}
\end{figure*}

\begin{figure*}
    \centering
    
\includegraphics[width=1\linewidth]{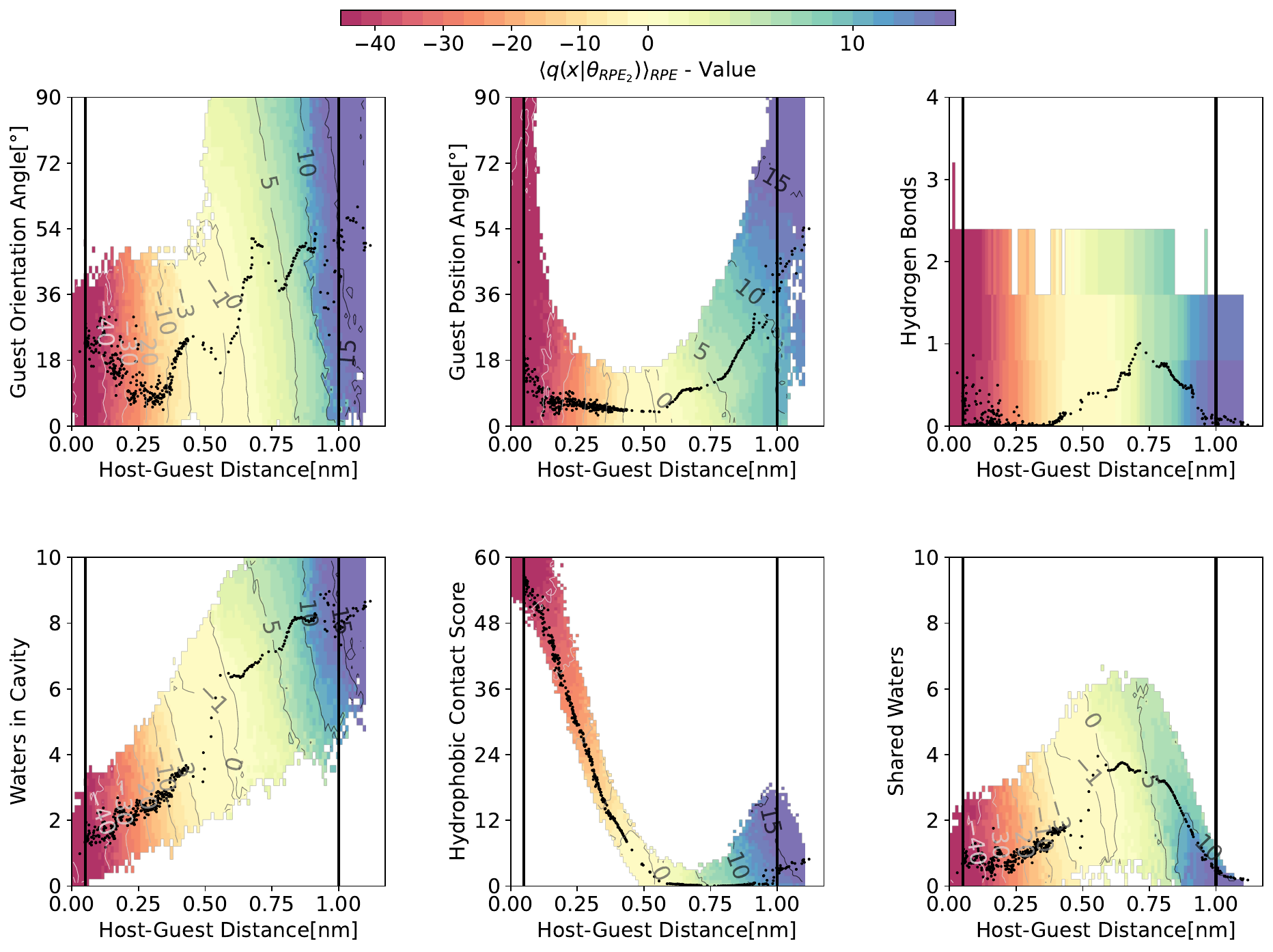}
    \caption{Logit committor projected onto the plane of six descriptors versus distance. Colored regions show the expectation value of the learned $q(x|\theta_{RPE_2})$-model, projected from RPE data. Stable states are defined as bound ($r_{CG}<0.05$nm) and unbound ($r_{CG}>1.0$nm). The black points show the mean path allong the model $q(x|\theta_{RPE_2})$ $\langle \xi\rangle|\tilde{q})$ for $\tilde{q}$ in $[-45,15]$.
    }
    \label{fig:SI:HG_result_it_2_committor_projection}
\end{figure*}

\textbf{RPE from $q(x|\theta_{RPE_1})$ based interfaces:}
For the second iteration of TIS sampling, interfaces are placed using $q(x \mid \theta_{\mathrm{RPE}_1})$ with a $50\%$ overlap between consecutive interfaces assuming an exact committor model starting with an initial interface at $q=-27.125$ and $q=10$ for forward and backward ensembles, respectively. For each interface again a gaussian selector is used. However since the steep barrier out of the bound state and the large spread in q-range because of this, a large gaussian selector width is used.
The gaussian selector for the first forward interfaces from $q=-27.125$ till $q=-3.62$ are set to $\alpha_g=1/128$ (gaussian width of 8 in $q$-space) and $l_0=\lambda_{q_i}$ and for the latter interfaces $\alpha_g=1/2$ (width of 1) and $l_0=\lambda_{q_i}$.
For all backward interfaces a gaussian selector with a $\alpha_g=1/2$ and $l_0=\lambda_{q_i}$ is used.
The resulting crossing probabilities are shown in Figure \ref{fig:A: Crossing Probabilities HG it 2}. The newly obtained crossing probabilities from the improved logit committor model exhibit a linear trend in the logarithm of crossing probability away from the transition state, highlighting the direct relationship between crossing probabilities and the committor as the model improves.
In Table \ref{tab:TIS_info_forward} and \ref{tab:TIS_info_backward} the acceptance ratio, path length and number of MC steps performed for each interfaces are shown.
To obtain the crossing probabilities and RPE for each interface ensemble the first 30 trajectories to decorrelate from the initial trajectory, which is the final AIMMD-TPS-sampled reactive path. 
In the WHAM reweighting a cut-off of $0.05$ is used for the forward ensembles and $0.1$ for the backward ensembles. The resulting RPE is shown in Fig~\ref{fig:RPE_HG_iteration_2} and the marginalized logit committor according in Fig~\ref{fig:SI:HG_RPE_q_marginalization}

\textbf{RPE training iteration 2:}
For the second iteration of RPE training the RPE dataset contained a total of $n_{points}\sim5.6\times 10^6$ datapoints again split into a train and test set with ratio 8/2, respectively. For the training a ratio between the log likelihood loss and smoothness loss was set such that the smoothness contributed 1/10th compared to the log-likelihood loss. 

\begin{figure*}
    \centering    \includegraphics[width=1\linewidth]{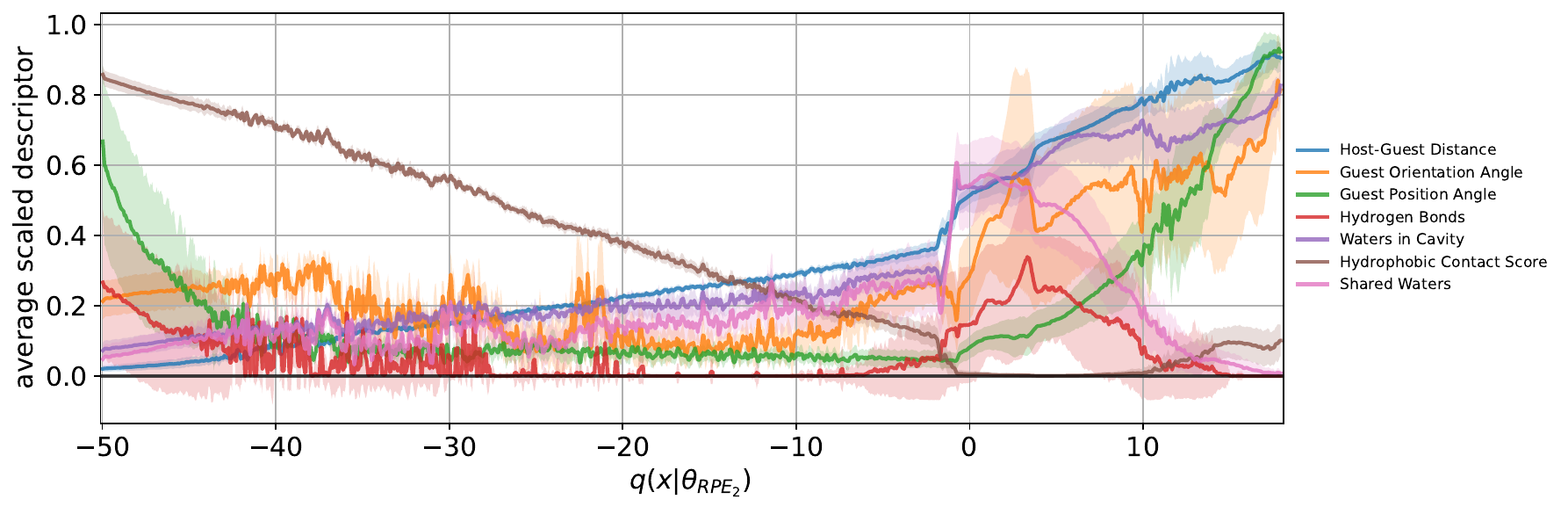}
    \caption{Average normalized descriptors along the $q(x|\theta_{RPE_2})$ model
    $\langle \xi(x)\rangle_{\text{RPE}|\tilde{q} }$. Descriptors are normalized such that their min,max values of the RPE fall within the range [0,1]. }
    \label{fig:SI:HG_iteration_2_average_descriptors}
\vspace{-0.0cm}
\end{figure*}

\begin{figure*}
    \centering
    \includegraphics[width=1\linewidth]{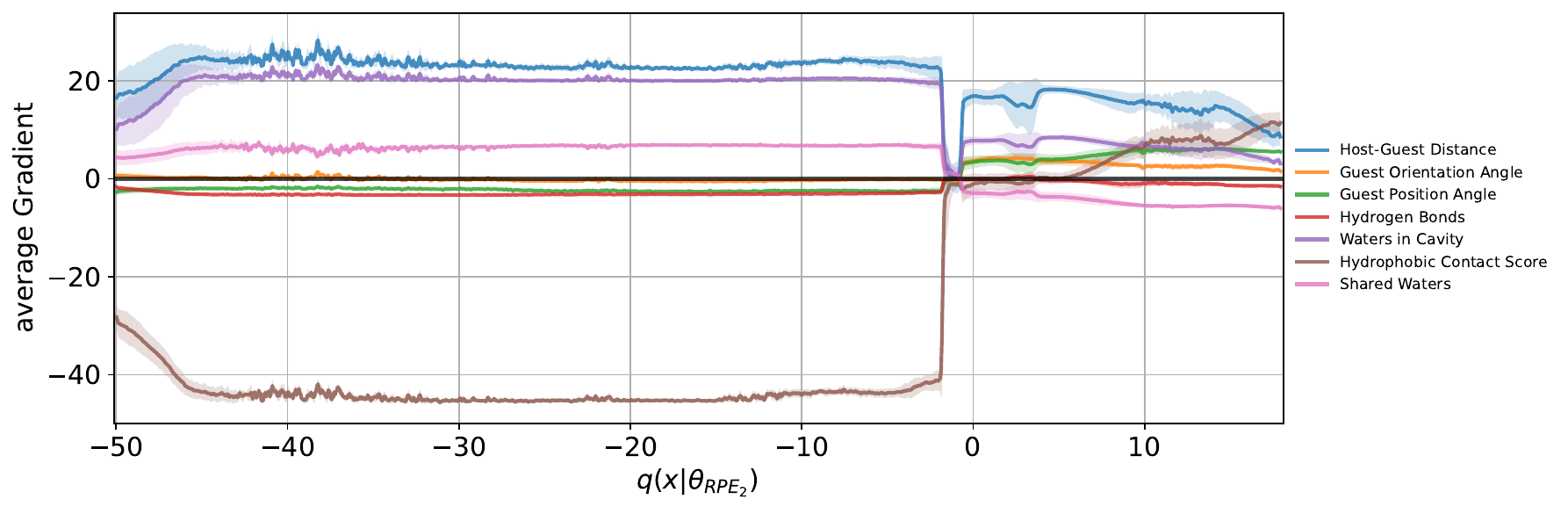}
    \caption{Average gradient along the $q(x|\theta_{RPE_2})$ model
    $\langle \nabla_{\xi} q(x|\theta_{RPE_2})\rangle_{\text{RPE}|\tilde{q}}$.
    }
    \label{fig:SI:HG_iteration_2_average_gradient}
    \vspace{-0.3cm}
\end{figure*}

In the Main Text Fig~4a-d and in Fig~\ref{fig:SI:HG_result_it_2_committor_projection} the resulting model projection along 
the key descriptors relative to the Center distance. The dotted path indicates the average descriptor position for a given q value based on the RPE distribution, showing  how the guest moves out of  (or into) the host cavity.  Figure~\ref{fig:SI:HG_iteration_2_average_descriptors} shows the 
normalized descriptors along the q model. In the main text Fig~4f the gradient of the model along q, indicates how important the key ingredients are  during the entire transition shown in terms of the unit gradient showcasing the relative importance between descriptors. In Fig~\ref{fig:SI:HG_iteration_2_average_gradient} the average model gradient is shown, indicating a metastable state where the gradient becomes zero.

\begin{figure}[h!]
    \centering
    \includegraphics[width=\linewidth]{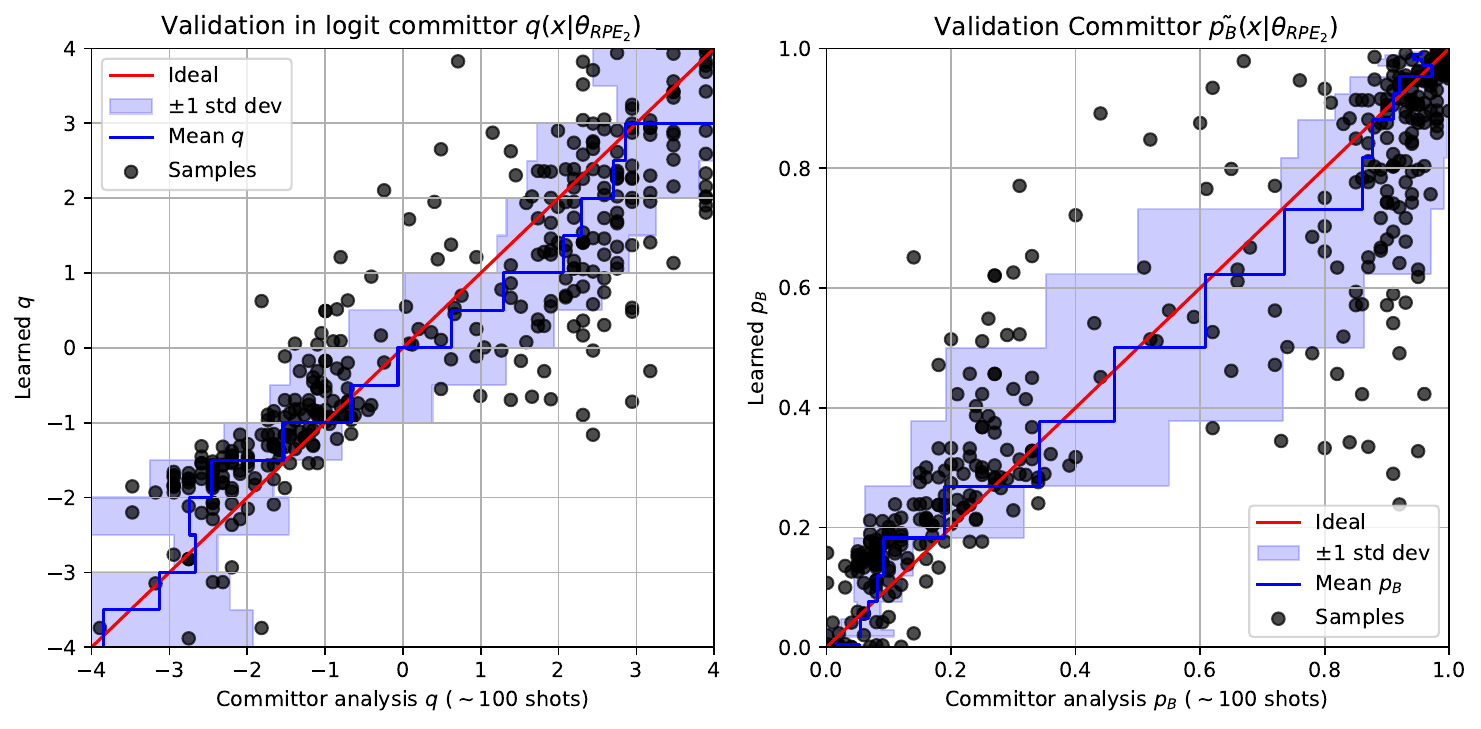}
    \caption{Validation of the learned committor. Comparison between the  predicted  $q(x|\theta_{RPE_2})$ model and the committor obtained by performing a committor analysis through repeated sampling from  given molecular configurations. The average of the sampled committors and their standard deviation were calculated in bins of the learned committor indicated by the vertical steps. For reference the red line indicates identity. Left: validation along Logit Committor $q$.  Right: validation along committor $p_B$.
    }
    \label{fig:A:committor model validation}
\end{figure}

\textbf{Committor validation:}
To strengthen the conclusion and to show that we have obtained a committor model that truthfully captures the true committor underlying the identified mechanism of the system, we conducted additional validation steps. 
We performed a committor analysis on 400 representative configurations sampling in the transition region (range $q\in[-4,4])$, generating 100 trajectories from each using Boltzmann-distributed initial velocities. The goal of the test is to verify whether the trained committor model correctly classifies these configurations and whether the model-predicted transition surface is thus properly determined. While the model gives committor values far lower than this range, committor analysis becomes increasingly expensive as the committor value becomes close to the states. The resulting cross-corelation is shown in Fig. \ref{fig:A:committor model validation} in which we see the learned model committor and sampled committor through a committor analysis based on 100 shots.
We conclude that the learned model can reproduce the sampled committor  values quite reasonably. While though there is  stile  some spread around the $q=0$ region, the average and standard deviation follows the ideal diagonal quite well. 

%% file: Sections/SI/Loss_function_theory_observation.tex
A theoretical analysis by Lechner et al. \cite{lechnerNonlinearReactionCoordinate2010} suggests that reweighting leads to a log-likelihood formulation over equilibrium configurations:
\begin{equation}
\mathcal{L}_{wl} = N \sum_x \rho_{\text{eq}}(x) \left[ p_A(x) \ln (1 - \tilde p_B(x)) + p_B(x) \ln \tilde p_B(x) \right],
\end{equation}
where $p_A(x)$ and $p_B(x)$ are the true committor, $\rho_{\text{eq}}(\mathbf{x}) = \frac{1}{Z} \sum_i^{\text{RPE}} w_i \delta(x_i - x)$ is the reweighted equilibrium density.
\begin{figure*}[t!]
   \centering
\includegraphics[width=0.8\linewidth]{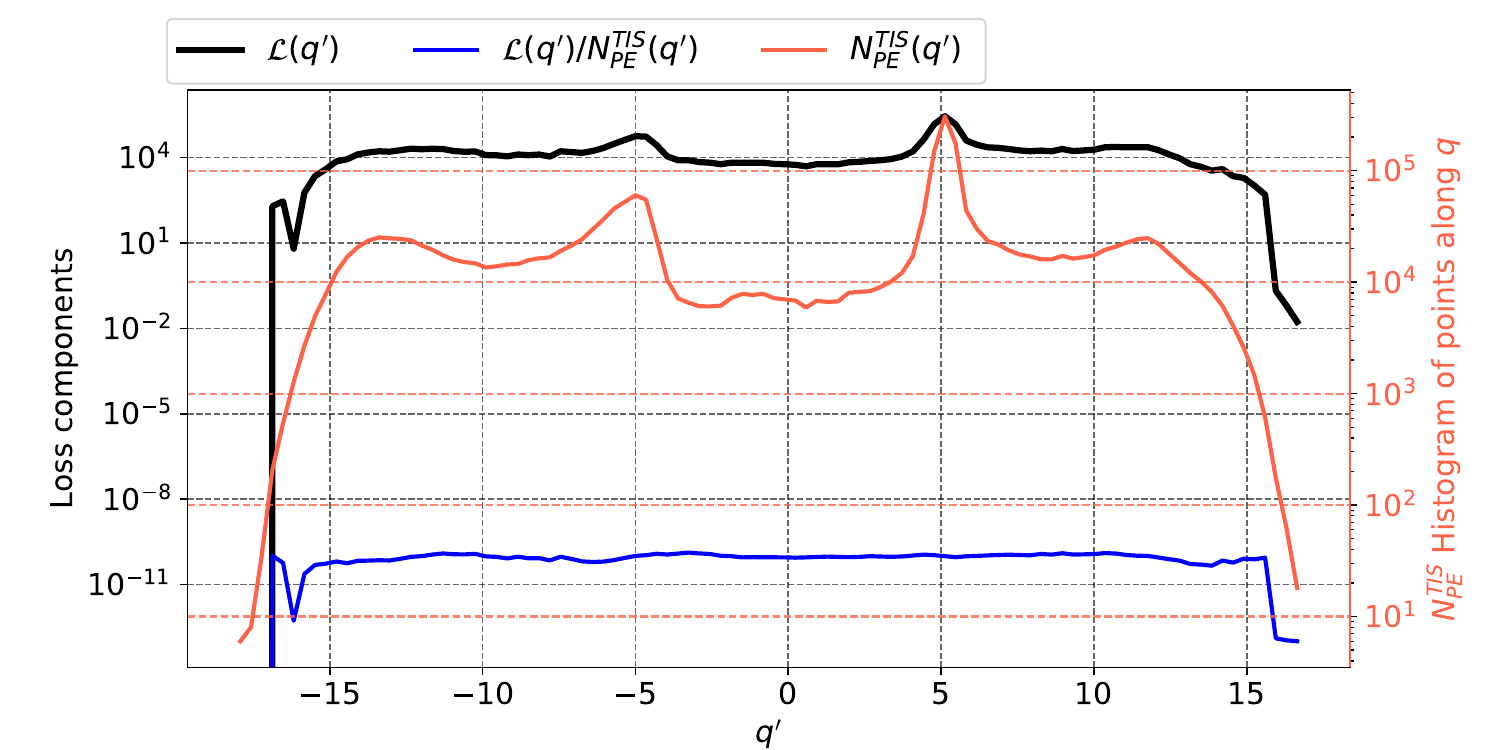}
    \caption{Average loss function contribution along the logit committor function q. The black curve gives the  total loss, the blue curve the distribution of sampled points. The contribution per point (blue curve)  remains relatively uniform, indicating that the reweighted path ensemble (RPE) balances training across the full transition pathway. Each configuration in the sampled TIS ensembles contributes equally to the loss.
    \label{FIG:A: average loss contribution}}
\end{figure*}
The informativeness of each configuration $x$ in this learning process is quantified through the information content:

\begin{equation}
I(x) = p_A(x) \ln p_A(x)+p_B(x) \ln  p_B(x) ,
\end{equation}
and the total contribution is given by the product of the equilibrium weight and this information content:

\begin{equation}
C(x) = \rho_{\text{eq}}(x) I(x),
\end{equation}
after reweighting, the contributions become nearly uniform across committor space. This balances the influence of stable regions and transition regions, correcting for the natural over-representation of the latter in path sampling and ensuring unbiased estimation of the committor.

\subsubsection{Balanced weighting across entire committor}

The above analysis does not take into account the presence of metastable states or regions where systems linger for a longer time.  
To refine the analysis, we consider the committor loss function projected along the reaction coordinate $q$:
\begin{equation}
\begin{split}
\mathcal{L}(q)=-\sum_{x_i}^{RPE}w_i\delta(q-\tilde{q}(x_i))(&n_i^A\ln(\tilde{p}_A(\tilde{q}(x_i)))\\
+n_i^B\ln(\tilde{p}_B(\tilde{q}(x_i)))),
\end{split}
\end{equation}
where the contribution of the loss at a given $q$ value is based on the logit model $\tilde{q}(x)$.
 Although the $\ln p_B$ term is balanced by the contribution to the Boltzmann distribution due to the RPE $w_i$  , 
 $\mathcal{L}(q)$ exhibits peaks in regions of metastability in between the defined A and B states. 
The peaks are caused by a larger number points in these regions occurring in the TIS path ensembles from which the RPE is made.
To correct for this, we can look at the number of points in this region for a given model $q$ value 
\begin{equation}
    N^{TIS}_{PE}(q')=\sum_{x_i}^{PE} (n_i^A+n_i^B)\delta(q-\tilde{q}(x_i))
\end{equation}
where we sum over all points in the TIS path ensembles, avoiding  double counting of points due to the use of two way shooting, both the forward and reverse ensembles and the stable state data (the latter  included in the RPE without reweighting ).
This gives us the distribution of data points along the committor model.
When looking at the average contribution to the loss $\mathcal{L}(q)$ for a given $q$ region of the model, shown in Figure \ref{FIG:A: average loss contribution} for the Wolfe-Quapp system for the second iteration model $q(x|\theta_{RPE_2})$, we can see (blue curve) that each point in the RPE contributes equally to the loss along all $q$ values. This indicates that points in the transition region as well as points close to the stable states give the same contribution to the loss. 
The weight attributed to each point compensates the contribution to the loss both near the stable states and near the transition regions, ensuring balanced training in all regions of $q$ space. 
Thus, the reweighted likelihood function ensures a well-balanced representation of the committor function.

%% file: main_arxiv.bbl
\begin{thebibliography}{30}%
\makeatletter
\providecommand \@ifxundefined [1]{%
 \@ifx{#1\undefined}
}%
\providecommand \@ifnum [1]{%
 \ifnum #1\expandafter \@firstoftwo
 \else \expandafter \@secondoftwo
 \fi
}%
\providecommand \@ifx [1]{%
 \ifx #1\expandafter \@firstoftwo
 \else \expandafter \@secondoftwo
 \fi
}%
\providecommand \natexlab [1]{#1}%
\providecommand \enquote  [1]{``#1''}%
\providecommand \bibnamefont  [1]{#1}%
\providecommand \bibfnamefont [1]{#1}%
\providecommand \citenamefont [1]{#1}%
\providecommand \href@noop [0]{\@secondoftwo}%
\providecommand \href [0]{\begingroup \@sanitize@url \@href}%
\providecommand \@href[1]{\@@startlink{#1}\@@href}%
\providecommand \@@href[1]{\endgroup#1\@@endlink}%
\providecommand \@sanitize@url [0]{\catcode `\\12\catcode `\$12\catcode `\&12\catcode `\#12\catcode `\^12\catcode `\_12\catcode `\%12\relax}%
\providecommand \@@startlink[1]{}%
\providecommand \@@endlink[0]{}%
\providecommand \url  [0]{\begingroup\@sanitize@url \@url }%
\providecommand \@url [1]{\endgroup\@href {#1}{\urlprefix }}%
\providecommand \urlprefix  [0]{URL }%
\providecommand \Eprint [0]{\href }%
\providecommand \doibase [0]{https://doi.org/}%
\providecommand \selectlanguage [0]{\@gobble}%
\providecommand \bibinfo  [0]{\@secondoftwo}%
\providecommand \bibfield  [0]{\@secondoftwo}%
\providecommand \translation [1]{[#1]}%
\providecommand \BibitemOpen [0]{}%
\providecommand \bibitemStop [0]{}%
\providecommand \bibitemNoStop [0]{.\EOS\space}%
\providecommand \EOS [0]{\spacefactor3000\relax}%
\providecommand \BibitemShut  [1]{\csname bibitem#1\endcsname}%
\let\auto@bib@innerbib\@empty
\bibitem [{\citenamefont {Bolhuis}\ \emph {et~al.}(2002)\citenamefont {Bolhuis}, \citenamefont {Chandler}, \citenamefont {Dellago},\ and\ \citenamefont {Geissler}}]{Bolhuis2002}%
  \BibitemOpen
  \bibfield  {author} {\bibinfo {author} {\bibfnamefont {P.~G.}\ \bibnamefont {Bolhuis}}, \bibinfo {author} {\bibfnamefont {D.}~\bibnamefont {Chandler}}, \bibinfo {author} {\bibfnamefont {C.}~\bibnamefont {Dellago}},\ and\ \bibinfo {author} {\bibfnamefont {P.}~\bibnamefont {Geissler}},\ }\href {https://doi.org/10.1146/annurev.physchem.53.082301.113146} {\bibfield  {journal} {\bibinfo  {journal} {Ann. Rev. Phys. Chem.}\ }\textbf {\bibinfo {volume} {53}},\ \bibinfo {pages} {291} (\bibinfo {year} {2002})}\BibitemShut {NoStop}%
\bibitem [{\citenamefont {Dellago}\ \emph {et~al.}(2002)\citenamefont {Dellago}, \citenamefont {Bolhuis},\ and\ \citenamefont {Geissler}}]{Dellago2002}%
  \BibitemOpen
  \bibfield  {author} {\bibinfo {author} {\bibfnamefont {C.}~\bibnamefont {Dellago}}, \bibinfo {author} {\bibfnamefont {P.~G.}\ \bibnamefont {Bolhuis}},\ and\ \bibinfo {author} {\bibfnamefont {P.~L.}\ \bibnamefont {Geissler}},\ }\href@noop {} {\bibfield  {journal} {\bibinfo  {journal} {Adv. Chem. Phys.}\ }\textbf {\bibinfo {volume} {123}},\ \bibinfo {pages} {1} (\bibinfo {year} {2002})}\BibitemShut {NoStop}%
\bibitem [{\citenamefont {Van~Erp}\ \emph {et~al.}(2003)\citenamefont {Van~Erp}, \citenamefont {Moroni},\ and\ \citenamefont {Bolhuis}}]{vanerpNovelPathSampling2003}%
  \BibitemOpen
  \bibfield  {author} {\bibinfo {author} {\bibfnamefont {T.~S.}\ \bibnamefont {Van~Erp}}, \bibinfo {author} {\bibfnamefont {D.}~\bibnamefont {Moroni}},\ and\ \bibinfo {author} {\bibfnamefont {P.~G.}\ \bibnamefont {Bolhuis}},\ }\href {https://doi.org/10.1063/1.1562614} {\bibfield  {journal} {\bibinfo  {journal} {The Journal of Chemical Physics}\ }\textbf {\bibinfo {volume} {118}},\ \bibinfo {pages} {7762} (\bibinfo {year} {2003})}\BibitemShut {NoStop}%
\bibitem [{\citenamefont {Cabriolu}\ \emph {et~al.}(2017)\citenamefont {Cabriolu}, \citenamefont {Refsnes}, \citenamefont {Bolhuis},\ and\ \citenamefont {van Erp}}]{Cabriolu2017}%
  \BibitemOpen
  \bibfield  {author} {\bibinfo {author} {\bibfnamefont {R.}~\bibnamefont {Cabriolu}}, \bibinfo {author} {\bibfnamefont {K.~M.~S.}\ \bibnamefont {Refsnes}}, \bibinfo {author} {\bibfnamefont {P.~G.}\ \bibnamefont {Bolhuis}},\ and\ \bibinfo {author} {\bibfnamefont {T.~S.}\ \bibnamefont {van Erp}},\ }\href {https://doi.org/10.1063/1.4989844} {\bibfield  {journal} {\bibinfo  {journal} {J. Chem. Phys.}\ }\textbf {\bibinfo {volume} {147}},\ \bibinfo {pages} {152722} (\bibinfo {year} {2017})}\BibitemShut {NoStop}%
\bibitem [{\citenamefont {Jung}\ \emph {et~al.}(2023)\citenamefont {Jung}, \citenamefont {Covino}, \citenamefont {Arjun}, \citenamefont {Leitold}, \citenamefont {Dellago}, \citenamefont {Bolhuis},\ and\ \citenamefont {Hummer}}]{Jung2023}%
  \BibitemOpen
  \bibfield  {author} {\bibinfo {author} {\bibfnamefont {H.}~\bibnamefont {Jung}}, \bibinfo {author} {\bibfnamefont {R.}~\bibnamefont {Covino}}, \bibinfo {author} {\bibfnamefont {A.}~\bibnamefont {Arjun}}, \bibinfo {author} {\bibfnamefont {C.}~\bibnamefont {Leitold}}, \bibinfo {author} {\bibfnamefont {C.}~\bibnamefont {Dellago}}, \bibinfo {author} {\bibfnamefont {P.~G.}\ \bibnamefont {Bolhuis}},\ and\ \bibinfo {author} {\bibfnamefont {G.}~\bibnamefont {Hummer}},\ }\href {https://doi.org/10.1038/s43588-023-00428-z} {\bibfield  {journal} {\bibinfo  {journal} {Nat. Comput. Sci.}\ }\textbf {\bibinfo {volume} {3}},\ \bibinfo {pages} {334–345} (\bibinfo {year} {2023})}\BibitemShut {NoStop}%
\bibitem [{\citenamefont {Peters}\ and\ \citenamefont {Trout}(2006)}]{Peters2006}%
  \BibitemOpen
  \bibfield  {author} {\bibinfo {author} {\bibfnamefont {B.}~\bibnamefont {Peters}}\ and\ \bibinfo {author} {\bibfnamefont {B.~L.}\ \bibnamefont {Trout}},\ }\href@noop {} {\bibfield  {journal} {\bibinfo  {journal} {J. Chem. Phys.}\ }\textbf {\bibinfo {volume} {125}},\ \bibinfo {pages} {054108} (\bibinfo {year} {2006})}\BibitemShut {NoStop}%
\bibitem [{\citenamefont {Hummer}(2004)}]{Hummer2004}%
  \BibitemOpen
  \bibfield  {author} {\bibinfo {author} {\bibfnamefont {G.}~\bibnamefont {Hummer}},\ }\href {https://doi.org/10.1063/1.1630572} {\bibfield  {journal} {\bibinfo  {journal} {The Journal of Chemical Physics}\ }\textbf {\bibinfo {volume} {120}},\ \bibinfo {pages} {516} (\bibinfo {year} {2004})}\BibitemShut {NoStop}%
\bibitem [{\citenamefont {Ma}\ and\ \citenamefont {Dinner}(2005)}]{maAutomaticMethodIdentifying2005}%
  \BibitemOpen
  \bibfield  {author} {\bibinfo {author} {\bibfnamefont {A.}~\bibnamefont {Ma}}\ and\ \bibinfo {author} {\bibfnamefont {A.~R.}\ \bibnamefont {Dinner}},\ }\href {https://doi.org/10.1021/jp045546c} {\bibfield  {journal} {\bibinfo  {journal} {The Journal of Physical Chemistry B}\ }\textbf {\bibinfo {volume} {109}},\ \bibinfo {pages} {6769} (\bibinfo {year} {2005})}\BibitemShut {NoStop}%
\bibitem [{\citenamefont {Jung}\ \emph {et~al.}(2019)\citenamefont {Jung}, \citenamefont {Covino},\ and\ \citenamefont {Hummer}}]{jungArtificialIntelligenceAssists2019}%
  \BibitemOpen
  \bibfield  {author} {\bibinfo {author} {\bibfnamefont {H.}~\bibnamefont {Jung}}, \bibinfo {author} {\bibfnamefont {R.}~\bibnamefont {Covino}},\ and\ \bibinfo {author} {\bibfnamefont {G.}~\bibnamefont {Hummer}},\ }\href@noop {} {\bibinfo {title} {Artificial {{Intelligence Assists Discovery}} of {{Reaction Coordinates}} and {{Mechanisms}} from {{Molecular Dynamics Simulations}}}} (\bibinfo {year} {2019}),\ \Eprint {https://arxiv.org/abs/1901.04595} {arXiv:1901.04595 [physics]} \BibitemShut {NoStop}%
\bibitem [{\citenamefont {Frassek}\ \emph {et~al.}(2021)\citenamefont {Frassek}, \citenamefont {Arjun},\ and\ \citenamefont {Bolhuis}}]{Frassek2021}%
  \BibitemOpen
  \bibfield  {author} {\bibinfo {author} {\bibfnamefont {M.}~\bibnamefont {Frassek}}, \bibinfo {author} {\bibfnamefont {A.}~\bibnamefont {Arjun}},\ and\ \bibinfo {author} {\bibfnamefont {P.~G.}\ \bibnamefont {Bolhuis}},\ }\href {http://dx.doi.org/10.1063/5.0058639} {\bibfield  {journal} {\bibinfo  {journal} {The Journal of Chemical Physics}\ }\textbf {\bibinfo {volume} {155}},\ \bibinfo {pages} {064103} (\bibinfo {year} {2021})}\BibitemShut {NoStop}%
\bibitem [{\citenamefont {Rogal}\ \emph {et~al.}(2010)\citenamefont {Rogal}, \citenamefont {Lechner}, \citenamefont {Juraszek}, \citenamefont {Ensing},\ and\ \citenamefont {Bolhuis}}]{Rogal2010}%
  \BibitemOpen
  \bibfield  {author} {\bibinfo {author} {\bibfnamefont {J.}~\bibnamefont {Rogal}}, \bibinfo {author} {\bibfnamefont {W.}~\bibnamefont {Lechner}}, \bibinfo {author} {\bibfnamefont {J.}~\bibnamefont {Juraszek}}, \bibinfo {author} {\bibfnamefont {B.}~\bibnamefont {Ensing}},\ and\ \bibinfo {author} {\bibfnamefont {P.~G.}\ \bibnamefont {Bolhuis}},\ }\href {https://doi.org/10.1063/1.3491817} {\bibfield  {journal} {\bibinfo  {journal} {The Journal of Chemical Physics}\ }\textbf {\bibinfo {volume} {133}},\ \bibinfo {pages} {174109} (\bibinfo {year} {2010})}\BibitemShut {NoStop}%
\bibitem [{\citenamefont {Ferrenberg}\ and\ \citenamefont {Swendsen}(1989)}]{ferrenbergOptimizedMonteCarlo1989}%
  \BibitemOpen
  \bibfield  {author} {\bibinfo {author} {\bibfnamefont {A.~M.}\ \bibnamefont {Ferrenberg}}\ and\ \bibinfo {author} {\bibfnamefont {R.~H.}\ \bibnamefont {Swendsen}},\ }\href {https://doi.org/10.1103/PhysRevLett.63.1195} {\bibfield  {journal} {\bibinfo  {journal} {Physical Review Letters}\ }\textbf {\bibinfo {volume} {63}},\ \bibinfo {pages} {1195} (\bibinfo {year} {1989})}\BibitemShut {NoStop}%
\bibitem [{\citenamefont {van Erp}\ \emph {et~al.}(2003)\citenamefont {van Erp}, \citenamefont {Moroni},\ and\ \citenamefont {Bolhuis}}]{vanErp2003}%
  \BibitemOpen
  \bibfield  {author} {\bibinfo {author} {\bibfnamefont {T.~S.}\ \bibnamefont {van Erp}}, \bibinfo {author} {\bibfnamefont {D.}~\bibnamefont {Moroni}},\ and\ \bibinfo {author} {\bibfnamefont {P.~G.}\ \bibnamefont {Bolhuis}},\ }\href@noop {} {\bibfield  {journal} {\bibinfo  {journal} {J. Chem. Phys.}\ }\textbf {\bibinfo {volume} {118}},\ \bibinfo {pages} {7762} (\bibinfo {year} {2003})}\BibitemShut {NoStop}%
\bibitem [{\citenamefont {E}\ and\ \citenamefont {{Vanden-Eijnden}}(2010)}]{eTransitionPathTheoryPathFinding2010}%
  \BibitemOpen
  \bibfield  {author} {\bibinfo {author} {\bibfnamefont {W.}~\bibnamefont {E}}\ and\ \bibinfo {author} {\bibfnamefont {E.}~\bibnamefont {{Vanden-Eijnden}}},\ }\href {https://doi.org/10.1146/annurev.physchem.040808.090412} {\bibfield  {journal} {\bibinfo  {journal} {Annual Review of Physical Chemistry}\ }\textbf {\bibinfo {volume} {61}},\ \bibinfo {pages} {391} (\bibinfo {year} {2010})}\BibitemShut {NoStop}%
\bibitem [{\citenamefont {Lazzeri}\ \emph {et~al.}(2023)\citenamefont {Lazzeri}, \citenamefont {Jung}, \citenamefont {Bolhuis},\ and\ \citenamefont {Covino}}]{lazzeriMolecularFreeEnergies2023a}%
  \BibitemOpen
  \bibfield  {author} {\bibinfo {author} {\bibfnamefont {G.}~\bibnamefont {Lazzeri}}, \bibinfo {author} {\bibfnamefont {H.}~\bibnamefont {Jung}}, \bibinfo {author} {\bibfnamefont {P.~G.}\ \bibnamefont {Bolhuis}},\ and\ \bibinfo {author} {\bibfnamefont {R.}~\bibnamefont {Covino}},\ }\href {https://doi.org/10.1021/acs.jctc.3c00821} {\bibfield  {journal} {\bibinfo  {journal} {Journal of Chemical Theory and Computation}\ }\textbf {\bibinfo {volume} {19}},\ \bibinfo {pages} {9060–9076} (\bibinfo {year} {2023})}\BibitemShut {NoStop}%
\bibitem [{\citenamefont {{Vanden-Eijnden}}\ \emph {et~al.}(2008)\citenamefont {{Vanden-Eijnden}}, \citenamefont {Venturoli}, \citenamefont {Ciccotti},\ and\ \citenamefont {Elber}}]{vanden-eijndenAssumptionsUnderlyingMilestoning2008}%
  \BibitemOpen
  \bibfield  {author} {\bibinfo {author} {\bibfnamefont {E.}~\bibnamefont {{Vanden-Eijnden}}}, \bibinfo {author} {\bibfnamefont {M.}~\bibnamefont {Venturoli}}, \bibinfo {author} {\bibfnamefont {G.}~\bibnamefont {Ciccotti}},\ and\ \bibinfo {author} {\bibfnamefont {R.}~\bibnamefont {Elber}},\ }\href {https://doi.org/10.1063/1.2996509} {\bibfield  {journal} {\bibinfo  {journal} {The Journal of Chemical Physics}\ }\textbf {\bibinfo {volume} {129}},\ \bibinfo {pages} {174102} (\bibinfo {year} {2008})}\BibitemShut {NoStop}%
\bibitem [{\citenamefont {Kingma}\ and\ \citenamefont {Ba}(2015)}]{Kingma2015Adam:Optimization}%
  \BibitemOpen
  \bibfield  {author} {\bibinfo {author} {\bibfnamefont {D.~P.}\ \bibnamefont {Kingma}}\ and\ \bibinfo {author} {\bibfnamefont {J.~L.}\ \bibnamefont {Ba}},\ }\href@noop {} {\bibfield  {journal} {\bibinfo  {journal} {3rd International Conference on Learning Representations, ICLR 2015 - Conference Track Proceedings}\ ,\ \bibinfo {pages} {1}} (\bibinfo {year} {2015})}\BibitemShut {NoStop}%
\bibitem [{\citenamefont {Swenson}\ \emph {et~al.}(2019)\citenamefont {Swenson}, \citenamefont {Prinz}, \citenamefont {Noe}, \citenamefont {Chodera},\ and\ \citenamefont {Bolhuis}}]{swensonOpenPathSamplingPythonFramework2019}%
  \BibitemOpen
  \bibfield  {author} {\bibinfo {author} {\bibfnamefont {D.~W.~H.}\ \bibnamefont {Swenson}}, \bibinfo {author} {\bibfnamefont {J.-H.}\ \bibnamefont {Prinz}}, \bibinfo {author} {\bibfnamefont {F.}~\bibnamefont {Noe}}, \bibinfo {author} {\bibfnamefont {J.~D.}\ \bibnamefont {Chodera}},\ and\ \bibinfo {author} {\bibfnamefont {P.~G.}\ \bibnamefont {Bolhuis}},\ }\href {https://doi.org/10.1021/acs.jctc.8b00626} {\bibfield  {journal} {\bibinfo  {journal} {Journal of Chemical Theory and Computation}\ }\textbf {\bibinfo {volume} {15}},\ \bibinfo {pages} {813} (\bibinfo {year} {2019})}\BibitemShut {NoStop}%
\bibitem [{\citenamefont {Wolfe}\ \emph {et~al.}(1975)\citenamefont {Wolfe}, \citenamefont {Schlegel}, \citenamefont {Csizmadia},\ and\ \citenamefont {Bernardi}}]{Wolfe1975}%
  \BibitemOpen
  \bibfield  {author} {\bibinfo {author} {\bibfnamefont {S.}~\bibnamefont {Wolfe}}, \bibinfo {author} {\bibfnamefont {H.~B.}\ \bibnamefont {Schlegel}}, \bibinfo {author} {\bibfnamefont {I.~G.}\ \bibnamefont {Csizmadia}},\ and\ \bibinfo {author} {\bibfnamefont {F.}~\bibnamefont {Bernardi}},\ }\href {https://doi.org/10.1021/ja00841a005} {\bibfield  {journal} {\bibinfo  {journal} {Journal of the American Chemical Society}\ }\textbf {\bibinfo {volume} {97}},\ \bibinfo {pages} {2020–2024} (\bibinfo {year} {1975})}\BibitemShut {NoStop}%
\bibitem [{\citenamefont {Quapp}(2005)}]{quappGrowingStringMethod2005}%
  \BibitemOpen
  \bibfield  {author} {\bibinfo {author} {\bibfnamefont {W.}~\bibnamefont {Quapp}},\ }\href {https://doi.org/10.1063/1.1885467} {\bibfield  {journal} {\bibinfo  {journal} {The Journal of Chemical Physics}\ }\textbf {\bibinfo {volume} {122}},\ \bibinfo {pages} {174106} (\bibinfo {year} {2005})}\BibitemShut {NoStop}%
\bibitem [{\citenamefont {Leimkuhler}\ and\ \citenamefont {Matthews}(2012)}]{leimkuhlerRationalConstructionStochastic2012}%
  \BibitemOpen
  \bibfield  {author} {\bibinfo {author} {\bibfnamefont {B.}~\bibnamefont {Leimkuhler}}\ and\ \bibinfo {author} {\bibfnamefont {C.}~\bibnamefont {Matthews}},\ }\href {https://doi.org/10.1093/amrx/abs010} {\bibfield  {journal} {\bibinfo  {journal} {Applied Mathematics Research eXpress}\ ,\ \bibinfo {pages} {abs010}} (\bibinfo {year} {2012})}\BibitemShut {NoStop}%
\bibitem [{\citenamefont {Covino}\ \emph {et~al.}(2019)\citenamefont {Covino}, \citenamefont {Woodside}, \citenamefont {Hummer}, \citenamefont {Szabo},\ and\ \citenamefont {Cossio}}]{covinoMolecularFreeEnergy2019}%
  \BibitemOpen
  \bibfield  {author} {\bibinfo {author} {\bibfnamefont {R.}~\bibnamefont {Covino}}, \bibinfo {author} {\bibfnamefont {M.~T.}\ \bibnamefont {Woodside}}, \bibinfo {author} {\bibfnamefont {G.}~\bibnamefont {Hummer}}, \bibinfo {author} {\bibfnamefont {A.}~\bibnamefont {Szabo}},\ and\ \bibinfo {author} {\bibfnamefont {P.}~\bibnamefont {Cossio}},\ }\href {https://doi.org/10.1063/1.5118362} {\bibfield  {journal} {\bibinfo  {journal} {The Journal of Chemical Physics}\ }\textbf {\bibinfo {volume} {151}},\ \bibinfo {pages} {154115} (\bibinfo {year} {2019})}\BibitemShut {NoStop}%
\bibitem [{\citenamefont {Moghaddam}\ \emph {et~al.}(2011)\citenamefont {Moghaddam}, \citenamefont {Yang}, \citenamefont {Rekharsky}, \citenamefont {Ko}, \citenamefont {Kim}, \citenamefont {Inoue},\ and\ \citenamefont {Gilson}}]{Moghaddam2011}%
  \BibitemOpen
  \bibfield  {author} {\bibinfo {author} {\bibfnamefont {S.}~\bibnamefont {Moghaddam}}, \bibinfo {author} {\bibfnamefont {C.}~\bibnamefont {Yang}}, \bibinfo {author} {\bibfnamefont {M.}~\bibnamefont {Rekharsky}}, \bibinfo {author} {\bibfnamefont {Y.~H.}\ \bibnamefont {Ko}}, \bibinfo {author} {\bibfnamefont {K.}~\bibnamefont {Kim}}, \bibinfo {author} {\bibfnamefont {Y.}~\bibnamefont {Inoue}},\ and\ \bibinfo {author} {\bibfnamefont {M.~K.}\ \bibnamefont {Gilson}},\ }\href {https://doi.org/10.1021/ja109904u} {\bibfield  {journal} {\bibinfo  {journal} {Journal of the American Chemical Society}\ }\textbf {\bibinfo {volume} {133}},\ \bibinfo {pages} {3570–3581} (\bibinfo {year} {2011})}\BibitemShut {NoStop}%
\bibitem [{\citenamefont {Fenley}\ \emph {et~al.}(2014)\citenamefont {Fenley}, \citenamefont {Henriksen}, \citenamefont {Muddana},\ and\ \citenamefont {Gilson}}]{Fenley2014}%
  \BibitemOpen
  \bibfield  {author} {\bibinfo {author} {\bibfnamefont {A.~T.}\ \bibnamefont {Fenley}}, \bibinfo {author} {\bibfnamefont {N.~M.}\ \bibnamefont {Henriksen}}, \bibinfo {author} {\bibfnamefont {H.~S.}\ \bibnamefont {Muddana}},\ and\ \bibinfo {author} {\bibfnamefont {M.~K.}\ \bibnamefont {Gilson}},\ }\href {https://doi.org/10.1021/ct5004109} {\bibfield  {journal} {\bibinfo  {journal} {Journal of Chemical Theory and Computation}\ }\textbf {\bibinfo {volume} {10}},\ \bibinfo {pages} {4069–4078} (\bibinfo {year} {2014})}\BibitemShut {NoStop}%
\bibitem [{\citenamefont {Henriksen}\ \emph {et~al.}(2015)\citenamefont {Henriksen}, \citenamefont {Fenley},\ and\ \citenamefont {Gilson}}]{Henriksen2015}%
  \BibitemOpen
  \bibfield  {author} {\bibinfo {author} {\bibfnamefont {N.~M.}\ \bibnamefont {Henriksen}}, \bibinfo {author} {\bibfnamefont {A.~T.}\ \bibnamefont {Fenley}},\ and\ \bibinfo {author} {\bibfnamefont {M.~K.}\ \bibnamefont {Gilson}},\ }\href {https://doi.org/10.1021/acs.jctc.5b00405} {\bibfield  {journal} {\bibinfo  {journal} {Journal of Chemical Theory and Computation}\ }\textbf {\bibinfo {volume} {11}},\ \bibinfo {pages} {4377–4394} (\bibinfo {year} {2015})}\BibitemShut {NoStop}%
\bibitem [{\citenamefont {Wang}\ \emph {et~al.}(2004)\citenamefont {Wang}, \citenamefont {Wolf}, \citenamefont {Caldwell}, \citenamefont {Kollman},\ and\ \citenamefont {Case}}]{Wang2004}%
  \BibitemOpen
  \bibfield  {author} {\bibinfo {author} {\bibfnamefont {J.}~\bibnamefont {Wang}}, \bibinfo {author} {\bibfnamefont {R.~M.}\ \bibnamefont {Wolf}}, \bibinfo {author} {\bibfnamefont {J.~W.}\ \bibnamefont {Caldwell}}, \bibinfo {author} {\bibfnamefont {P.~A.}\ \bibnamefont {Kollman}},\ and\ \bibinfo {author} {\bibfnamefont {D.~A.}\ \bibnamefont {Case}},\ }\href {https://doi.org/10.1002/jcc.20035} {\bibfield  {journal} {\bibinfo  {journal} {Journal of Computational Chemistry}\ }\textbf {\bibinfo {volume} {25}},\ \bibinfo {pages} {1157–1174} (\bibinfo {year} {2004})}\BibitemShut {NoStop}%
\bibitem [{\citenamefont {Chodera}\ \emph {et~al.}(2018)\citenamefont {Chodera}, \citenamefont {Rizzi}, \citenamefont {Naden}, \citenamefont {Beauchamp}, \citenamefont {Grinaway}, \citenamefont {Fass}, \citenamefont {Rustenburg}, \citenamefont {Ross}, \citenamefont {Simmonett},\ and\ \citenamefont {Swenson}}]{chodera2018choderalab}%
  \BibitemOpen
  \bibfield  {author} {\bibinfo {author} {\bibfnamefont {J.}~\bibnamefont {Chodera}}, \bibinfo {author} {\bibfnamefont {A.}~\bibnamefont {Rizzi}}, \bibinfo {author} {\bibfnamefont {L.}~\bibnamefont {Naden}}, \bibinfo {author} {\bibfnamefont {K.}~\bibnamefont {Beauchamp}}, \bibinfo {author} {\bibfnamefont {P.}~\bibnamefont {Grinaway}}, \bibinfo {author} {\bibfnamefont {J.}~\bibnamefont {Fass}}, \bibinfo {author} {\bibfnamefont {B.}~\bibnamefont {Rustenburg}}, \bibinfo {author} {\bibfnamefont {G.~A.}\ \bibnamefont {Ross}}, \bibinfo {author} {\bibfnamefont {A.}~\bibnamefont {Simmonett}},\ and\ \bibinfo {author} {\bibfnamefont {D.~W.}\ \bibnamefont {Swenson}},\ }\href@noop {} {\bibfield  {journal} {\bibinfo  {journal} {doi. org/10.5281/zenod o}\ }\textbf {\bibinfo {volume} {11611}},\ \bibinfo {pages} {49} (\bibinfo {year} {2018})}\BibitemShut {NoStop}%
\bibitem [{\citenamefont {Eastman}\ \emph {et~al.}(2024)\citenamefont {Eastman}, \citenamefont {Galvelis}, \citenamefont {Pel{\'a}ez}, \citenamefont {Abreu}, \citenamefont {Farr}, \citenamefont {Gallicchio}, \citenamefont {Gorenko}, \citenamefont {Henry}, \citenamefont {Hu}, \citenamefont {Huang}, \citenamefont {Kr{\"a}mer}, \citenamefont {Michel}, \citenamefont {Mitchell}, \citenamefont {Pande}, \citenamefont {Rodrigues}, \citenamefont {{Rodriguez-Guerra}}, \citenamefont {Simmonett}, \citenamefont {Singh}, \citenamefont {Swails}, \citenamefont {Turner}, \citenamefont {Wang}, \citenamefont {Zhang}, \citenamefont {Chodera}, \citenamefont {De~Fabritiis},\ and\ \citenamefont {Markland}}]{eastmanOpenMM8Molecular2024}%
  \BibitemOpen
  \bibfield  {author} {\bibinfo {author} {\bibfnamefont {P.}~\bibnamefont {Eastman}}, \bibinfo {author} {\bibfnamefont {R.}~\bibnamefont {Galvelis}}, \bibinfo {author} {\bibfnamefont {R.~P.}\ \bibnamefont {Pel{\'a}ez}}, \bibinfo {author} {\bibfnamefont {C.~R.~A.}\ \bibnamefont {Abreu}}, \bibinfo {author} {\bibfnamefont {S.~E.}\ \bibnamefont {Farr}}, \bibinfo {author} {\bibfnamefont {E.}~\bibnamefont {Gallicchio}}, \bibinfo {author} {\bibfnamefont {A.}~\bibnamefont {Gorenko}}, \bibinfo {author} {\bibfnamefont {M.~M.}\ \bibnamefont {Henry}}, \bibinfo {author} {\bibfnamefont {F.}~\bibnamefont {Hu}}, \bibinfo {author} {\bibfnamefont {J.}~\bibnamefont {Huang}}, \bibinfo {author} {\bibfnamefont {A.}~\bibnamefont {Kr{\"a}mer}}, \bibinfo {author} {\bibfnamefont {J.}~\bibnamefont {Michel}}, \bibinfo {author} {\bibfnamefont {J.~A.}\ \bibnamefont {Mitchell}}, \bibinfo {author} {\bibfnamefont {V.~S.}\ \bibnamefont {Pande}}, \bibinfo {author} {\bibfnamefont {J.~P.}\ \bibnamefont {Rodrigues}}, \bibinfo {author}
  {\bibfnamefont {J.}~\bibnamefont {{Rodriguez-Guerra}}}, \bibinfo {author} {\bibfnamefont {A.~C.}\ \bibnamefont {Simmonett}}, \bibinfo {author} {\bibfnamefont {S.}~\bibnamefont {Singh}}, \bibinfo {author} {\bibfnamefont {J.}~\bibnamefont {Swails}}, \bibinfo {author} {\bibfnamefont {P.}~\bibnamefont {Turner}}, \bibinfo {author} {\bibfnamefont {Y.}~\bibnamefont {Wang}}, \bibinfo {author} {\bibfnamefont {I.}~\bibnamefont {Zhang}}, \bibinfo {author} {\bibfnamefont {J.~D.}\ \bibnamefont {Chodera}}, \bibinfo {author} {\bibfnamefont {G.}~\bibnamefont {De~Fabritiis}},\ and\ \bibinfo {author} {\bibfnamefont {T.~E.}\ \bibnamefont {Markland}},\ }\href {https://doi.org/10.1021/acs.jpcb.3c06662} {\bibfield  {journal} {\bibinfo  {journal} {The Journal of Physical Chemistry B}\ }\textbf {\bibinfo {volume} {128}},\ \bibinfo {pages} {109} (\bibinfo {year} {2024})}\BibitemShut {NoStop}%
\bibitem [{\citenamefont {Wernet}\ \emph {et~al.}(2004)\citenamefont {Wernet}, \citenamefont {Nordlund}, \citenamefont {Bergmann}, \citenamefont {Cavalleri}, \citenamefont {Odelius}, \citenamefont {Ogasawara}, \citenamefont {Naslund}, \citenamefont {Hirsch}, \citenamefont {Ojamae}, \citenamefont {Glatzel}, \citenamefont {Pettersson},\ and\ \citenamefont {Nilsson}}]{wernetStructureFirstCoordination2004}%
  \BibitemOpen
  \bibfield  {author} {\bibinfo {author} {\bibfnamefont {P.}~\bibnamefont {Wernet}}, \bibinfo {author} {\bibfnamefont {D.}~\bibnamefont {Nordlund}}, \bibinfo {author} {\bibfnamefont {U.}~\bibnamefont {Bergmann}}, \bibinfo {author} {\bibfnamefont {M.}~\bibnamefont {Cavalleri}}, \bibinfo {author} {\bibfnamefont {M.}~\bibnamefont {Odelius}}, \bibinfo {author} {\bibfnamefont {H.}~\bibnamefont {Ogasawara}}, \bibinfo {author} {\bibfnamefont {L.~{\AA}.}\ \bibnamefont {Naslund}}, \bibinfo {author} {\bibfnamefont {T.~K.}\ \bibnamefont {Hirsch}}, \bibinfo {author} {\bibfnamefont {L.}~\bibnamefont {Ojamae}}, \bibinfo {author} {\bibfnamefont {P.}~\bibnamefont {Glatzel}}, \bibinfo {author} {\bibfnamefont {L.~G.~M.}\ \bibnamefont {Pettersson}},\ and\ \bibinfo {author} {\bibfnamefont {A.}~\bibnamefont {Nilsson}},\ }\href@noop {} {\bibfield  {journal} {\bibinfo  {journal} {Science}\ } (\bibinfo {year} {2004})}\BibitemShut {NoStop}%
\bibitem [{\citenamefont {Lechner}\ \emph {et~al.}(2010)\citenamefont {Lechner}, \citenamefont {Rogal}, \citenamefont {Juraszek}, \citenamefont {Ensing},\ and\ \citenamefont {Bolhuis}}]{lechnerNonlinearReactionCoordinate2010}%
  \BibitemOpen
  \bibfield  {author} {\bibinfo {author} {\bibfnamefont {W.}~\bibnamefont {Lechner}}, \bibinfo {author} {\bibfnamefont {J.}~\bibnamefont {Rogal}}, \bibinfo {author} {\bibfnamefont {J.}~\bibnamefont {Juraszek}}, \bibinfo {author} {\bibfnamefont {B.}~\bibnamefont {Ensing}},\ and\ \bibinfo {author} {\bibfnamefont {P.~G.}\ \bibnamefont {Bolhuis}},\ }\href {https://doi.org/10.1063/1.3491818} {\bibfield  {journal} {\bibinfo  {journal} {The Journal of Chemical Physics}\ }\textbf {\bibinfo {volume} {133}},\ \bibinfo {pages} {174110} (\bibinfo {year} {2010})}\BibitemShut {NoStop}%
\end{thebibliography}%


\begin{thebibliography}{40}%
\makeatletter
\providecommand \@ifxundefined [1]{%
 \@ifx{#1\undefined}
}%
\providecommand \@ifnum [1]{%
 \ifnum #1\expandafter \@firstoftwo
 \else \expandafter \@secondoftwo
 \fi
}%
\providecommand \@ifx [1]{%
 \ifx #1\expandafter \@firstoftwo
 \else \expandafter \@secondoftwo
 \fi
}%
\providecommand \natexlab [1]{#1}%
\providecommand \enquote  [1]{``#1''}%
\providecommand \bibnamefont  [1]{#1}%
\providecommand \bibfnamefont [1]{#1}%
\providecommand \citenamefont [1]{#1}%
\providecommand \href@noop [0]{\@secondoftwo}%
\providecommand \href [0]{\begingroup \@sanitize@url \@href}%
\providecommand \@href[1]{\@@startlink{#1}\@@href}%
\providecommand \@@href[1]{\endgroup#1\@@endlink}%
\providecommand \@sanitize@url [0]{\catcode `\\12\catcode `\$12\catcode `\&12\catcode `\#12\catcode `\^12\catcode `\_12\catcode `\%12\relax}%
\providecommand \@@startlink[1]{}%
\providecommand \@@endlink[0]{}%
\providecommand \url  [0]{\begingroup\@sanitize@url \@url }%
\providecommand \@url [1]{\endgroup\@href {#1}{\urlprefix }}%
\providecommand \urlprefix  [0]{URL }%
\providecommand \Eprint [0]{\href }%
\providecommand \doibase [0]{https://doi.org/}%
\providecommand \selectlanguage [0]{\@gobble}%
\providecommand \bibinfo  [0]{\@secondoftwo}%
\providecommand \bibfield  [0]{\@secondoftwo}%
\providecommand \translation [1]{[#1]}%
\providecommand \BibitemOpen [0]{}%
\providecommand \bibitemStop [0]{}%
\providecommand \bibitemNoStop [0]{.\EOS\space}%
\providecommand \EOS [0]{\spacefactor3000\relax}%
\providecommand \BibitemShut  [1]{\csname bibitem#1\endcsname}%
\let\auto@bib@innerbib\@empty
\bibitem [{\citenamefont {Chandler}(1998)}]{Chandler1998}%
  \BibitemOpen
  \bibfield  {author} {\bibinfo {author} {\bibfnamefont {D.}~\bibnamefont {Chandler}},\ }\bibinfo {title} {Classical and quantum dynamics in condensed phase simulations}\ (\bibinfo  {publisher} {World Scientific},\ \bibinfo {year} {1998})\ Chap.\ \bibinfo {chapter} {Barrier crossings: classical theory of rare but important events}, pp.\ \bibinfo {pages} {3--23}\BibitemShut {NoStop}%
\bibitem [{\citenamefont {Bolhuis}\ \emph {et~al.}(2000)\citenamefont {Bolhuis}, \citenamefont {Dellago},\ and\ \citenamefont {Chandler}}]{Bolhuis2000}%
  \BibitemOpen
  \bibfield  {author} {\bibinfo {author} {\bibfnamefont {P.~G.}\ \bibnamefont {Bolhuis}}, \bibinfo {author} {\bibfnamefont {C.}~\bibnamefont {Dellago}},\ and\ \bibinfo {author} {\bibfnamefont {D.}~\bibnamefont {Chandler}},\ }\href {https://doi.org/10.1073/pnas.100127697} {\bibfield  {journal} {\bibinfo  {journal} {Proceedings of the National Academy of Sciences}\ }\textbf {\bibinfo {volume} {97}},\ \bibinfo {pages} {5877–5882} (\bibinfo {year} {2000})}\BibitemShut {NoStop}%
\bibitem [{\citenamefont {E}\ and\ \citenamefont {Vanden-Eijnden}(2010)}]{E2010}%
  \BibitemOpen
  \bibfield  {author} {\bibinfo {author} {\bibfnamefont {W.}~\bibnamefont {E}}\ and\ \bibinfo {author} {\bibfnamefont {E.}~\bibnamefont {Vanden-Eijnden}},\ }\href {https://doi.org/10.1146/annurev.physchem.040808.090412} {\bibfield  {journal} {\bibinfo  {journal} {Annu. Rev. Phys. Chem.}\ }\textbf {\bibinfo {volume} {61}},\ \bibinfo {pages} {391} (\bibinfo {year} {2010})}\BibitemShut {NoStop}%
\bibitem [{\citenamefont {Lazzeri}\ \emph {et~al.}(2025)\citenamefont {Lazzeri}, \citenamefont {Bolhuis},\ and\ \citenamefont {Covino}}]{Lazzeri2025}%
  \BibitemOpen
  \bibfield  {author} {\bibinfo {author} {\bibfnamefont {G.}~\bibnamefont {Lazzeri}}, \bibinfo {author} {\bibfnamefont {P.~G.}\ \bibnamefont {Bolhuis}},\ and\ \bibinfo {author} {\bibfnamefont {R.}~\bibnamefont {Covino}},\ }\href {https://arxiv.org/abs/2503.21037} {\bibinfo {title} {Optimal rejection-free path sampling}} (\bibinfo {year} {2025}),\ \Eprint {https://arxiv.org/abs/2503.21037} {arXiv:2503.21037 [physics.chem-ph]} \BibitemShut {NoStop}%
\bibitem [{\citenamefont {Dellago}\ \emph {et~al.}(1998)\citenamefont {Dellago}, \citenamefont {Bolhuis}, \citenamefont {Csajka},\ and\ \citenamefont {Chandler}}]{Dellago1998}%
  \BibitemOpen
  \bibfield  {author} {\bibinfo {author} {\bibfnamefont {C.}~\bibnamefont {Dellago}}, \bibinfo {author} {\bibfnamefont {P.~G.}\ \bibnamefont {Bolhuis}}, \bibinfo {author} {\bibfnamefont {F.~S.}\ \bibnamefont {Csajka}},\ and\ \bibinfo {author} {\bibfnamefont {D.}~\bibnamefont {Chandler}},\ }\href {https://doi.org/10.1063/1.475562} {\bibfield  {journal} {\bibinfo  {journal} {J. Chem. Phys.}\ }\textbf {\bibinfo {volume} {108}},\ \bibinfo {pages} {1964} (\bibinfo {year} {1998})}\BibitemShut {NoStop}%
\bibitem [{\citenamefont {Dellago}\ \emph {et~al.}(2002)\citenamefont {Dellago}, \citenamefont {Bolhuis},\ and\ \citenamefont {Geissler}}]{Dellago2002}%
  \BibitemOpen
  \bibfield  {author} {\bibinfo {author} {\bibfnamefont {C.}~\bibnamefont {Dellago}}, \bibinfo {author} {\bibfnamefont {P.~G.}\ \bibnamefont {Bolhuis}},\ and\ \bibinfo {author} {\bibfnamefont {P.~L.}\ \bibnamefont {Geissler}},\ }\href@noop {} {\bibfield  {journal} {\bibinfo  {journal} {Adv. Chem. Phys.}\ }\textbf {\bibinfo {volume} {123}},\ \bibinfo {pages} {1} (\bibinfo {year} {2002})}\BibitemShut {NoStop}%
\bibitem [{\citenamefont {Bolhuis}\ \emph {et~al.}(2002)\citenamefont {Bolhuis}, \citenamefont {Chandler}, \citenamefont {Dellago},\ and\ \citenamefont {Geissler}}]{Bolhuis2002}%
  \BibitemOpen
  \bibfield  {author} {\bibinfo {author} {\bibfnamefont {P.~G.}\ \bibnamefont {Bolhuis}}, \bibinfo {author} {\bibfnamefont {D.}~\bibnamefont {Chandler}}, \bibinfo {author} {\bibfnamefont {C.}~\bibnamefont {Dellago}},\ and\ \bibinfo {author} {\bibfnamefont {P.}~\bibnamefont {Geissler}},\ }\href {https://doi.org/10.1146/annurev.physchem.53.082301.113146} {\bibfield  {journal} {\bibinfo  {journal} {Ann. Rev. Phys. Chem.}\ }\textbf {\bibinfo {volume} {53}},\ \bibinfo {pages} {291} (\bibinfo {year} {2002})}\BibitemShut {NoStop}%
\bibitem [{\citenamefont {Peters}\ and\ \citenamefont {Trout}(2006)}]{Peters2006}%
  \BibitemOpen
  \bibfield  {author} {\bibinfo {author} {\bibfnamefont {B.}~\bibnamefont {Peters}}\ and\ \bibinfo {author} {\bibfnamefont {B.~L.}\ \bibnamefont {Trout}},\ }\href@noop {} {\bibfield  {journal} {\bibinfo  {journal} {J. Chem. Phys.}\ }\textbf {\bibinfo {volume} {125}},\ \bibinfo {pages} {054108} (\bibinfo {year} {2006})}\BibitemShut {NoStop}%
\bibitem [{\citenamefont {Lechner}\ \emph {et~al.}(2010)\citenamefont {Lechner}, \citenamefont {Rogal}, \citenamefont {Juraszek}, \citenamefont {Ensing},\ and\ \citenamefont {Bolhuis}}]{lechnerNonlinearReactionCoordinate2010}%
  \BibitemOpen
  \bibfield  {author} {\bibinfo {author} {\bibfnamefont {W.}~\bibnamefont {Lechner}}, \bibinfo {author} {\bibfnamefont {J.}~\bibnamefont {Rogal}}, \bibinfo {author} {\bibfnamefont {J.}~\bibnamefont {Juraszek}}, \bibinfo {author} {\bibfnamefont {B.}~\bibnamefont {Ensing}},\ and\ \bibinfo {author} {\bibfnamefont {P.~G.}\ \bibnamefont {Bolhuis}},\ }\href {https://doi.org/10.1063/1.3491818} {\bibfield  {journal} {\bibinfo  {journal} {The Journal of Chemical Physics}\ }\textbf {\bibinfo {volume} {133}},\ \bibinfo {pages} {174110} (\bibinfo {year} {2010})}\BibitemShut {NoStop}%
\bibitem [{\citenamefont {Mullen}\ \emph {et~al.}(2014)\citenamefont {Mullen}, \citenamefont {Shea},\ and\ \citenamefont {Peters}}]{Mullen2014}%
  \BibitemOpen
  \bibfield  {author} {\bibinfo {author} {\bibfnamefont {R.~G.}\ \bibnamefont {Mullen}}, \bibinfo {author} {\bibfnamefont {J.-E.}\ \bibnamefont {Shea}},\ and\ \bibinfo {author} {\bibfnamefont {B.}~\bibnamefont {Peters}},\ }\href {http://dx.doi.org/10.1021/ct4009798} {\bibfield  {journal} {\bibinfo  {journal} {Journal of Chemical Theory and Computation}\ }\textbf {\bibinfo {volume} {10}},\ \bibinfo {pages} {659–667} (\bibinfo {year} {2014})}\BibitemShut {NoStop}%
\bibitem [{\citenamefont {Peters}(2016)}]{Peters2016}%
  \BibitemOpen
  \bibfield  {author} {\bibinfo {author} {\bibfnamefont {B.}~\bibnamefont {Peters}},\ }\href {https://doi.org/10.1146/annurev-physchem-040215-112215} {\bibfield  {journal} {\bibinfo  {journal} {Annual Review of Physical Chemistry}\ }\textbf {\bibinfo {volume} {67}},\ \bibinfo {pages} {669–690} (\bibinfo {year} {2016})}\BibitemShut {NoStop}%
\bibitem [{\citenamefont {Ma}\ and\ \citenamefont {Dinner}(2005)}]{maAutomaticMethodIdentifying2005}%
  \BibitemOpen
  \bibfield  {author} {\bibinfo {author} {\bibfnamefont {A.}~\bibnamefont {Ma}}\ and\ \bibinfo {author} {\bibfnamefont {A.~R.}\ \bibnamefont {Dinner}},\ }\href {https://doi.org/10.1021/jp045546c} {\bibfield  {journal} {\bibinfo  {journal} {The Journal of Physical Chemistry B}\ }\textbf {\bibinfo {volume} {109}},\ \bibinfo {pages} {6769} (\bibinfo {year} {2005})}\BibitemShut {NoStop}%
\bibitem [{\citenamefont {van Erp}\ \emph {et~al.}(2016)\citenamefont {van Erp}, \citenamefont {Moqadam}, \citenamefont {Riccardi},\ and\ \citenamefont {Lervik}}]{vanErp2016}%
  \BibitemOpen
  \bibfield  {author} {\bibinfo {author} {\bibfnamefont {T.~S.}\ \bibnamefont {van Erp}}, \bibinfo {author} {\bibfnamefont {M.}~\bibnamefont {Moqadam}}, \bibinfo {author} {\bibfnamefont {E.}~\bibnamefont {Riccardi}},\ and\ \bibinfo {author} {\bibfnamefont {A.}~\bibnamefont {Lervik}},\ }\href {https://doi.org/10.1021/acs.jctc.6b00642} {\bibfield  {journal} {\bibinfo  {journal} {Journal of Chemical Theory and Computation}\ }\textbf {\bibinfo {volume} {12}},\ \bibinfo {pages} {5398–5410} (\bibinfo {year} {2016})}\BibitemShut {NoStop}%
\bibitem [{\citenamefont {Khoo}\ \emph {et~al.}(2018)\citenamefont {Khoo}, \citenamefont {Lu},\ and\ \citenamefont {Ying}}]{Khoo2018}%
  \BibitemOpen
  \bibfield  {author} {\bibinfo {author} {\bibfnamefont {Y.}~\bibnamefont {Khoo}}, \bibinfo {author} {\bibfnamefont {J.}~\bibnamefont {Lu}},\ and\ \bibinfo {author} {\bibfnamefont {L.}~\bibnamefont {Ying}},\ }\href {http://dx.doi.org/10.1007/s40687-018-0160-2} {\bibfield  {journal} {\bibinfo  {journal} {Research in the Mathematical Sciences}\ }\textbf {\bibinfo {volume} {6}},\ \bibinfo {pages} {1} (\bibinfo {year} {2018})}\BibitemShut {NoStop}%
\bibitem [{\citenamefont {Frassek}\ \emph {et~al.}(2021)\citenamefont {Frassek}, \citenamefont {Arjun},\ and\ \citenamefont {Bolhuis}}]{Frassek2021}%
  \BibitemOpen
  \bibfield  {author} {\bibinfo {author} {\bibfnamefont {M.}~\bibnamefont {Frassek}}, \bibinfo {author} {\bibfnamefont {A.}~\bibnamefont {Arjun}},\ and\ \bibinfo {author} {\bibfnamefont {P.~G.}\ \bibnamefont {Bolhuis}},\ }\href {http://dx.doi.org/10.1063/5.0058639} {\bibfield  {journal} {\bibinfo  {journal} {The Journal of Chemical Physics}\ }\textbf {\bibinfo {volume} {155}},\ \bibinfo {pages} {064103} (\bibinfo {year} {2021})}\BibitemShut {NoStop}%
\bibitem [{\citenamefont {Mori}\ \emph {et~al.}(2020)\citenamefont {Mori}, \citenamefont {Okazaki}, \citenamefont {Mori}, \citenamefont {Kim},\ and\ \citenamefont {Matubayasi}}]{Mori2020}%
  \BibitemOpen
  \bibfield  {author} {\bibinfo {author} {\bibfnamefont {Y.}~\bibnamefont {Mori}}, \bibinfo {author} {\bibfnamefont {K.-i.}\ \bibnamefont {Okazaki}}, \bibinfo {author} {\bibfnamefont {T.}~\bibnamefont {Mori}}, \bibinfo {author} {\bibfnamefont {K.}~\bibnamefont {Kim}},\ and\ \bibinfo {author} {\bibfnamefont {N.}~\bibnamefont {Matubayasi}},\ }\bibfield  {journal} {\bibinfo  {journal} {The Journal of Chemical Physics}\ }\textbf {\bibinfo {volume} {153}},\ \href {https://doi.org/10.1063/5.0009066} {10.1063/5.0009066} (\bibinfo {year} {2020})\BibitemShut {NoStop}%
\bibitem [{\citenamefont {Wang}\ \emph {et~al.}(2020)\citenamefont {Wang}, \citenamefont {Lamim~Ribeiro},\ and\ \citenamefont {Tiwary}}]{Wang2020}%
  \BibitemOpen
  \bibfield  {author} {\bibinfo {author} {\bibfnamefont {Y.}~\bibnamefont {Wang}}, \bibinfo {author} {\bibfnamefont {J.~M.}\ \bibnamefont {Lamim~Ribeiro}},\ and\ \bibinfo {author} {\bibfnamefont {P.}~\bibnamefont {Tiwary}},\ }\href {https://doi.org/10.1016/j.sbi.2019.12.016} {\bibfield  {journal} {\bibinfo  {journal} {Current Opinion in Structural Biology}\ }\textbf {\bibinfo {volume} {61}},\ \bibinfo {pages} {139–145} (\bibinfo {year} {2020})}\BibitemShut {NoStop}%
\bibitem [{\citenamefont {Rotskoff}\ \emph {et~al.}(2022)\citenamefont {Rotskoff}, \citenamefont {Mitchell},\ and\ \citenamefont {Vanden-Eijnden}}]{Rotskoff2022}%
  \BibitemOpen
  \bibfield  {author} {\bibinfo {author} {\bibfnamefont {G.~M.}\ \bibnamefont {Rotskoff}}, \bibinfo {author} {\bibfnamefont {A.~R.}\ \bibnamefont {Mitchell}},\ and\ \bibinfo {author} {\bibfnamefont {E.}~\bibnamefont {Vanden-Eijnden}},\ }\href@noop {} {\bibfield  {journal} {\bibinfo  {journal} {Proceedings of the 2nd Mathematical and Scientific Machine Learning Conference, PMLR}\ }\textbf {\bibinfo {volume} {145}},\ \bibinfo {pages} {757} (\bibinfo {year} {2022})}\BibitemShut {NoStop}%
\bibitem [{\citenamefont {Jung}\ \emph {et~al.}(2023)\citenamefont {Jung}, \citenamefont {Covino}, \citenamefont {Arjun}, \citenamefont {Leitold}, \citenamefont {Dellago}, \citenamefont {Bolhuis},\ and\ \citenamefont {Hummer}}]{Jung2023}%
  \BibitemOpen
  \bibfield  {author} {\bibinfo {author} {\bibfnamefont {H.}~\bibnamefont {Jung}}, \bibinfo {author} {\bibfnamefont {R.}~\bibnamefont {Covino}}, \bibinfo {author} {\bibfnamefont {A.}~\bibnamefont {Arjun}}, \bibinfo {author} {\bibfnamefont {C.}~\bibnamefont {Leitold}}, \bibinfo {author} {\bibfnamefont {C.}~\bibnamefont {Dellago}}, \bibinfo {author} {\bibfnamefont {P.~G.}\ \bibnamefont {Bolhuis}},\ and\ \bibinfo {author} {\bibfnamefont {G.}~\bibnamefont {Hummer}},\ }\href {https://doi.org/10.1038/s43588-023-00428-z} {\bibfield  {journal} {\bibinfo  {journal} {Nat. Comput. Sci.}\ }\textbf {\bibinfo {volume} {3}},\ \bibinfo {pages} {334–345} (\bibinfo {year} {2023})}\BibitemShut {NoStop}%
\bibitem [{\citenamefont {Lazzeri}\ \emph {et~al.}(2023)\citenamefont {Lazzeri}, \citenamefont {Jung}, \citenamefont {Bolhuis},\ and\ \citenamefont {Covino}}]{lazzeriMolecularFreeEnergies2023a}%
  \BibitemOpen
  \bibfield  {author} {\bibinfo {author} {\bibfnamefont {G.}~\bibnamefont {Lazzeri}}, \bibinfo {author} {\bibfnamefont {H.}~\bibnamefont {Jung}}, \bibinfo {author} {\bibfnamefont {P.~G.}\ \bibnamefont {Bolhuis}},\ and\ \bibinfo {author} {\bibfnamefont {R.}~\bibnamefont {Covino}},\ }\href {https://doi.org/10.1021/acs.jctc.3c00821} {\bibfield  {journal} {\bibinfo  {journal} {Journal of Chemical Theory and Computation}\ }\textbf {\bibinfo {volume} {19}},\ \bibinfo {pages} {9060–9076} (\bibinfo {year} {2023})}\BibitemShut {NoStop}%
\bibitem [{\citenamefont {Kang}\ \emph {et~al.}(2024)\citenamefont {Kang}, \citenamefont {Trizio},\ and\ \citenamefont {Parrinello}}]{kangComputingCommittorCommittor2024}%
  \BibitemOpen
  \bibfield  {author} {\bibinfo {author} {\bibfnamefont {P.}~\bibnamefont {Kang}}, \bibinfo {author} {\bibfnamefont {E.}~\bibnamefont {Trizio}},\ and\ \bibinfo {author} {\bibfnamefont {M.}~\bibnamefont {Parrinello}},\ }\href {https://doi.org/10.1038/s43588-024-00645-0} {\bibfield  {journal} {\bibinfo  {journal} {Nature Computational Science}\ }\textbf {\bibinfo {volume} {4}},\ \bibinfo {pages} {451} (\bibinfo {year} {2024})}\BibitemShut {NoStop}%
\bibitem [{\citenamefont {Chen}\ \emph {et~al.}(2023)\citenamefont {Chen}, \citenamefont {Roux},\ and\ \citenamefont {Chipot}}]{Chen2023}%
  \BibitemOpen
  \bibfield  {author} {\bibinfo {author} {\bibfnamefont {H.}~\bibnamefont {Chen}}, \bibinfo {author} {\bibfnamefont {B.}~\bibnamefont {Roux}},\ and\ \bibinfo {author} {\bibfnamefont {C.}~\bibnamefont {Chipot}},\ }\href {https://doi.org/10.1021/acs.jctc.3c00028} {\bibfield  {journal} {\bibinfo  {journal} {Journal of Chemical Theory and Computation}\ }\textbf {\bibinfo {volume} {19}},\ \bibinfo {pages} {4414–4426} (\bibinfo {year} {2023})}\BibitemShut {NoStop}%
\bibitem [{\citenamefont {Mitchell}\ and\ \citenamefont {Rotskoff}(2024)}]{Mitchell2024}%
  \BibitemOpen
  \bibfield  {author} {\bibinfo {author} {\bibfnamefont {A.~R.}\ \bibnamefont {Mitchell}}\ and\ \bibinfo {author} {\bibfnamefont {G.~M.}\ \bibnamefont {Rotskoff}},\ }\href {https://doi.org/10.1021/acs.jctc.4c00997} {\bibfield  {journal} {\bibinfo  {journal} {Journal of Chemical Theory and Computation}\ }\textbf {\bibinfo {volume} {20}},\ \bibinfo {pages} {9378–9393} (\bibinfo {year} {2024})}\BibitemShut {NoStop}%
\bibitem [{\citenamefont {Megías}\ \emph {et~al.}(2025)\citenamefont {Megías}, \citenamefont {Contreras~Arredondo}, \citenamefont {Chen}, \citenamefont {Tang}, \citenamefont {Roux},\ and\ \citenamefont {Chipot}}]{Megas2025}%
  \BibitemOpen
  \bibfield  {author} {\bibinfo {author} {\bibfnamefont {A.}~\bibnamefont {Megías}}, \bibinfo {author} {\bibfnamefont {S.}~\bibnamefont {Contreras~Arredondo}}, \bibinfo {author} {\bibfnamefont {C.~G.}\ \bibnamefont {Chen}}, \bibinfo {author} {\bibfnamefont {C.}~\bibnamefont {Tang}}, \bibinfo {author} {\bibfnamefont {B.}~\bibnamefont {Roux}},\ and\ \bibinfo {author} {\bibfnamefont {C.}~\bibnamefont {Chipot}},\ }\href {http://dx.doi.org/10.1038/s43588-025-00828-3} {\bibfield  {journal} {\bibinfo  {journal} {Nature Computational Science}\ } (\bibinfo {year} {2025})}\BibitemShut {NoStop}%
\bibitem [{\citenamefont {van Erp}\ \emph {et~al.}(2003)\citenamefont {van Erp}, \citenamefont {Moroni},\ and\ \citenamefont {Bolhuis}}]{vanErp2003}%
  \BibitemOpen
  \bibfield  {author} {\bibinfo {author} {\bibfnamefont {T.~S.}\ \bibnamefont {van Erp}}, \bibinfo {author} {\bibfnamefont {D.}~\bibnamefont {Moroni}},\ and\ \bibinfo {author} {\bibfnamefont {P.~G.}\ \bibnamefont {Bolhuis}},\ }\href@noop {} {\bibfield  {journal} {\bibinfo  {journal} {J. Chem. Phys.}\ }\textbf {\bibinfo {volume} {118}},\ \bibinfo {pages} {7762} (\bibinfo {year} {2003})}\BibitemShut {NoStop}%
\bibitem [{\citenamefont {Cabriolu}\ \emph {et~al.}(2017)\citenamefont {Cabriolu}, \citenamefont {Refsnes}, \citenamefont {Bolhuis},\ and\ \citenamefont {van Erp}}]{Cabriolu2017}%
  \BibitemOpen
  \bibfield  {author} {\bibinfo {author} {\bibfnamefont {R.}~\bibnamefont {Cabriolu}}, \bibinfo {author} {\bibfnamefont {K.~M.~S.}\ \bibnamefont {Refsnes}}, \bibinfo {author} {\bibfnamefont {P.~G.}\ \bibnamefont {Bolhuis}},\ and\ \bibinfo {author} {\bibfnamefont {T.~S.}\ \bibnamefont {van Erp}},\ }\href {https://doi.org/10.1063/1.4989844} {\bibfield  {journal} {\bibinfo  {journal} {J. Chem. Phys.}\ }\textbf {\bibinfo {volume} {147}},\ \bibinfo {pages} {152722} (\bibinfo {year} {2017})}\BibitemShut {NoStop}%
\bibitem [{\citenamefont {Swenson}\ \emph {et~al.}(2019)\citenamefont {Swenson}, \citenamefont {Prinz}, \citenamefont {Noe}, \citenamefont {Chodera},\ and\ \citenamefont {Bolhuis}}]{swensonOpenPathSamplingPythonFramework2019}%
  \BibitemOpen
  \bibfield  {author} {\bibinfo {author} {\bibfnamefont {D.~W.~H.}\ \bibnamefont {Swenson}}, \bibinfo {author} {\bibfnamefont {J.-H.}\ \bibnamefont {Prinz}}, \bibinfo {author} {\bibfnamefont {F.}~\bibnamefont {Noe}}, \bibinfo {author} {\bibfnamefont {J.~D.}\ \bibnamefont {Chodera}},\ and\ \bibinfo {author} {\bibfnamefont {P.~G.}\ \bibnamefont {Bolhuis}},\ }\href {https://doi.org/10.1021/acs.jctc.8b00626} {\bibfield  {journal} {\bibinfo  {journal} {Journal of Chemical Theory and Computation}\ }\textbf {\bibinfo {volume} {15}},\ \bibinfo {pages} {813} (\bibinfo {year} {2019})}\BibitemShut {NoStop}%
\bibitem [{\citenamefont {Rogal}\ \emph {et~al.}(2010)\citenamefont {Rogal}, \citenamefont {Lechner}, \citenamefont {Juraszek}, \citenamefont {Ensing},\ and\ \citenamefont {Bolhuis}}]{Rogal2010}%
  \BibitemOpen
  \bibfield  {author} {\bibinfo {author} {\bibfnamefont {J.}~\bibnamefont {Rogal}}, \bibinfo {author} {\bibfnamefont {W.}~\bibnamefont {Lechner}}, \bibinfo {author} {\bibfnamefont {J.}~\bibnamefont {Juraszek}}, \bibinfo {author} {\bibfnamefont {B.}~\bibnamefont {Ensing}},\ and\ \bibinfo {author} {\bibfnamefont {P.~G.}\ \bibnamefont {Bolhuis}},\ }\href {https://doi.org/10.1063/1.3491817} {\bibfield  {journal} {\bibinfo  {journal} {The Journal of Chemical Physics}\ }\textbf {\bibinfo {volume} {133}},\ \bibinfo {pages} {174109} (\bibinfo {year} {2010})}\BibitemShut {NoStop}%
\bibitem [{\citenamefont {Ferrenberg}\ and\ \citenamefont {Swendsen}(1989)}]{ferrenbergOptimizedMonteCarlo1989}%
  \BibitemOpen
  \bibfield  {author} {\bibinfo {author} {\bibfnamefont {A.~M.}\ \bibnamefont {Ferrenberg}}\ and\ \bibinfo {author} {\bibfnamefont {R.~H.}\ \bibnamefont {Swendsen}},\ }\href {https://doi.org/10.1103/PhysRevLett.63.1195} {\bibfield  {journal} {\bibinfo  {journal} {Physical Review Letters}\ }\textbf {\bibinfo {volume} {63}},\ \bibinfo {pages} {1195} (\bibinfo {year} {1989})}\BibitemShut {NoStop}%
\bibitem [{\citenamefont {Shirts}\ and\ \citenamefont {Chodera}(2008)}]{shirtsStatisticallyOptimalAnalysis2008}%
  \BibitemOpen
  \bibfield  {author} {\bibinfo {author} {\bibfnamefont {M.~R.}\ \bibnamefont {Shirts}}\ and\ \bibinfo {author} {\bibfnamefont {J.~D.}\ \bibnamefont {Chodera}},\ }\href {https://doi.org/10.1063/1.2978177} {\bibfield  {journal} {\bibinfo  {journal} {The Journal of Chemical Physics}\ }\textbf {\bibinfo {volume} {129}},\ \bibinfo {pages} {124105} (\bibinfo {year} {2008})}\BibitemShut {NoStop}%
\bibitem [{Note1()}]{Note1}%
  \BibitemOpen
  \bibinfo {note} {This term is reminiscent of the variational principle for solving the backward Kolmogorov equation in the overdamped regime \cite {E2010,kangComputingCommittorCommittor2024}.}\BibitemShut {Stop}%
\bibitem [{\citenamefont {Covino}\ \emph {et~al.}(2019)\citenamefont {Covino}, \citenamefont {Woodside}, \citenamefont {Hummer}, \citenamefont {Szabo},\ and\ \citenamefont {Cossio}}]{covinoMolecularFreeEnergy2019}%
  \BibitemOpen
  \bibfield  {author} {\bibinfo {author} {\bibfnamefont {R.}~\bibnamefont {Covino}}, \bibinfo {author} {\bibfnamefont {M.~T.}\ \bibnamefont {Woodside}}, \bibinfo {author} {\bibfnamefont {G.}~\bibnamefont {Hummer}}, \bibinfo {author} {\bibfnamefont {A.}~\bibnamefont {Szabo}},\ and\ \bibinfo {author} {\bibfnamefont {P.}~\bibnamefont {Cossio}},\ }\href {https://doi.org/10.1063/1.5118362} {\bibfield  {journal} {\bibinfo  {journal} {The Journal of Chemical Physics}\ }\textbf {\bibinfo {volume} {151}},\ \bibinfo {pages} {154115} (\bibinfo {year} {2019})}\BibitemShut {NoStop}%
\bibitem [{\citenamefont {Wolfe}\ \emph {et~al.}(1975)\citenamefont {Wolfe}, \citenamefont {Schlegel}, \citenamefont {Csizmadia},\ and\ \citenamefont {Bernardi}}]{Wolfe1975}%
  \BibitemOpen
  \bibfield  {author} {\bibinfo {author} {\bibfnamefont {S.}~\bibnamefont {Wolfe}}, \bibinfo {author} {\bibfnamefont {H.~B.}\ \bibnamefont {Schlegel}}, \bibinfo {author} {\bibfnamefont {I.~G.}\ \bibnamefont {Csizmadia}},\ and\ \bibinfo {author} {\bibfnamefont {F.}~\bibnamefont {Bernardi}},\ }\href {https://doi.org/10.1021/ja00841a005} {\bibfield  {journal} {\bibinfo  {journal} {Journal of the American Chemical Society}\ }\textbf {\bibinfo {volume} {97}},\ \bibinfo {pages} {2020–2024} (\bibinfo {year} {1975})}\BibitemShut {NoStop}%
\bibitem [{\citenamefont {Quapp}(2005)}]{quappGrowingStringMethod2005}%
  \BibitemOpen
  \bibfield  {author} {\bibinfo {author} {\bibfnamefont {W.}~\bibnamefont {Quapp}},\ }\href {https://doi.org/10.1063/1.1885467} {\bibfield  {journal} {\bibinfo  {journal} {The Journal of Chemical Physics}\ }\textbf {\bibinfo {volume} {122}},\ \bibinfo {pages} {174106} (\bibinfo {year} {2005})}\BibitemShut {NoStop}%
\bibitem [{\citenamefont {Moghaddam}\ \emph {et~al.}(2011)\citenamefont {Moghaddam}, \citenamefont {Yang}, \citenamefont {Rekharsky}, \citenamefont {Ko}, \citenamefont {Kim}, \citenamefont {Inoue},\ and\ \citenamefont {Gilson}}]{Moghaddam2011}%
  \BibitemOpen
  \bibfield  {author} {\bibinfo {author} {\bibfnamefont {S.}~\bibnamefont {Moghaddam}}, \bibinfo {author} {\bibfnamefont {C.}~\bibnamefont {Yang}}, \bibinfo {author} {\bibfnamefont {M.}~\bibnamefont {Rekharsky}}, \bibinfo {author} {\bibfnamefont {Y.~H.}\ \bibnamefont {Ko}}, \bibinfo {author} {\bibfnamefont {K.}~\bibnamefont {Kim}}, \bibinfo {author} {\bibfnamefont {Y.}~\bibnamefont {Inoue}},\ and\ \bibinfo {author} {\bibfnamefont {M.~K.}\ \bibnamefont {Gilson}},\ }\href {https://doi.org/10.1021/ja109904u} {\bibfield  {journal} {\bibinfo  {journal} {Journal of the American Chemical Society}\ }\textbf {\bibinfo {volume} {133}},\ \bibinfo {pages} {3570–3581} (\bibinfo {year} {2011})}\BibitemShut {NoStop}%
\bibitem [{\citenamefont {Fenley}\ \emph {et~al.}(2014)\citenamefont {Fenley}, \citenamefont {Henriksen}, \citenamefont {Muddana},\ and\ \citenamefont {Gilson}}]{Fenley2014}%
  \BibitemOpen
  \bibfield  {author} {\bibinfo {author} {\bibfnamefont {A.~T.}\ \bibnamefont {Fenley}}, \bibinfo {author} {\bibfnamefont {N.~M.}\ \bibnamefont {Henriksen}}, \bibinfo {author} {\bibfnamefont {H.~S.}\ \bibnamefont {Muddana}},\ and\ \bibinfo {author} {\bibfnamefont {M.~K.}\ \bibnamefont {Gilson}},\ }\href {https://doi.org/10.1021/ct5004109} {\bibfield  {journal} {\bibinfo  {journal} {Journal of Chemical Theory and Computation}\ }\textbf {\bibinfo {volume} {10}},\ \bibinfo {pages} {4069–4078} (\bibinfo {year} {2014})}\BibitemShut {NoStop}%
\bibitem [{\citenamefont {Henriksen}\ \emph {et~al.}(2015)\citenamefont {Henriksen}, \citenamefont {Fenley},\ and\ \citenamefont {Gilson}}]{Henriksen2015}%
  \BibitemOpen
  \bibfield  {author} {\bibinfo {author} {\bibfnamefont {N.~M.}\ \bibnamefont {Henriksen}}, \bibinfo {author} {\bibfnamefont {A.~T.}\ \bibnamefont {Fenley}},\ and\ \bibinfo {author} {\bibfnamefont {M.~K.}\ \bibnamefont {Gilson}},\ }\href {https://doi.org/10.1021/acs.jctc.5b00405} {\bibfield  {journal} {\bibinfo  {journal} {Journal of Chemical Theory and Computation}\ }\textbf {\bibinfo {volume} {11}},\ \bibinfo {pages} {4377–4394} (\bibinfo {year} {2015})}\BibitemShut {NoStop}%
\bibitem [{\citenamefont {Van~Erp}(2007)}]{vanerpReactionRateCalculation2007}%
  \BibitemOpen
  \bibfield  {author} {\bibinfo {author} {\bibfnamefont {T.~S.}\ \bibnamefont {Van~Erp}},\ }\href {https://doi.org/10.1103/PhysRevLett.98.268301} {\bibfield  {journal} {\bibinfo  {journal} {Physical Review Letters}\ }\textbf {\bibinfo {volume} {98}},\ \bibinfo {pages} {268301} (\bibinfo {year} {2007})}\BibitemShut {NoStop}%
\bibitem [{\citenamefont {Bolhuis}(2008)}]{bolhuisRareEventsMultiple2008}%
  \BibitemOpen
  \bibfield  {author} {\bibinfo {author} {\bibfnamefont {P.~G.}\ \bibnamefont {Bolhuis}},\ }\href {https://doi.org/10.1063/1.2976011} {\bibfield  {journal} {\bibinfo  {journal} {The Journal of Chemical Physics}\ }\textbf {\bibinfo {volume} {129}},\ \bibinfo {pages} {114108} (\bibinfo {year} {2008})}\BibitemShut {NoStop}%
\bibitem [{\citenamefont {Roet}\ \emph {et~al.}(2022)\citenamefont {Roet}, \citenamefont {Zhang},\ and\ \citenamefont {van Erp}}]{Roet2022}%
  \BibitemOpen
  \bibfield  {author} {\bibinfo {author} {\bibfnamefont {S.}~\bibnamefont {Roet}}, \bibinfo {author} {\bibfnamefont {D.~T.}\ \bibnamefont {Zhang}},\ and\ \bibinfo {author} {\bibfnamefont {T.~S.}\ \bibnamefont {van Erp}},\ }\href {https://doi.org/10.1021/acs.jpca.2c06004} {\bibfield  {journal} {\bibinfo  {journal} {The Journal of Physical Chemistry A}\ }\textbf {\bibinfo {volume} {126}},\ \bibinfo {pages} {8878–8886} (\bibinfo {year} {2022})}\BibitemShut {NoStop}%
\end{thebibliography}%
